%% file: main.tex
\documentclass[conference,compsoc]{IEEEtran}
\PassOptionsToPackage{hyperfootnotes=false}{hyperref}
\usepackage[T1]{fontenc}
\usepackage[utf8]{inputenc}
\usepackage{cite}
\usepackage{xurl}
\usepackage{xspace}
\usepackage[table]{xcolor}
\usepackage{enumitem}
\usepackage{array}
\usepackage{booktabs}
\usepackage{multirow}
\usepackage{amsmath}
\usepackage{graphicx}
\usepackage{tikz}
\usepackage{acro}
\usepackage[colorlinks=true,linkcolor=cyan,citecolor=cyan,urlcolor=cyan]{hyperref}
\usepackage{mdframed}
\mdfdefinestyle{insightstyle}{linewidth=0.7pt,innertopmargin=5pt,innerbottommargin=5pt,innerleftmargin=7pt,innerrightmargin=7pt,skipabove=\smallskipamount,skipbelow=\smallskipamount}
\newenvironment{insight}[1]{\begin{mdframed}[style=insightstyle]\noindent\textbf{Insight\,#1:}\enspace\ignorespaces}{\end{mdframed}}
\mdfdefinestyle{remarksty}{linewidth=0.5pt,linecolor=black!35,innertopmargin=5pt,innerbottommargin=5pt,innerleftmargin=7pt,innerrightmargin=7pt,skipabove=\medskipamount,skipbelow=\medskipamount,backgroundcolor=black!2}
\newenvironment{designremark}[1]{\begin{mdframed}[style=remarksty]\noindent\textbf{\underline{Q}: #1}\par\smallskip\noindent\textbf{\underline{A}:}\enspace\ignorespaces}{\end{mdframed}}
\usetikzlibrary{arrows.meta,calc}
\newcommand{\circled}[1]{\tikz[baseline=(c.base)]{\node[shape=circle,draw,inner sep=0.5pt,line width=0.5pt,font=\small](c){#1};}}
\newcommand{\circledtab}[1]{\tikz[baseline=(c.base)]{\node[shape=circle,draw,inner sep=0.2pt,line width=0.4pt,font=\tiny](c){#1};}}
\DeclareAcronym{llm}{
  short = LLM,
  long = large language model
}
\DeclareAcronym{cvss}{
  short = CVSS,
  long = Common Vulnerability Scoring System
}
\DeclareAcronym{poc}{
  short = PoC,
  long = proof of concept
}
\DeclareAcronym{cwe}{
  short = CWE,
  long = Common Weakness Enumeration
}
\DeclareAcronym{oss}{
  short = OSS,
  long = open source software
}
\DeclareAcronym{saas}{
  short = SaaS,
  long = software as a service
}

\newcommand{\tool}{\textsc{EvoHunt}\xspace}
\newcommand{\codexgpt}{Codex\slash GPT5.4-xhigh\xspace}
\newcommand{\JudgeModel}{GPT5.4-xhigh\xspace}
\newcommand{\glmopen}{OpenCode\slash GLM5.1\xspace}
\newcommand{\codexsecurity}{OpenAI Codex Security\xspace}
\newcommand{\qwenmoe}{Qwen3.6-35B-A3B\xspace}
\newcommand{\qwendense}{Qwen3.6-27B\xspace}
\setlist[itemize]{noitemsep,nolistsep,left=0pt}
\newcommand{\empirical}[1]{#1\xspace}
\definecolor{revisionBlue}{RGB}{0,76,170}

\newcommand{\resval}[1]{#1\xspace}
\newcommand{\maskid}{\raisebox{0.4ex}{\rule{3ex}{1ex}}}
\newcommand{\ghsa}[1]{\href{https://github.com/advisories/#1}{#1}}
\xspaceaddexceptions{\%}
\definecolor{legendBlue}{RGB}{65,105,225}
\definecolor{legendOrange}{RGB}{230,145,56}
\definecolor{legendTeal}{RGB}{0,128,128}
\definecolor{legendRed}{RGB}{190,50,50}
\definecolor{legendPurple}{RGB}{130,80,180}
\definecolor{legendGray}{RGB}{120,120,120}
\DeclareRobustCommand{\LegendSwatch}[1]{%
  \begingroup
  \raisebox{0.08ex}{\textcolor{#1}{\rule{1.15ex}{1.15ex}}}%
  \endgroup
}

\newcommand{\TrainingAdvisoryCount}{\empirical{813}}
\newcommand{\TestingAdvisoryCount}{\empirical{371}}

\newcommand{\TrainingStartMonth}{\empirical{2023-01}}
\newcommand{\TrainingEndMonth}{\empirical{2025-12}}
\newcommand{\TestingStartMonth}{\empirical{2026-01}}
\newcommand{\TestingEndMonth}{\empirical{2026-04}}
\newcommand{\ConfirmedVulnerabilityCount}{\empirical{28}}
\newcommand{\GPTKnowledgeCutoff}{\resval{2025-08}}
\newcommand{\QwenKnowledgeCutoff}{\resval{2025-04}}
\newcommand{\GLMKnowledgeCutoff}{\resval{2026-01}}
\newcommand{\TrainingCriticalCount}{\empirical{181}}
\newcommand{\TrainingCriticalPercent}{\empirical{22.3\%}}
\newcommand{\TrainingHighCount}{\empirical{632}}
\newcommand{\TrainingHighPercent}{\empirical{77.7\%}}
\newcommand{\TestingCriticalCount}{\empirical{84}}
\newcommand{\TestingCriticalPercent}{\empirical{22.6\%}}
\newcommand{\TestingHighCount}{\empirical{287}}
\newcommand{\TestingHighPercent}{\empirical{77.4\%}}
\newcommand{\TrainingUniquePackageCount}{\empirical{707}}
\newcommand{\TestingUniquePackageCount}{\empirical{236}}
\newcommand{\TrainingUniqueRepoCount}{\empirical{541}}
\newcommand{\TestingUniqueRepoCount}{\empirical{201}}
\newcommand{\TestingPackageOverlapCount}{\empirical{50}}

\newcommand{\TestingRepoOverlapCount}{\empirical{51}}
\newcommand{\TestingRepoOverlapPercent}{\empirical{25.4\%}}

\newcommand{\TestingTopFiveRepoAdvisoryCount}{\empirical{63}}
\newcommand{\TestingTopFiveRepoAdvisoryPercent}{\empirical{17.0\%}}
\newcommand{\EvalMachineCount}{\resval{2}}
\newcommand{\EvalMachineCpuCores}{\resval{10}}
\newcommand{\EvalMachineMemoryGiB}{\resval{16}}
\newcommand{\EvalMacOSVersion}{\resval{15.6}}

\newcommand{\EvalCaseCount}{\resval{371}}
\newcommand{\CodexSecurityJudgedFindings}{\resval{1601}}
\newcommand{\CodexSecurityQualifiedFindings}{\resval{194}}

\newcommand{\CodexSecurityNoMatchFindings}{\resval{160}}
\newcommand{\CodexSecurityTOne}{\resval{13}}
\newcommand{\CodexSecurityTTwo}{\resval{4}}
\newcommand{\CodexSecurityTThree}{\resval{17}}
\newcommand{\CodexSecurityTargetMatch}{\resval{34}}

\newcommand{\GPTMinJudgedFindings}{\resval{317}}
\newcommand{\GPTMinQualifiedFindings}{\resval{34}}

\newcommand{\GPTMinNoMatchFindings}{\resval{28}}
\newcommand{\GPTMinTOne}{\resval{4}}
\newcommand{\GPTMinTTwo}{\resval{2}}
\newcommand{\GPTMinTThree}{\resval{0}}
\newcommand{\GPTMinTargetMatch}{\resval{6}}

\newcommand{\GPTEvolvedJudgedFindings}{\resval{442}}
\newcommand{\GPTEvolvedQualifiedFindings}{\resval{120}}

\newcommand{\GPTEvolvedNoMatchFindings}{\resval{97}}
\newcommand{\GPTEvolvedTOne}{\resval{23}}
\newcommand{\GPTEvolvedTTwo}{\resval{0}}
\newcommand{\GPTEvolvedTThree}{\resval{0}}
\newcommand{\GPTEvolvedTargetMatch}{\resval{23}}

\newcommand{\GLMMinJudgedFindings}{\resval{447}}
\newcommand{\GLMMinQualifiedFindings}{\resval{247}}

\newcommand{\GLMMinNoMatchFindings}{\resval{211}}
\newcommand{\GLMMinTOne}{\resval{18}}
\newcommand{\GLMMinTTwo}{\resval{0}}
\newcommand{\GLMMinTThree}{\resval{18}}
\newcommand{\GLMMinTargetMatch}{\resval{36}}

\newcommand{\GLMEvolvedJudgedFindings}{\resval{645}}
\newcommand{\GLMEvolvedQualifiedFindings}{\resval{412}}

\newcommand{\GLMEvolvedNoMatchFindings}{\resval{370}}
\newcommand{\GLMEvolvedTOne}{\resval{31}}
\newcommand{\GLMEvolvedTTwo}{\resval{8}}
\newcommand{\GLMEvolvedTThree}{\resval{3}}
\newcommand{\GLMEvolvedTargetMatch}{\resval{42}}

\newcommand{\QwenMoEMinJudgedFindings}{\resval{538}}
\newcommand{\QwenMoEMinQualifiedFindings}{\resval{36}}

\newcommand{\QwenMoEMinNoMatchFindings}{\resval{32}}
\newcommand{\QwenMoEMinTOne}{\resval{4}}
\newcommand{\QwenMoEMinTTwo}{\resval{0}}
\newcommand{\QwenMoEMinTThree}{\resval{0}}
\newcommand{\QwenMoEMinTargetMatch}{\resval{4}}

\newcommand{\QwenMoETransferJudgedFindings}{\resval{415}}
\newcommand{\QwenMoETransferQualifiedFindings}{\resval{42}}

\newcommand{\QwenMoETransferNoMatchFindings}{\resval{35}}
\newcommand{\QwenMoETransferTOne}{\resval{3}}
\newcommand{\QwenMoETransferTTwo}{\resval{0}}
\newcommand{\QwenMoETransferTThree}{\resval{4}}
\newcommand{\QwenMoETransferTargetMatch}{\resval{7}}

\newcommand{\QwenDenseMinJudgedFindings}{\resval{256}}
\newcommand{\QwenDenseMinQualifiedFindings}{\resval{25}}

\newcommand{\QwenDenseMinNoMatchFindings}{\resval{16}}
\newcommand{\QwenDenseMinTOne}{\resval{8}}
\newcommand{\QwenDenseMinTTwo}{\resval{0}}
\newcommand{\QwenDenseMinTThree}{\resval{1}}
\newcommand{\QwenDenseMinTargetMatch}{\resval{9}}

\newcommand{\QwenDenseTransferJudgedFindings}{\resval{235}}
\newcommand{\QwenDenseTransferQualifiedFindings}{\resval{67}}

\newcommand{\QwenDenseTransferNoMatchFindings}{\resval{48}}
\newcommand{\QwenDenseTransferTOne}{\resval{17}}
\newcommand{\QwenDenseTransferTTwo}{\resval{2}}
\newcommand{\QwenDenseTransferTThree}{\resval{0}}
\newcommand{\QwenDenseTransferTargetMatch}{\resval{19}}

\newcommand{\GLMDenseTransferJudgedFindings}{\resval{406}}
\newcommand{\GLMDenseTransferQualifiedFindings}{\resval{218}}

\newcommand{\GLMDenseTransferNoMatchFindings}{\resval{194}}
\newcommand{\GLMDenseTransferTOne}{\resval{15}}
\newcommand{\GLMDenseTransferTTwo}{\resval{2}}
\newcommand{\GLMDenseTransferTThree}{\resval{7}}
\newcommand{\GLMDenseTransferTargetMatch}{\resval{24}}
\newcommand{\GLMMoETransferJudgedFindings}{\resval{577}}
\newcommand{\GLMMoETransferQualifiedFindings}{\resval{179}}

\newcommand{\GLMMoETransferNoMatchFindings}{\resval{162}}
\newcommand{\TotalJudgedFindings}{\resval{5{,}879}}
\newcommand{\TotalQualifiedFindings}{\resval{1{,}574}}
\newcommand{\GLMMoETransferTOne}{\resval{7}}
\newcommand{\GLMMoETransferTTwo}{\resval{4}}
\newcommand{\GLMMoETransferTThree}{\resval{6}}
\newcommand{\GLMMoETransferTargetMatch}{\resval{17}}

\newcommand{\GPTMinTOneRate}{\empirical{1.1\%}}
\newcommand{\GPTEvolvedTOneRate}{\empirical{6.2\%}}

\newcommand{\CodexSecurityTargetMatchRate}{\empirical{9.2\%}}
\newcommand{\GPTMinTargetMatchRate}{\empirical{1.6\%}}
\newcommand{\GPTEvolvedTargetMatchRate}{\empirical{6.2\%}}
\newcommand{\GLMMinTargetMatchRate}{\empirical{9.7\%}}
\newcommand{\GLMEvolvedTargetMatchRate}{\empirical{11.3\%}}
\newcommand{\QwenMoEMinTargetMatchRate}{\empirical{1.1\%}}
\newcommand{\QwenMoETransferTargetMatchRate}{\empirical{1.9\%}}
\newcommand{\GLMMoETransferTargetMatchRate}{\empirical{4.6\%}}
\newcommand{\QwenDenseMinTargetMatchRate}{\empirical{2.4\%}}
\newcommand{\QwenDenseTransferTargetMatchRate}{\empirical{5.1\%}}
\newcommand{\GLMDenseTransferTargetMatchRate}{\empirical{6.5\%}}
\newcommand{\CodexSecurityQualifiedRate}{\empirical{12.1\%}}
\newcommand{\GPTMinQualifiedRate}{\empirical{10.7\%}}
\newcommand{\GPTEvolvedQualifiedRate}{\empirical{27.1\%}}
\newcommand{\GLMMinQualifiedRate}{\empirical{55.3\%}}
\newcommand{\GLMEvolvedQualifiedRate}{\empirical{63.9\%}}
\newcommand{\QwenMoEMinQualifiedRate}{\empirical{6.7\%}}
\newcommand{\QwenMoETransferQualifiedRate}{\empirical{10.1\%}}
\newcommand{\GLMMoETransferQualifiedRate}{\empirical{31.0\%}}
\newcommand{\QwenDenseMinQualifiedRate}{\empirical{9.8\%}}
\newcommand{\QwenDenseTransferQualifiedRate}{\empirical{28.5\%}}
\newcommand{\GLMDenseTransferQualifiedRate}{\empirical{53.7\%}}

\newcommand{\BenchmarkEcosystemCount}{\empirical{10}}

\newcommand{\FPRCodexDryrun}{\resval{29.4\%}}
\newcommand{\FPRCodexRun}{\resval{0.0\%}}

\newcommand{\FPRGLMDryrun}{\resval{44.4\%}}
\newcommand{\FPRGLMRun}{\resval{44.8\%}}
\newcommand{\FPRQwenAThreeBDryrun}{\resval{70.6\%}}
\newcommand{\FPRQwenAThreeBGPTRun}{\resval{44.4\%}}
\newcommand{\FPRQwenAThreeBGLMRun}{\resval{65.4\%}}
\newcommand{\FPRQwenTwentySevenBDryrun}{\resval{20.0\%}}
\newcommand{\FPRQwenTwentySevenBGPTRun}{\resval{28.6\%}}
\newcommand{\FPRQwenTwentySevenBGLMRun}{\resval{25.9\%}}
\newcommand{\BenchmarkTopCWECount}{\empirical{112}}

\newcommand{\Pmin}{P_{\emptyset}}
\newcommand{\Pgptstar}{P_{\mathrm{GPT}}^\star}
\newcommand{\Pglmstar}{P_{\mathrm{GLM}}^\star}
\newcommand{\ExpCondition}[2]{\mathrm{Exp}_{\mathrm{#1}}^{\mathrm{#2}}}
\newcommand{\Bcs}{\ExpCondition{GPT}{OCS}}
\newcommand{\Bgpt}{\ExpCondition{GPT}{Empty}}
\newcommand{\Bglm}{\ExpCondition{GLM}{Empty}}
\newcommand{\Egpt}{\ExpCondition{GPT}{Evolve}}
\newcommand{\Eglm}{\ExpCondition{GLM}{Evolve}}
\newcommand{\Ctest}{C_{\mathrm{test}}}

\newcommand{\StudentBaseline}[1]{\ExpCondition{#1}{Empty}}
\newcommand{\StudentTransfer}[1]{\ExpCondition{#1}{Tr/GPT}}
\newcommand{\StudentGLMTransfer}[1]{\ExpCondition{#1}{Tr/GLM}}

\IEEEoverridecommandlockouts
\begin{document}

\title{Transferable Self-Evolving Playbooks for Agentic Security Auditing\thanks{\textbf{Author contributions:} Liyi supervised the project; Ziyue led implementation. Ziyue and Liyi jointly contributed to research design, analysis, and writing. Maurice and Chenchen helped run experiments and check results. Maurice, Strick, and Chenchen handled manual validation and responsible disclosure. Kaihua contributed framing, interpretation, and revision.}\thanks{\textbf{Funding:} Supported by the Australian Research Council Discovery Early Career Researcher Award (ARC DECRA) under grant \href{https://rms.arc.gov.au/RMS/Report/Download/Report/1b0c8b2e-7bb0-4f2d-8f52-ad207cfbb41d/285}{DE260101642}.}}
\author{
\IEEEauthorblockN{Ziyue Wang\IEEEauthorrefmark{1},
Cheuk Wang Maurice Ng\IEEEauthorrefmark{1},
Chenchen Yu\IEEEauthorrefmark{2},
Strick Sheng\IEEEauthorrefmark{1},
Kaihua Qin\IEEEauthorrefmark{3}\IEEEauthorrefmark{4},
Liyi Zhou\IEEEauthorrefmark{1}\IEEEauthorrefmark{4}}
\IEEEauthorblockA{\IEEEauthorrefmark{1}The University of Sydney \quad
\IEEEauthorrefmark{2}Independent \quad
\IEEEauthorrefmark{3}University of Warwick \quad
\IEEEauthorrefmark{4}UC Berkeley RDI}
}
\maketitle

\begin{abstract}
An LLM agent for vulnerability discovery and validation is more than a model. It combines three components: (i) an underlying LLM for code analysis, (ii) a general-purpose agent harness, such as Codex or OpenCode, for repository navigation, tool use, context, and long-horizon execution, and (iii) an audit ``playbook'', domain-specific procedural knowledge that guides the LLM and harness toward effective vulnerability discovery. Prior work relies on human-supplied ``playbooks'' in several forms, including prompt engineering, role play, manually designed audit workflows, curated vulnerability knowledge bases (e.g., via RAG), and heuristics. This raises two research questions: (RQ1) Acquisition -- Is human curation necessary? Can playbook creation be fully automated? (RQ2) Transfer -- Can an evolved playbook transfer the audit procedure to weaker agents, improving their capability?

We present \tool, which instantiates a playbook evolution environment over open-source repositories for security auditing. Three agents drive the evolution loop: (i) an audit agent rolls out the current playbook and produces findings and evidence; (ii) an evaluator scores outcomes against ground truth; and (iii) a reviser commits updates to the playbook based on failure analysis. The playbook format is unconstrained: starting empty, \tool freely adds or removes workflows, heuristics, vulnerability knowledge, or any domain-specific content. The evolved playbook requires only minor adaptation to run under a different LLM or harness.

We evaluate \tool on \TrainingAdvisoryCount{} open-source security advisories for evolution and \TestingAdvisoryCount{} held-out advisories for testing. For acquisition, playbook evolution raises end-to-end exploits for \codexgpt{} $6{\times}$ (\GPTMinTOneRate{}$\to$\GPTEvolvedTOneRate{}), and the evolved \glmopen{} playbook surpasses \codexsecurity{} on every metric (\GLMEvolvedTargetMatchRate{} vs.\ \CodexSecurityTargetMatchRate{}), showing open-source evolution can outperform a dedicated commercial product. For transfer, the GLM-evolved playbook gives the strongest student lift (27B: \QwenDenseMinTargetMatchRate{}$\to$\GLMDenseTransferTargetMatchRate{}; A3B: \QwenMoEMinTargetMatchRate{}$\to$\GLMMoETransferTargetMatchRate{}) and yields $2.4{\times}$ more A3B matches than GPT transfer.
\end{abstract}
\acresetall

\section{Introduction}\label{sec:introduction}

\begin{figure}[!t]
\centering
\includegraphics[width=\columnwidth]{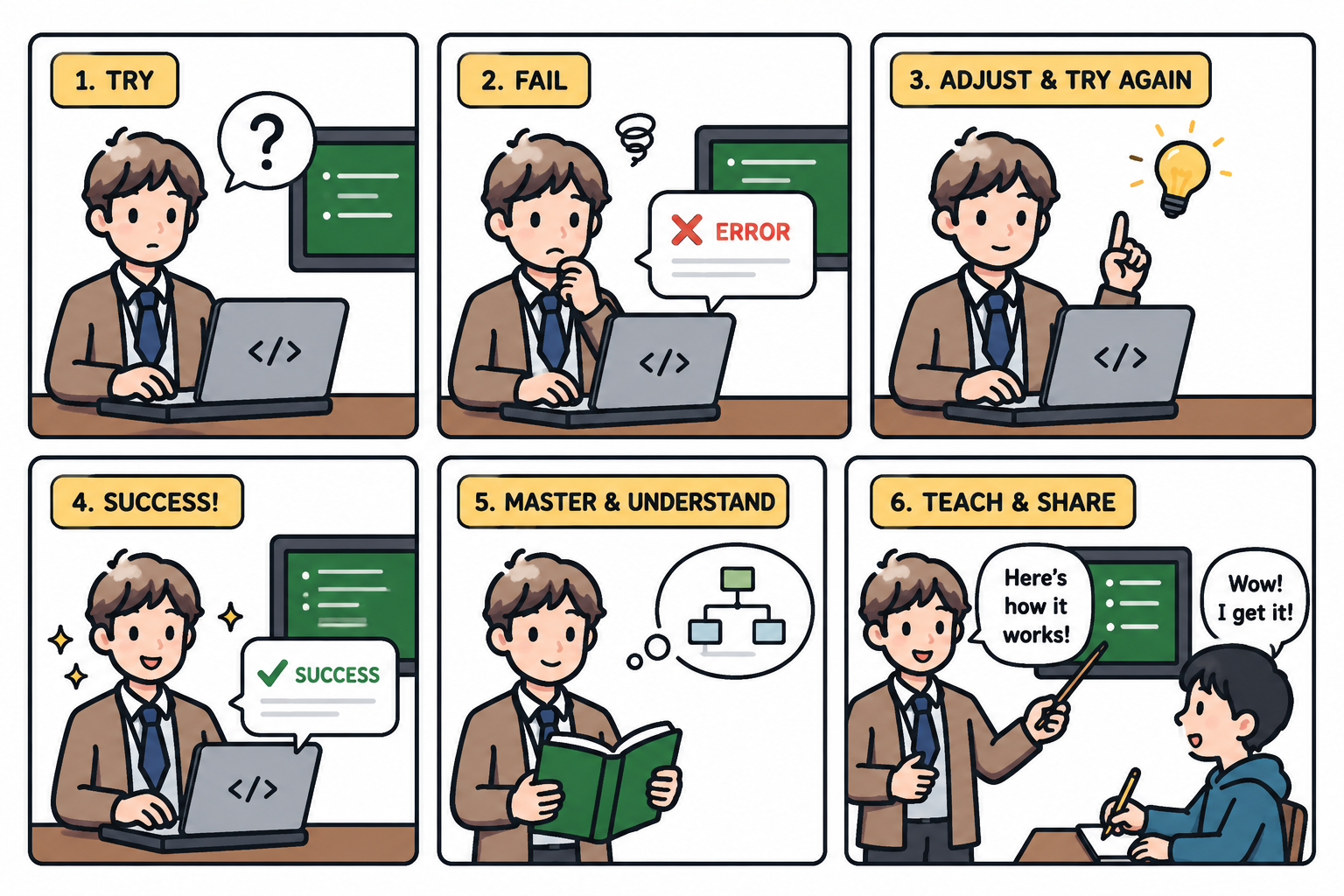}
\caption{\tool overview. A teacher agent iterates on source audit cases to revise an audit playbook \(P_s^\star\); the playbook transfers to a student agent via a lightweight adapter \(A_t\), without changing model weights or harness. Sections~\ref{sec:background}--\ref{sec:models} define background and scope; Sections~\ref{sec:methodology}--\ref{sec:implementation} formalize and implement evolution and transfer; Sections~\ref{sec:benchmark}--\ref{sec:evaluation} cover benchmark, conditions, and results; Sections~\ref{sec:related-work}--\ref{sec:conclusion} situate, discuss, and conclude.}
\label{fig:overview}
\end{figure}

\Ac{llm} agents are moving from assistants to operators that inspect code, run tools, modify files, drive browsers, and maintain multistep state over long horizons. Agentic operation is consequential for application security. Security auditing is not a single classification decision: an agent must enumerate attack surfaces, infer trust boundaries, form exploit hypotheses, trace source-to-sink flows, build or run targets, and decide whether runtime evidence supports a root cause claim. This shift is visible across recent security research: benchmarks test agents in exploit reproduction and cyber task environments~\cite{zhang2025cybench,wang2026cybergym,wang2026exploitgym}; specialized agents pair \ac{llm}s with scanners, browsers, fuzzers, or program analysis tools for penetration testing, mobile and web analysis, protocol testing, and firmware auditing~\cite{deng2024pentestgpt,wang2025a2,allen2024webrr,song2026protocolguard,ji2026firmagent}; and repository-level systems study vulnerability detection as code auditing with program context, validation, memory, or multiple agent reasoning~\cite{ding2025primevul,wang2024reposvul,guo2025repoaudit,peng2026reasonvul}. Industry deployments such as Project Glasswing, Daybreak, and Codex Security show that frontier agents are being used for vulnerability discovery, validation, and remediation~\cite{anthropic2026glasswing,openai2026daybreak,openai2026codexsecurity}. Yet progress is often framed as a race centered on model capability: use a stronger model, a larger context window, or a detailed workflow written by humans.

Treating security auditing performance as primarily a property of the underlying model misses a procedural dimension. Many audit failures are failures of method: the agent searches the wrong attack surface, accepts a nearby bug that is not exploitable, stops before building a verifier, or drifts away from the root cause. An auditor with weaker model capability may still approach the performance of a stronger agent by following a disciplined procedure and using the available tools reliably. We therefore ask how much of the security auditing task is intrinsic model capability, and how much can be externalized as a reusable procedure.

The procedure question mirrors Sutton's ``bitter lesson'': durable progress often comes from systems that learn and search, rather than systems that freeze expert rules~\cite{sutton2019bitterlesson}. We do not claim that model weights or agent harnesses should be frozen in general; both are powerful, general-purpose components that will continue to improve. In our setting, however, fixing the model and harness isolates the effect of the external audit procedure: improvements cannot be attributed to a stronger model, a different orchestration policy, or hidden changes in tool use. This isolation tests whether security auditing has a procedural engineering core: can disciplined coverage, validation, and evidence construction be learned as a reusable method, or are gains dominated by irreducible model capability? Given the same pretrained model and the same externally supplied harness, can repeated attempts, grounded evaluation, and revision distill a reusable playbook? The learned artifact is not merely a prompt or memory log. It is a procedure specific to the task, outside the model and harness, that can be inspected, versioned, improved, and transferred across agents.

We present \tool, a playbook evolution framework. The framework keeps the underlying model and harness fixed. The learned object is a versioned text repository containing workflow guidance, vulnerability class audit strategies, validation gates, false-positive controls, and reproduction requirements. During evolution, a discovery agent runs on audit cases whose source is available; an evaluator compares its output against ground truth and executable evidence; and a reviser edits the playbook through changes backed by evidence. Playbook evolution produces a sequence of commits that are measurable on held-out cases and inspectable for learned procedure. Our paper is organized around two research questions:

\noindent\textbf{(RQ1) Acquisition.} Can an agent acquire an audit procedure through repeated attempts, grounded evaluation, and revision, without changing its model or harness? We instantiate \tool across two agent environments: a closed-source \codexgpt agent and an open-source \glmopen agent~\cite{zai2026glm51,opencode2026}. For each environment, the controlled variable is not model capability but the playbook: empty versus evolved by the loop. We also include \codexsecurity as a product-style baseline for open-source software vulnerability scanning, since the Codex Security workflow builds security context specific to each repository, validates candidate issues, and proposes fixes~\cite{openai2026codexsecurity}.

\noindent\textbf{(RQ2) Transfer.} Can a playbook evolved by a stronger teacher improve weaker student models? We ask whether Qwen family student models~\cite{qwen2026qwen36a3b,qwen2026qwen3627b} improve when given playbooks evolved by stronger GPT and GLM teachers. The transfer study separates three quantities that are usually entangled: model capability, playbook quality, and the compatibility between a model and a procedure.

We evaluate these questions on a temporally separated benchmark of locally reproducible, high- and critical-severity open-source advisories, spanning three years of evolution (\TrainingStartMonth{}--\TrainingEndMonth{}) and four months of held-out testing (\TestingStartMonth{}--\TestingEndMonth{}). Our primary contributions are:

\begin{itemize}
    \item \textbf{An autonomous playbook evolution framework.}
    Prior work supplies agents with security knowledge through human authored prompts, curated vulnerability databases, or manually designed audit workflows. We take the opposite approach: give the agent \textbf{an empty playbook and an unconstrained search space}, and only provide a minimal evolution loop (discovery, evaluation, revision) to ground updates in evidence. The agent decides what to learn, how to structure the procedure, and what security knowledge to accumulate. Starting from empty, both evolved playbooks reach 1,616 and 2,177 lines of agent-authored audit procedure across 38 accepted revisions.

    \item \textbf{A temporally separated, reproducible \ac{oss} advisory benchmark.}
    We construct a fully reproducible benchmark from the GitHub Advisory Database: \TrainingAdvisoryCount{} training and \TestingAdvisoryCount{} held-out advisories across \BenchmarkEcosystemCount{} ecosystems and \BenchmarkTopCWECount{} \acs{cwe} classes, each reproducible in a local Docker environment. \acs{cvss} v4.0 reachability and low attack complexity filters ensure each vulnerability is genuinely exploitable in a controlled setting, requiring no external assumptions or vendor confirmation to verify. Every case thus admits objective, executable verification.

    \item \textbf{Procedure acquisition and transfer.}
    In RQ1, playbook evolution raises \codexgpt{} end-to-end exploits $6{\times}$ (\GPTMinTOne{}$\to$\GPTEvolvedTOne{}, \GPTMinTOneRate{}$\to$\GPTEvolvedTOneRate{}); the evolved \glmopen{} playbook surpasses \codexsecurity{} on every metric (\GLMEvolvedTOne{} vs.\ \CodexSecurityTOne{} exploits; \GLMEvolvedTargetMatchRate{} vs.\ \CodexSecurityTargetMatchRate{} match rate), showing \textbf{open-source evolution can outperform a dedicated commercial product}. In RQ2, GLM-teacher transfer improves both Qwen students: \QwenDenseMinTargetMatchRate{}$\to$\GLMDenseTransferTargetMatchRate{} ($2.7{\times}$) for \qwendense{} and \QwenMoEMinTargetMatchRate{}$\to$\GLMMoETransferTargetMatchRate{} ($4{\times}$) for \qwenmoe{}.

    \item \textbf{Vulnerability disclosures.}
    \tool has produced \ConfirmedVulnerabilityCount{} confirmed vulnerability disclosures across \empirical{18} open-source projects, acknowledged via GitHub Security Advisories, CVEs, and a \$1{,}500 bug bounty award (cf. Appendix~\ref{app:reported-vulnerabilities}).

    \item \textbf{Public artifacts.}
    {\raggedright To support public verifiability and reproducibility, we release the paper artifacts, benchmark materials, and scoring records at \href{https://github.com/evohunt-project/artifacts}{\mbox{\footnotesize\texttt{https://github.com/evohunt-project/artifacts}}}.\par}

\end{itemize}

\section{Background}\label{sec:background}

We introduce the background used throughout the paper.

\noindent\textbf{Severity.} \acs{cvss} is an open framework for representing vulnerability characteristics and severity~\cite{firstCVSSv40}. A \acs{cvss} v4.0 record contains a numerical score and a vector of metrics. The qualitative scale maps the base score to None (0.0), Low (0.1--3.9), Medium (4.0--6.9), High (7.0--8.9), or Critical (9.0--10.0); our benchmark focuses on High and Critical. The vector fields matter more for our threat model than the final label. Exploitability metrics describe how an attacker reaches the vulnerability (AV, AC, AT, PR, UI); impact metrics describe confidentiality, integrity, and availability effects. Section~\ref{sec:threat-model} uses these fields to define which vulnerabilities are reachable and locally auditable.

\noindent\textbf{Weakness classes.} \acs{cwe} is a community-maintained taxonomy of software and hardware weakness types~\cite{mitreCWE}. A \acs{cwe} class describes the underlying weakness pattern rather than a specific vulnerable product. For example, CWE-22 denotes path traversal, CWE-89 denotes SQL injection, and CWE-79 denotes cross-site scripting. We use \acs{cwe} labels to analyze benchmark diversity and playbook coverage.

\noindent\textbf{Advisories.} Public vulnerability advisories are structured records about known vulnerabilities in released software. In the GitHub Advisory Database, an advisory may include identifiers, affected packages and version ranges, references, severity labels, \acs{cvss} vectors, \acs{cwe} identifiers, and links to commits, releases, or upstream reports~\cite{githubAdvisoryDatabase}. In our benchmark, advisories provide the public disclosure record and metadata used for case qualification and temporal splitting.

\noindent\textbf{Model.} The model is the underlying \ac{llm} that performs code and security reasoning. Model capability affects semantic understanding, long-horizon reasoning, tool-use recovery, and the ability to connect dispersed evidence across a repository. The model set referenced later includes GPT5.4-xhigh (cutoff \GPTKnowledgeCutoff{}), whose parameter count is not publicly disclosed; GLM5.1 (cutoff \GLMKnowledgeCutoff{}), a mixture-of-experts model with 744B total parameters and 40B active parameters~\cite{zai2026glm51}; Qwen3.6-35B-A3B (cutoff \QwenKnowledgeCutoff{}), a mixture-of-experts model with 35B total parameters and 3B active parameters~\cite{qwen2026qwen36a3b}; and Qwen3.6-27B (cutoff \QwenKnowledgeCutoff{}), a dense model with 27B parameters~\cite{qwen2026qwen3627b}.

\noindent\textbf{Harness.} The harness is the runtime substrate that supplies repository state, tool access, shell and browser execution, sandboxing, context management, memory or compaction, and optional subagent invocation. The harness provides execution infrastructure, but it does not define the audit method. The harnesses referenced later are Codex and OpenCode~\cite{opencode2026}; \codexgpt and \glmopen denote the corresponding model-harness environments.

\noindent\textbf{Playbook.} The playbook is the external audit procedure. Decisions such as which attack surfaces to inspect first, when to move from discovery to validation, how to assign subagents, which vulnerabilities to prioritize, how to reject non-exploitable adjacent bugs, and what evidence is required before reporting belong to the playbook. External security knowledge not supplied by model weights, including vulnerability class guides, source-to-sink patterns, validation and reproduction strategies, is also playbook content.

\noindent\textbf{\codexsecurity.} A commercial cloud workflow for repository vulnerability scanning and remediation~\cite{openai2026codexsecurity}; used in this study as a reference point (see Table~\ref{tab:conditions}).

\begin{figure*}[t]
\centering
\begin{minipage}[t]{0.48\textwidth}
\centering
\includegraphics[width=\linewidth]{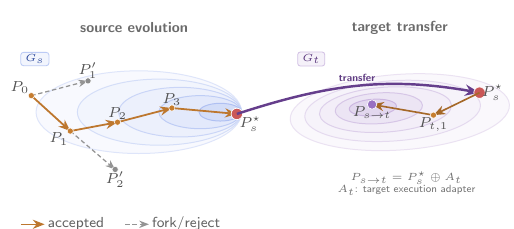}
\end{minipage}%
\hfill
\begin{minipage}[t]{0.48\textwidth}
\centering
\includegraphics[width=\linewidth]{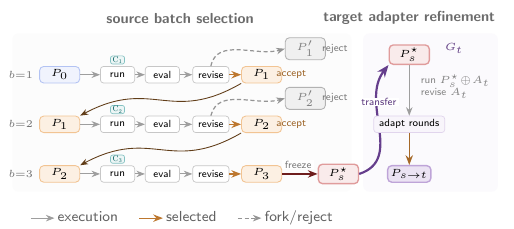}
\end{minipage}
\caption{Procedure learning in playbook space.
\textbf{Left}: abstract trajectory through a shared space of audit playbooks.
\LegendSwatch{legendBlue!25}~\(G_s\) (source); \LegendSwatch{legendPurple!25}~\(G_t\) (target); \LegendSwatch{legendOrange}~accepted revisions; \LegendSwatch{legendGray}~forked/rejected; \LegendSwatch{legendRed}~frozen \(P_s^\star\).
Nearby points are semantically similar procedures; darker contours indicate higher audit utility.
Starting from \(P_0\), the learner keeps \(P_1,P_2,P_3\), discards \(P_1'\) and \(P_2'\), freezes \(P_s^\star\), and transfers it to \(G_t\), where adapter \(A_t\) yields \(P_{s\to t}\).
\textbf{Right}: the same trajectory as concrete batch execution.
\(C_b = N_b \cup R_b\) is the case batch for round \(b\), combining new cases \(N_b\) with replay cases \(R_b\).
\(C_1,C_2,C_3\) are successive audit workloads, not model checkpoints; after the final source round \(P_s^\star\) is frozen and the target adapter refined to yield \(P_{s\to t}\).}
\label{fig:playbook-landscape}
\end{figure*}

\section{Models}\label{sec:models}

This section defines the system model, threat model, audit scope, and auditor model used throughout the paper.

\noindent\textbf{System model.}
The target system is an \acs{oss} repository or package available to an auditor with source code access. The auditor may use any local analysis method, including code inspection, dependency and configuration analysis, local tools, builds, tests, scripts, service execution, and HTTP or browser interaction.
The audit is open world: a repository may contain multiple vulnerabilities, and the complete vulnerability set is generally unknown. The auditor may report any in-scope vulnerability it can substantiate, regardless of whether that vulnerability is the known target used for benchmark scoring. We model an audit case as:
\[\mathsf{case}=(\mathsf{repo},\mathsf{env},\mathsf{setup})\],
where \(\mathsf{repo}\) is the repository snapshot, \(\mathsf{env}\) is the local execution environment, and \(\mathsf{setup}\) is setup metadata. The local execution environment includes dependency manifests, scripts, containers, service configurations, test harnesses, and local mocks. It need not reproduce a full production deployment. It must, however, make the relevant attack surface reachable and allow runtime evidence collection.

\noindent\textbf{Threat model.}
The attacker can obtain the target software, deploy it in an isolated local environment, and operate against it as an external party. Depending on the target type, the attacker controls inputs through network interfaces (HTTP requests, remote API calls, and custom protocols), local interfaces (command-line arguments, files, environment variables, and inter-process communication), or programmatic interfaces (function arguments, serialized data, and configuration passed to library APIs). The attacker is subject to the \acs{cvss} attack vector and privilege assumptions. It lacks high privileges (unless the vulnerability is reachable through a privileged interface), maintainer or commit access, deployment shell access, privileged cloud credentials, and trusted-infrastructure control.

\noindent\textbf{Audit scope.}\label{sec:threat-model}
The following three criteria select vulnerabilities that are both impactful and reproducible under the threat model. Severity filters for meaningful impact; reachability and local verifiability together ensure exploitation can be demonstrated in an isolated environment, a necessary condition for automated evaluation with concrete evidence. An in-scope case must satisfy all three:
\begin{itemize}
    \item \textbf{Severity.} The advisory must be rated High or Critical (\acs{cvss} base score $\geq7.0$). Low and Medium severity cases are more ambiguous, introduce significant noise, and are harder to validate with concrete evidence; we focus on cases where impact is clear and significant.
    \item \textbf{Reachability.} The vulnerability must be exercisable within the attacker capability defined above. The \acs{cvss} v4.0 qualification profile is~\cite{firstCVSSv40}:
    \[
    \begin{aligned}
    &\mathrm{AV}\in\{\mathrm{Network},\mathrm{Local}\},\quad
    \mathrm{AC}=\mathrm{Low},\quad
    \mathrm{AT}=\mathrm{None},\\
    &\mathrm{PR}\in\{\mathrm{None},\mathrm{Low}\},\quad
    \mathrm{UI}\in\{\mathrm{None},\mathrm{Passive}\},\\
    &\exists I\in\{\mathrm{VC},\mathrm{VI},\mathrm{VA},\mathrm{SC},\mathrm{SI},\mathrm{SA}\}:
    \operatorname{val}(I)\neq\mathrm{None}.
    \end{aligned}
    \]
    AV, AC, AT, PR, and UI are the \acs{cvss} v4.0 exploitability metrics: attack vector, complexity, requirements, privileges required, and user interaction. We restrict to Network or Local attack vectors (exercisable in the audit workspace; programmatic interfaces such as library APIs fall under AV:Local), Low complexity and no attack requirements (no special conditions or deployment prerequisites), and None or Low privileges with None or Passive user interaction, so that exploitation is reproducible by a low-privilege external attacker without active target behavior. The six impact metrics measure confidentiality, integrity, and availability across two scopes: the directly vulnerable system (V prefix, e.g., VC) and subsequently affected systems (S prefix, e.g., SC). The final condition requires at least one of these six to be non-None, excluding vulnerabilities that cause no measurable consequence.
    \item \textbf{Local verifiability.} The exploit predicate must be locally reproducible inside Docker; vulnerabilities requiring live third-party services, cloud infrastructure, physical access, or production-only conditions are out of scope.
\end{itemize}

\noindent\textbf{Auditor model.}
\tool externalizes audit procedure into a text playbook \(P\), keeping the agent environment \(G\) fixed across experimental conditions. A playbook version is
\[
P=(\mathsf{workflow},\mathsf{knowledge},\mathsf{validation}),
\]
where \(\mathsf{workflow}\) specifies how to cover a repository and form exploit hypotheses, \(\mathsf{knowledge}\) encodes security-domain guidance, and \(\mathsf{validation}\) defines what evidence is sufficient for a reportable finding. This external playbook \(P\) is the variable object in our controlled comparisons. The agent environment \(G\) comprises the underlying \acs{llm}, tool interface, and orchestration code; it is fixed within each experimental condition. A discovery run is
\[
\mathsf{run}(G,P,\mathsf{case})\rightarrow(\mathsf{trace},\mathsf{artifacts},\mathsf{evidence}),
\]
where \(\mathsf{trace}\) is the execution trace, \(\mathsf{artifacts}\) the finding set, and \(\mathsf{evidence}\) collects runtime evidence (command outputs, verifier results, logs, and \acs{poc} outputs).

A reportable finding \(f\in\mathsf{artifacts}\) is a tuple
\[
f=(\mathsf{root},\mathsf{impact},\mathsf{predicate},\mathsf{support}),
\]
where \(\mathsf{root}\) is a root-cause claim, \(\mathsf{impact}\) is an impact claim, \(\mathsf{predicate}\) is the exploit predicate connecting them, and \(\mathsf{support}\subseteq\mathsf{evidence}\) is supporting evidence. Crashes, error messages, or anomalous behavior alone are insufficient unless the evidence establishes \(\mathsf{predicate}\). To prevent shortcut answer lookup, the auditor may not use external search or retrieval to obtain vulnerability information about the target, such as advisory text, patch diffs, public \acsp{poc}, issue discussions, or writeups; our experiments enforce this policy by checking execution traces.

\section{Methodology}\label{sec:methodology}

\begin{table}[t]
\centering
\caption{Text-space analogues for procedure acquisition (top) and playbook transfer (bottom).}
\label{tab:analogues}
\resizebox{\columnwidth}{!}{%
\begin{tabular}{ll}
\toprule
Notion & \tool analogue \\
\midrule
\rowcolor{black!7}\multicolumn{2}{c}{\textbf{Continual procedure learning}} \\
\midrule
Trainable state      & External playbook \(P\) \\
Data stream          & Temporal audit batches \(N_1,\ldots,N_B\) \\
Replay memory        & Replay cases \(R_b\) \\
Trajectory batch     & Rollout records \(\mathcal{X}_{P,C_b}\) \\
Selection signal     & \(S(P,C_b)\) and \(F(P,C_b)\) \\
Gradient step        & Textual revision \(\mathsf{Revise}(P,F)\) \\
Checkpoint           & Playbook revision commit \\
Model selection      & Same-batch playbook comparison \\
Deployment           & Frozen \(P^\star\) on held-out cases \\
\midrule
\rowcolor{black!7}\multicolumn{2}{c}{\textbf{Transfer procedure learning}} \\
\midrule
Pretrained backbone  & Evolved source playbook \(P_s^\star\) \\
Adapter parameters   & Execution adapter \(A_t\), initialized empty \\
Fine-tuned model     & Combined procedure \(P_s^\star\oplus A_t\) \\
Loss signal          & Rollout failure report under \(G_t\) \\
Gradient direction   & Abstracted failure pattern \\
Budget constraint    & Maximum adapter size \(B\) \\
Early stopping       & Score saturation or gap exhaustion \\
No-transfer baseline & Empty playbook \(\Pmin\) \\
Transfer gain        & Paired lift \(\Delta_{t\leftarrow s}\) \\
Residual gap         & Remaining model or harness capability \\
\bottomrule
\end{tabular}%
}
\end{table}

This section formulates \tool as a model-agnostic procedure learning framework; model and harness choices are deferred to the evaluation. Inspired by continual~\cite{mccloskey1989catastrophic,parisi2019continual} and transfer learning~\cite{pan2010transfer}, we externalize the learning signal into a revisable text procedure rather than model weights: \textit{continual procedure learning} for acquisition and \textit{transfer procedure learning} for generalization.

\subsection{Continual Procedure Learning}

Acquisition asks whether audited experience can improve a reusable audit procedure, i.e., update \(P\) so that \(\mathsf{run}(G,P,\mathsf{case})\) produces better-validated findings (cf.\ Section~\ref{sec:models}). We model acquisition as a sequence of training batches \(N_1,\ldots,N_B\); at round \(b\) the learner may augment \(N_b\) with replay cases \(R_b\) from earlier rounds:
\[
C_b=N_b\cup R_b.
\]
Replay is theoretically optional in the abstract method, but it captures the continual learning intuition: a procedure should improve on new failures without forgetting earlier audit lessons. For a candidate playbook \(P\), running the fixed agent on \(C_b\) produces execution records
\[
\mathcal{X}_{P,C_b}
=
\{\mathsf{run}(G,P,\mathsf{case}) : \mathsf{case}\in C_b\}.
\]
A grounded evaluator maps those records and the withheld case labels \(\mathcal{L}_{C_b}\) (ground-truth vulnerability identities and evidence thresholds, not shown to the agent) to a scalar score for selection and a structured failure report for revision:
\[
\mathsf{Eval}(P,C_b,\mathcal{X}_{P,C_b},\mathcal{L}_{C_b})
\rightarrow
(S(P,C_b),F(P,C_b)).
\]
The score \(S\) selects between playbook versions; the failure report \(F\) is the learning signal for procedure revision. The reviser converts failures into edits of the playbook:
\[
\mathsf{Revise}(P,F)\rightarrow P',
\]
where \(P'\) is the proposed new playbook. This is the text-space analogue of a gradient step: the edit should explain what went wrong across runs, generalize the failure into reusable audit guidance, and avoid memorizing repository names, advisory details, or exploit strings. Table~\ref{tab:analogues} (top) summarizes the continual learning analogues.

\noindent\textbf{Selection and stopping.}
Procedure learning uses a selection gate before deployment. Let \(P_b^{\mathrm{best}}\) denote the current selected playbook and \(P_b^{\mathrm{cand}}\) a proposed alternative. A round compares playbooks on the same fixed batch and selects
\[
P_b^\star
=
\mathop{\mathrm{arg\,max}}_{P\in\{P_b^{\mathrm{best}},P_b^{\mathrm{cand}}\}}
S(P,C_b).
\]
The selected playbook becomes the base for the next revision. After the finite training window ends, the learner freezes the selected playbook \(P^\star\). Held-out testing runs \(P^\star\) without further revision. This separation is essential: the method studies whether prior validated experience can produce a reusable audit procedure, not whether an agent can keep adapting on the test set.

\begin{designremark}{Why process each data point once, in chronological order (continual learning), rather than re-running it until \tool stops improving (typical training)?}
While there are many possible designs, our goal is to make \tool as production ready as possible. In reality, all published advisories seed the initial playbook, and new disclosures simply extend the same stream as they arrive. In other words, \textbf{the system self-improves indefinitely without retraining using continual learning}. The temporal train/test split used in this paper is an artificial construct introduced solely to measure held-out performance; the underlying mechanism is identical.
\end{designremark}

\subsection{Transfer Procedure Learning}

Transfer asks whether a learned audit procedure can move across agent environments, and whether adaptation can close the residual gap. The source environment \(G_s\) produces a frozen playbook \(P_s^\star\). For the target environment \(G_t\), we introduce an adapter \(A_t\), initialized as empty, that records target-specific execution guidance rather than new source-task knowledge. The combined procedure is
\[
P_{s\to t} = P_s^\star \oplus A_t,
\]
where \(\oplus\) denotes prompt-level concatenation: both documents are presented as distinct tagged sections in the agent's context, with \(P_s^\star\) preceding \(A_t\). The source playbook \(P_s^\star\) is never modified during adaptation; only \(A_t\) evolves. The adapter encodes execution gaps between \(G_s\) and \(G_t\). Table~\ref{tab:analogues} (bottom) summarizes the transfer learning analogues.

The adapter is evolved through a revision loop on training cases run under \(G_t\). Each round collects rollout traces, abstracts them into recurring target-execution failure patterns rather than case-specific fixes, and proposes an edit to \(A_t\). The failure diagnosis asks not what is wrong with the source playbook, but where the target agent diverges from the source procedure. The reviser may only modify \(A_t\); a valid adapter edit adds or rewrites target-environment execution guidance without altering \(P_s^\star\). This separation keeps the transfer experiment interpretable: gains from \(P_s^\star\) alone measure how much source procedure is directly executable by the target, while the additional gain from \(A_t\) measures how much of the remaining gap is closeable by adaptation.

\noindent\textbf{Adapter selection and budget.}
Let \(A_t^{\mathrm{best}}\) denote the current selected adapter and \(A_t^{\mathrm{cand}}\) a proposed revision. A candidate is accepted when it does not regress beyond tolerance \(\epsilon\) on a small held-out validation set \(C_{\mathrm{val}}\):
\[
S(P_s^\star\oplus A_t^{\mathrm{cand}},C_{\mathrm{val}})
\;\geq\;
S(P_s^\star\oplus A_t^{\mathrm{best}},C_{\mathrm{val}}) - \epsilon.
\] A budget constraint \(B\) limits adapter length, preventing the reviser from duplicating source playbook content or memorizing individual training cases. The loop stops when failure abstraction yields no new recurring transfer gaps, the score saturates, or the round budget is exhausted. The frozen adapter \(A_t^\star\) is then finalized as part of the combined procedure \(P_{s\to t}\) for deployment. The transfer effect decomposes into direct and adapted components:
\[
\Delta_{t\leftarrow s}^{\mathrm{direct}}
= S(G_t,P_s^\star,\Ctest) - S(G_t,\Pmin,\Ctest),
\]
\[
\Delta_{t\leftarrow s}^{\mathrm{adapted}}
= S(G_t,P_{s\to t},\Ctest) - S(G_t,\Pmin,\Ctest).
\]
A positive \(\Delta_{t\leftarrow s}^{\mathrm{direct}}\) means the source playbook produces value in the target environment as-is, without any adaptation; \(\Delta_{t\leftarrow s}^{\mathrm{adapted}}\) means that target adaptation recovers additional value on top. The residual gap \(S(G_s,P_s^\star,\Ctest) - S(G_t,P_{s\to t},\Ctest)\) isolates what procedure transfer cannot explain and must be attributed to model / harness capability.

\section{Implementation}\label{sec:implementation}

This section describes the concrete system design of \tool: how the playbook is stored and versioned, how agents are orchestrated, how revisions are gated, and how candidate playbooks are selected across rounds.

\noindent\textbf{Playbook repository.}
\tool stores the playbook as a Git repository rather than a prompt string. The repository stores the playbook as versioned text files: workflow rules, vulnerability-class guides, validation gates, evidence checklists, and optional subagent instructions. At session start, the playbook's entry-point document and any target adapter file present on the checked-out branch are pre-injected into the orchestrator's initial prompt as distinct tagged sections; the orchestrator reads additional files such as vulnerability-class guides and subagent prompts on demand during the audit run. Git history is part of the artifact: accepted, rejected, and superseded variants remain inspectable.

\begin{figure}[t]
\centering
\includegraphics[width=\columnwidth]{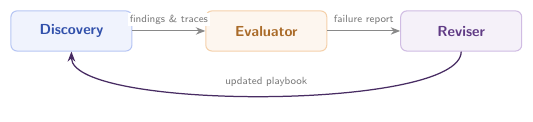}
\caption{The \tool evolution loop. A \emph{discovery agent} audits the repository and produces findings and traces; an \emph{evaluator} scores the outcome against ground truth; a \emph{reviser} turns the failure report into a playbook commit, completing the round.}
\label{fig:evolution-loop}
\end{figure}

\noindent\textbf{Agent framework.}
Figure~\ref{fig:evolution-loop} shows the evolution loop: three agents iterate in sequence using fixed, hand-authored prompts. These prompts are the only point at which human knowledge enters the system; everything the agents learn about vulnerability classes and audit procedure is accumulated in the playbook through iteration. The discovery, revision, and evaluation agents run under one of two coding agent harnesses (Codex or OpenCode), depending on the experimental condition. Both harnesses follow playbook instructions, invoke local tools, and maintain a shared context window. Harness and model pairings for each condition are specified in the evaluation.

\noindent\textbf{Replay memory.}
When constructing \(C_b\), the implementation combines newly introduced cases with replay cases drawn from earlier batches. Replay selection tokenizes advisory metadata and ranks historical cases with BM25, then selects a configured fraction of the new-case count to preserve vulnerability-class coverage across batches.

\noindent\textbf{Discovery rollouts.}
For each playbook and case pair, the discovery agent inspects the repository in a sandbox, invokes local tools, runs commands, executes tests or reproduction scripts, and may delegate scoped tasks to subagents. The orchestrator consolidates duplicate hypotheses, decides reportability, and emits the final artifact set. Each rollout produces the \(\mathsf{trace}\), \(\mathsf{artifacts}\), and \(\mathsf{evidence}\) for the evaluator.

\noindent\textbf{Evaluation feedback.}
The evaluator produces per-run feedback for revision, including target-match rationales, reproduction assessments, evidence references, and failure summaries. The concrete evaluator is separated from the discovery run: after a session finishes, an adjudicator reads the recorded artifacts, withheld advisory metadata, and execution evidence, then assigns the run to the target-outcome ladder defined in the evaluation. A consistency check rejects internally incoherent judgments before they are used for branch selection or held-out scoring.

\noindent\textbf{Git branch tournament.}
\tool maintains two branch roles: \(P_{\mathrm{best}}\), the selected playbook so far, and \(P_{\mathrm{cand}}\), the proposed alternative. In bootstrap mode, the system evaluates the current playbook and asks the reviser to create the first candidate. In standard mode, each round follows $\mathrm{Run}\to\mathrm{Evaluate}\to\mathrm{Select}\to\mathrm{Revise}$.
The runner executes \(P_{\mathrm{best}}\) and \(P_{\mathrm{cand}}\) on the same batch \(C_b\). The evaluator scores both, and the pipeline advances \(P_{\mathrm{best}}\) to the winning commit. The reviser checks out the winner and creates the next candidate. The new candidate enters the next tournament rather than on the batch that produced it, preventing an untested revision from overwriting the selected playbook.

\noindent\textbf{Revision gate.}
The reviser edits the playbook repository and must commit the result on top of the selected base. The implementation checks that the commit descends from the base, at least one tracked file changed, no uncommitted changes remain, and the revision report satisfies the schema. If a revision fails these checks, the gate requests repair and eventually fails the round. These checks keep the procedure auditable and prevent silent changes outside the playbook.

\noindent\textbf{Hyperparameters.}
Each rollout runs for at most 20 turns with a wall-clock timeout of 2 hours. At most 2 cases run concurrently per playbook branch. Each batch draws up to 10 new cases plus replay at a 0.25 ratio. The selection score weights discovery correctness at 0.7 and reproduction effectiveness at 0.3, each on a \(\{0, 0.5, 1\}\) ordinal scale. The revision gate allows up to 3 repair attempts per round.

\noindent\textbf{Transfer mode.}
Transfer runs in two phases. In the direct phase, \(P_s^\star\) is checked out from \(P_{\mathrm{best}}\) without an adapter and run in \(G_t\) to measure \(\Delta_{t\leftarrow s}^{\mathrm{direct}}\). In the adaptation phase, \(A_t\) is a Markdown file co-committed alongside \(P_s^\star\); the harness pre-injects both as tagged sections, realizing \(P_s^\star\oplus A_t\). Adapter evolution reuses the same revision gate and branch tournament, with the reviser editing only the adapter file.

\section{Benchmark}\label{sec:benchmark}

\noindent\textbf{Source and temporal split.}
We construct the benchmark from the GitHub Advisory Database~\cite{githubAdvisoryDatabase}, using GitHub-reviewed advisories. The training split contains high- and critical-severity advisories from \TrainingStartMonth{} through \TrainingEndMonth{}; the held-out testing split covers \TestingStartMonth{} through \TestingEndMonth{}. Every testing advisory is published after the training window, matching the deployment question of whether a learned procedure improves discovery on later vulnerabilities.

\noindent\textbf{Filtering.}
Cases follow the qualification criteria in Section~\ref{sec:threat-model}. We exclude duplicates, withdrawn advisories, and cases requiring live services, cloud infrastructure, or physical access.

\noindent\textbf{Distribution.}
Table~\ref{tab:benchmark-summary} summarizes the two splits. The testing split is not single-project dominated: the top five repositories account for \TestingTopFiveRepoAdvisoryCount{} advisories (\TestingTopFiveRepoAdvisoryPercent{} of the split).

\begin{table}[t]
\caption{Temporal Benchmark Split Summary}
\label{tab:benchmark-summary}
\centering
\footnotesize
\setlength{\tabcolsep}{4pt}
\begin{tabular}{@{}>{\raggedright\arraybackslash}l>{\raggedleft\arraybackslash}r>{\raggedleft\arraybackslash}r@{}}
\toprule
Property & Training & Testing \\
\midrule
Advisories & \TrainingAdvisoryCount{} & \TestingAdvisoryCount{} \\
Month range & \TrainingStartMonth{}\textasciitilde\TrainingEndMonth{} & \TestingStartMonth{}\textasciitilde\TestingEndMonth{} \\
High severity & \TrainingHighCount{} (\TrainingHighPercent{}) & \TestingHighCount{} (\TestingHighPercent{}) \\
Critical severity & \TrainingCriticalCount{} (\TrainingCriticalPercent{}) & \TestingCriticalCount{} (\TestingCriticalPercent{}) \\
Unique packages / repos. & \TrainingUniquePackageCount{} / \TrainingUniqueRepoCount{} & \TestingUniquePackageCount{} / \TestingUniqueRepoCount{} \\
Overlap packages / repos. & --- & \TestingPackageOverlapCount{} / \TestingRepoOverlapCount{} \\
\bottomrule
\end{tabular}
\end{table}

\begin{figure}[t]
  \centering
  \includegraphics[width=\columnwidth]{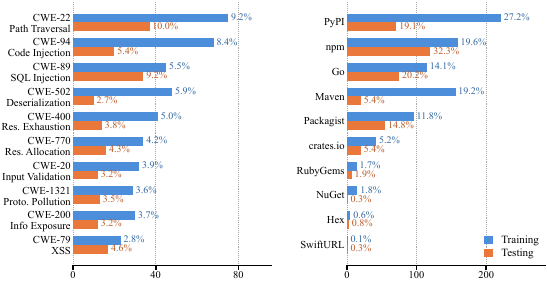}
  \caption{Vulnerability-type (CWE) and ecosystem distribution of the training and testing splits. Top 10 categories by combined count; bar-end labels show percentage of each split.}
  \label{fig:benchmark-distribution}
\end{figure}

\begin{itemize}[leftmargin=*,topsep=2pt,itemsep=2pt]
  \item \textbf{Ecosystem distribution.} As shown in Figure~\ref{fig:benchmark-distribution}, the benchmark is application-security heavy: PyPI, npm, Maven, Go, and Packagist dominate both splits, making it a test of multi-language auditing rather than any single ecosystem. One notable distribution shift is Maven, which accounts for \empirical{19.2\%} of training but only \empirical{5.5\%} of testing, reflecting advisory publication timing and potentially limiting generalization to Java-heavy targets.

  \item \textbf{Vulnerability types.} The \acs{cwe} distribution is broad but not uniform (Figure~\ref{fig:benchmark-distribution}): the top five classes cover \empirical{34.1\%} of training and \empirical{33.4\%} of testing, with no single type dominating. There is moderate distribution shift between splits: SQL injection (CWE-89) rises from \empirical{5.5\%} in training to \empirical{9.2\%} in testing, while deserialization (CWE-502) falls from \empirical{5.9\%} to \empirical{2.7\%}. A playbook evolved primarily on the training mix may underweight SQL injection patterns relative to the test distribution.

  \item \textbf{CVSS profile.} The \acs{cvss} metric distribution confirms that the benchmark targets reachable, externally auditable vulnerabilities: 95.7\% of testing cases have network attack vector, 69.8\% require no privileges, and 89.8\% require no user interaction. Critically, the impact profile spans confidentiality, integrity, and availability rather than availability alone, requiring an agent to reason about diverse exploit effects rather than optimizing for only crashes or hangs.

  \item \textbf{Project overlap.} \TestingRepoOverlapCount{} (\TestingRepoOverlapPercent{}) of testing repositories also appear in training, but the splits are disjoint at the advisory level: every testing vulnerability was published after the training window, and its root cause, patch, and reproduction oracle are withheld from the agent. Repository recurrence is realistic --- practitioners routinely audit codebases they have previously examined --- but the agent does not see the specific bug under test. We report both aggregate and recurrence-stratified performance to characterize any advantage from prior codebase exposure.

\end{itemize}

\section{Experiments}\label{sec:experiments}

Each condition is written as \(\mathrm{Exp}_{x}^{r}\): \(x\) names the agent environment and \(r\) the role (Empty, Evolve, Tr/GPT, Tr/GLM, or OCS). Scoring is defined in Section~\ref{sec:evaluation}.

\noindent\textbf{Execution environment.}
All \tool jobs ran on \EvalMachineCount{} fixed Apple M4 machines (macOS \EvalMacOSVersion{}, \EvalMachineCpuCores{} CPU cores, \EvalMachineMemoryGiB{}\,GiB memory). \tool conditions share the benchmark checkout, harness code, output schema, and scoring; \(\Bcs\) shares benchmark inputs and scoring but uses the \codexsecurity{} product workflow.

\noindent\textbf{Harness configuration.}
Both harnesses use default settings (Codex v0.136.0; OpenCode v1.15.7). Codex uses a 272k-token input limit and compacts at ${\approx}245{,}000$ tokens ($90\%$), preserving an encrypted summary plus up to 20k tokens of recent context; OpenCode uses a 192k-token limit with compaction at the same threshold and an 8,192-token output cap, via \ac{llm}-assisted compaction. The pre-compaction window is approximately 28\% wider under \codexgpt{}; this context asymmetry, alongside model capability differences, should be considered when interpreting performance gaps between Codex and OpenCode. The open-source models (GLM and Qwen) all run in FP8.

\begin{table}[t]
\caption{Experimental Conditions}
\label{tab:conditions}
\centering
\scriptsize
\resizebox{\columnwidth}{!}{%
\begin{tabular}{@{}llll@{}}
\hline
Condition & Agent environment & Procedure & Source \\
\hline
\rowcolor{black!7}\multicolumn{4}{c}{\textbf{Experiment 1: procedure acquisition}} \\
\midrule
\(\Bcs\) & \codexsecurity & --- & --- \\
\midrule
\(\Bgpt\) & \codexgpt & \(\Pmin\) & --- \\
\(\Egpt\) & \codexgpt & \(\Pgptstar\) & --- \\
\midrule
\(\Bglm\) & \glmopen & \(\Pmin\) & --- \\
\(\Eglm\) & \glmopen & \(\Pglmstar\) & --- \\
\midrule
\rowcolor{black!7}\multicolumn{4}{c}{\textbf{Experiment 2: teacher playbook transfer}} \\
\midrule
\(\StudentBaseline{27B}\) & \qwendense/OpenCode & \(\Pmin\) & --- \\
\(\StudentTransfer{27B}\) & \qwendense/OpenCode & \(P_{\mathrm{GPT}\to 27B}\) & \(\Pgptstar\) \\
\(\StudentGLMTransfer{27B}\) & \qwendense/OpenCode & \(P_{\mathrm{GLM}\to 27B}\) & \(\Pglmstar\) \\
\midrule
\(\StudentBaseline{A3B}\) & \qwenmoe/OpenCode & \(\Pmin\) & --- \\
\(\StudentTransfer{A3B}\) & \qwenmoe/OpenCode & \(P_{\mathrm{GPT}\to A3B}\) & \(\Pgptstar\) \\
\(\StudentGLMTransfer{A3B}\) & \qwenmoe/OpenCode & \(P_{\mathrm{GLM}\to A3B}\) & \(\Pglmstar\) \\
\hline
\end{tabular}
}
\par\vspace{3pt}
\begin{minipage}{\columnwidth}
\scriptsize
\emph{Note:} \(\Bcs\) is \codexsecurity, an independently developed cloud workflow that scans repositories, builds repository-specific security context, validates findings in an isolated environment, and surfaces ranked results with patch options~\cite{openai2026codexsecurity}. It is not an \tool implementation; we include it as a product-style reference because relying solely on self-comparison would not be representative. Source is the frozen teacher procedure imported by each transfer condition.
\end{minipage}
\end{table}

\noindent\textbf{Experiment 1: procedure acquisition.}
Acquisition tests whether repeated evaluated attempts can produce a better audit procedure for the same environment. \codexgpt evolves playbook \(\Pgptstar\) and \glmopen evolves \(\Pglmstar\) on the training split only; both are frozen at the end of training before held-out testing. Conditions are listed in Table~\ref{tab:conditions}. The controlled comparisons are \(\Egpt\) versus \(\Bgpt\) and \(\Eglm\) versus \(\Bglm\); in each pair, model, harness, test cases, and task interface are fixed; only the procedure changes.

\noindent\textbf{Experiment 2: teacher playbook transfer.}
Transfer tests whether a procedure evolved by a stronger teacher benefits weaker student models when combined with bounded target-environment adaptation. We test two source playbooks: the frozen \(\Pgptstar\) and the frozen \(\Pglmstar\). For each teacher, adaptation evolves one adapter per student while leaving the source playbook unchanged. Conditions are listed in Table~\ref{tab:conditions}. The causal transfer comparisons are each transferred condition against the same student's empty-playbook baseline; \(\Bcs\) and \(\Egpt\) are reported in RQ1 for context and are not transfer comparisons.

\section{Evaluation}\label{sec:evaluation}

This section defines the scoring protocol and reports results for experiments in Section~\ref{sec:experiments}. Each run is scored at two levels: whether the agent identified the target vulnerability across \EvalCaseCount{} held-out advisories, and how its reported findings distributed across qualification, target match, and off-target categories (defined in Section~\ref{sec:eval-protocol}).

\subsection{Scoring Challenges}

Our evaluation poses four challenges:

\noindent\textbf{Challenge 1. Scale.} The experiments produce \TotalJudgedFindings{} findings across \EvalCaseCount{} held-out cases, repositories, and vulnerability classes; fully manual scoring is not practical.

\noindent\textbf{Challenge 2. Cross-domain consistency.} Human reviewer expertise varies across ecosystems and vulnerability families, making consistent scoring difficult.

\noindent\textbf{Challenge 3. Judge calibration.} Evaluation strictness varies with judge capability: a stricter judge rejects borderline findings that a more lenient judge accepts, biasing cross-condition score comparisons.

\noindent\textbf{Challenge 4. Finding quality heterogeneity.} Not every finding is the same quality: an agent that produces a confirmed end-to-end exploit and one that reports a static hypothesis about the same root cause represent fundamentally different capability levels. Aggregating these into a single count obscures where agent capability actually lies.

We address these challenges with the following design:

\noindent\textbf{Step 1. Single LLM judge.} We designate \JudgeModel as the sole scoring judge for all experimental conditions, applying the rules defined in Section~\ref{sec:eval-protocol} (\textit{Scoring Protocol}), mitigating cross-reviewer inconsistency and reporting bias. For each judge run, we also collect all available agent-produced evidence (session artifacts, execution logs, and reproduction scripts) to support the manual review and public release in Steps 2 and 3.

\noindent\textbf{Step 2. Manual review.} For every judge-qualified high/critical finding (\TotalQualifiedFindings{} findings across all conditions), we checked consistency between evidence tier, target-match verdict, and scoring rules without re-reading the full trace. Aggregate judge counts remain unchanged; Section~\ref{sec:fpr} reports sampled false-positive estimates.

\noindent\textbf{Step 3. Artifact release.} All session artifacts, judge traces, and scoring records are released with the paper for independent reader verification.

\noindent\textbf{Step 4. Tiered evidence scoring.} We adopt a three-tier scheme (detailed in \circled{3}, Section~\ref{sec:eval-protocol}) that distinguishes end-to-end exploit success (T1), bug activation with a runnable PoC (T2), and speculative identification (T3). Reporting all three counts separately reveals the full capability profile of each condition rather than collapsing it to a single number.

\subsection{Scoring Protocol}\label{sec:eval-protocol}

Scoring proceeds in four steps: \circled{1}~qualification, \circled{2}~target matching, \circled{3}~evidence tiering, and \circled{4}~outcome classification.

\noindent\textbf{\circled{1}~Qualification.}
The judge reassesses \acs{cvss} v4.0 metrics from the available evidence under our system and threat model (cf. Section~\ref{sec:threat-model}). For example, a finding is qualified only if this reassessment assigns High or Critical severity.

\noindent\textbf{\circled{2}~Target matching.}
The judge compares each qualified finding from \circled{1} with the held-out advisory. A finding receives target credit only if it identifies the same vulnerable component and root-cause mechanism as the advisory. Matching the same repository, package, file, endpoint family, broad \acs{cwe}, or vulnerability class is supporting context but is not sufficient. A real vulnerability that differs from the held-out advisory is recorded as an off-target finding and receives no target credit.

\noindent\textbf{\circled{3}~Evidence tiering.}
The judge assigns the highest evidence tier supported by the matched finding, executing reproduction artifacts in an isolated workspace to verify T1 and T2 claims. The key distinction is whether the security impact itself was observed (T1), or only the vulnerable execution path was triggered without reaching full impact (T2), or no runtime evidence was available at all (T3):
\begin{itemize}
    \item \textbf{T1} --- \textit{End-to-end exploit success.} Runtime evidence demonstrates the full security consequence (e.g., unauthorized read or write, command execution, authentication bypass, path traversal, or SSRF).
    \item \textbf{T2} --- \textit{Bug activation with runnable PoC.} Runnable evidence shows the vulnerable logic was reached but the security impact was not observed (e.g., attacker-controlled input reaching the vulnerable sink, or an unsafe deserialization call invoked without achieving code execution).
    \item \textbf{T3} --- \textit{Speculative match.} The finding matches the advisory root cause but T1/T2 evidence is unavailable (static-only report, missing reproduction bundle, or PoC that fails before activating vulnerable logic).
\end{itemize}

\noindent\textbf{\circled{4}~Judged no-match.}
\tool produced at least one qualified finding under \circled{2}, but none matched the advisory.

\subsection{RQ1: Procedure Acquisition}\label{sec:rq1}

Evolving the playbook for \codexgpt{} raises the target-match rate from \GPTMinTargetMatchRate{} (\GPTMinTargetMatch{}/\EvalCaseCount{}) to \GPTEvolvedTargetMatchRate{} (\GPTEvolvedTargetMatch{}/\EvalCaseCount{}), with qualification rate rising from \GPTMinQualifiedRate{} (\GPTMinQualifiedFindings{}/\GPTMinJudgedFindings{}) to \GPTEvolvedQualifiedRate{} (\GPTEvolvedQualifiedFindings{}/\GPTEvolvedJudgedFindings{}). For GLM5.1, the match count rises from \GLMMinTargetMatch{} to \GLMEvolvedTargetMatch{} (\GLMMinTargetMatchRate{} $\to$ \GLMEvolvedTargetMatchRate{}). The two lifts are structurally different: the GPT gain is large in relative terms but from a low base (\GPTMinTargetMatchRate{}, \GPTMinTargetMatch{}/\EvalCaseCount{}); the GLM gain is smaller in absolute terms because the empty baseline already performs at a competitive level (detailed in Insight~2). The acquisition comparisons are based on the paired empty-versus-evolved conditions (\(\Egpt\) vs.\ \(\Bgpt\) and \(\Eglm\) vs.\ \(\Bglm\)), where model, harness, test cases, and task interface are fixed; \(\Bcs\) is a product-style reference line (cf.\ Table~\ref{tab:rq1-finding-outcomes}).

\begin{table}[t]
\caption{RQ1 Post-Judge High/Critical Finding Aggregate}
\label{tab:rq1-finding-outcomes}
\centering
\scriptsize
\resizebox{\columnwidth}{!}{%
\begin{tabular}{@{}lrrrrrr@{}}
\midrule
Condition & \circledtab{1} & \circledtab{2} & T1 & T2 & T3 & \circledtab{4} \\
\midrule
\(\Bcs\) & \CodexSecurityQualifiedFindings{}/\CodexSecurityJudgedFindings{} (\CodexSecurityQualifiedRate{}) & \CodexSecurityTargetMatch{}/\EvalCaseCount{} (\CodexSecurityTargetMatchRate{}) & \CodexSecurityTOne{} & \CodexSecurityTTwo{} & \CodexSecurityTThree{} & \CodexSecurityNoMatchFindings{} \\
\midrule
\(\Bgpt\) & \GPTMinQualifiedFindings{}/\GPTMinJudgedFindings{} (\GPTMinQualifiedRate{}) & \GPTMinTargetMatch{}/\EvalCaseCount{} (\GPTMinTargetMatchRate{}) & \GPTMinTOne{} & \GPTMinTTwo{} & \GPTMinTThree{} & \GPTMinNoMatchFindings{} \\
\(\Egpt\) & \GPTEvolvedQualifiedFindings{}/\GPTEvolvedJudgedFindings{} (\GPTEvolvedQualifiedRate{}) & \GPTEvolvedTargetMatch{}/\EvalCaseCount{} (\GPTEvolvedTargetMatchRate{}) & \GPTEvolvedTOne{} & \GPTEvolvedTTwo{} & \GPTEvolvedTThree{} & \GPTEvolvedNoMatchFindings{} \\
\midrule
\(\Bglm\) & \GLMMinQualifiedFindings{}/\GLMMinJudgedFindings{} (\GLMMinQualifiedRate{}) & \GLMMinTargetMatch{}/\EvalCaseCount{} (\GLMMinTargetMatchRate{}) & \GLMMinTOne{} & \GLMMinTTwo{} & \GLMMinTThree{} & \GLMMinNoMatchFindings{} \\
\(\Eglm\) & \GLMEvolvedQualifiedFindings{}/\GLMEvolvedJudgedFindings{} (\GLMEvolvedQualifiedRate{}) & \GLMEvolvedTargetMatch{}/\EvalCaseCount{} (\GLMEvolvedTargetMatchRate{}) & \GLMEvolvedTOne{} & \GLMEvolvedTTwo{} & \GLMEvolvedTThree{} & \GLMEvolvedNoMatchFindings{} \\
\bottomrule
\end{tabular}
}
\par\vspace{3pt}
\begin{minipage}{\columnwidth}
\scriptsize
\emph{Note:} \circledtab{1}: retained high/critical qualified findings / judged findings after case-level collapse of duplicated matched findings where applicable (see \circledtab{1}, Section~\ref{sec:eval-protocol}). \circledtab{2}: target-matched / \EvalCaseCount{} held-out cases. \circledtab{3}: evidence tier (T1 end-to-end exploit, T2 PoC activation, T3 speculative). \circledtab{4}: retained qualified no-match findings.
E.g., \(\Bcs\) contributes \CodexSecurityJudgedFindings{} judged finding entries under this accounting, of which \CodexSecurityQualifiedFindings{} are retained as high/critical qualified findings (\circledtab{1}); \CodexSecurityTargetMatch{}/\EvalCaseCount{} match a target advisory (\circledtab{2}) with T1/T2/T3 split \CodexSecurityTOne{}/\CodexSecurityTTwo{}/\CodexSecurityTThree{} (\circledtab{3}); the remaining \CodexSecurityNoMatchFindings{} findings are no-match (\circledtab{4}).
\end{minipage}
\end{table}

\begin{insight}{1}
Evolution teaches the agent \textbf{how to judge its findings}, not merely where to look. For \codexgpt{}, qualification rate nearly triples (\GPTMinQualifiedRate{} $\to$ \GPTEvolvedQualifiedRate{}, $+16$ pp) while target-match rate increases by only 4.6 pp. Most of the acquisition gain is signal quality, not discovery of new vulnerability locations.
\end{insight}

\begin{insight}{2}
With minimal guidance, \glmopen{} reaches \GLMMinTargetMatchRate{} target-match rate (\GLMMinTargetMatch{}/\EvalCaseCount{}), comparable to \codexsecurity{} (\CodexSecurityTargetMatchRate{}) while generating $4{\times}$ fewer judged findings. The conditions are not controlled (model, harness, and workflow all differ), but reaching comparable performance with an empty playbook reflects strong inherent model capability.
\end{insight}

\subsubsection{What the Playbooks Learn}

Because the evolved playbooks are versioned text artifacts, what the agent learns is directly inspectable. Table~\ref{tab:playbook-content-summary} and Appendix~\ref{app:playbook-pointer-histories} summarize the two playbooks and their revision histories. Both runs start from an empty-README seed over the same number of rounds, but diverge structurally from the first accepted revision: \codexgpt{} immediately creates a workflow, an index, and 9 modular vulnerability guides; \glmopen{} creates a workflow and one centralized vulnerability-class document.

\begin{table}[t]
\caption{Selected Playbook and Pointer-History Summary}
\label{tab:playbook-content-summary}
\centering
\scriptsize
\begin{tabular}{lrr}
\hline
Property & \(\Pgptstar\) & \(\Pglmstar\) \\
\hline
Retained candidate revisions & 82 & 82 \\
Pointer rows in history & 40 & 40 \\
Rows with multiple candidates & 29 & 18 \\
Max candidates per incumbent & 5 & 7 \\
Accepted best-pointer moves & 38 & 38 \\
First accepted update (ins.) & 564 & 231 \\
First accepted update (files) & 13 & 2 \\
Final selected playbook size & 1,616 lines & 2,177 lines \\
Workflow text & 234 lines & 284 lines \\
Knowledge text & 1,340 lines & 1,892 lines \\
Knowledge organization & 19 guides & 74 sections \\
Moves touching workflow \& knowledge & 38/38 & 36/38 \\
\hline
\end{tabular}
\par\vspace{3pt}
\begin{minipage}{\columnwidth}
\scriptsize
\emph{Note:} Candidate histories are shown as best-pointer rows in Appendix~\ref{app:playbook-pointer-histories}; the final selected playbook is the blue P40 incumbent in each history.
\end{minipage}
\end{table}

Two observations emerge from these numbers. First, the learned playbook is not a growing checklist of vulnerability patterns: in both runs, nearly every accepted revision modifies workflow and knowledge files together, confirming that what is learned is procedural (how to conduct an audit), not merely factual (cf. Table~\ref{tab:playbook-content-summary}). Second, surprisingly, the structural divergence is an early choice, not a late convergence: each run commits to its form in the first accepted revision and preserves it. GPT5.4 adopts a modular representation; GLM5.1 adopts a monolithic class catalogue.

\begin{figure*}[!t]
\centering
\includegraphics[width=0.98\textwidth]{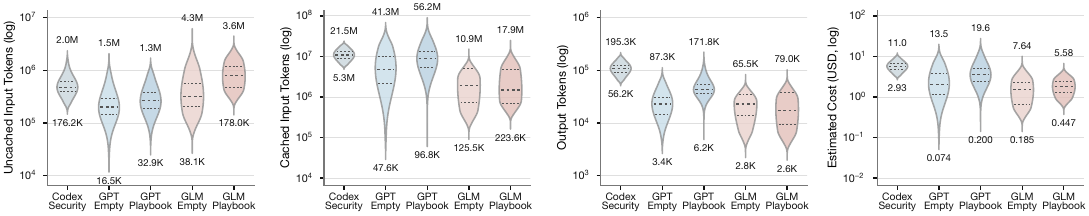}
\caption{Token usage and estimated API cost distributions for \codexsecurity{}, \codexgpt{} Empty and Playbook conditions, and \glmopen{} Empty and Playbook conditions. Cost estimates apply provider-list pricing at the time of the experiment: \codexgpt{} at \$2.50\,/\,\$15\,/\,\$0.25 per million input\,/\,output\,/\,cache-read tokens (OpenAI); \glmopen{} at \$1.40\,/\,\$4.40\,/\,\$0.26 (Z.ai). \codexsecurity{} costs reflect product-level usage and are not derived from per-token rates.}
\label{fig:token-cost-4panel}
\end{figure*}

\begin{insight}{3}
The playbooks converge on \textbf{opposite quality versus coverage tradeoffs}. Every target-matched finding from \(\Egpt\) is T1 (\GPTEvolvedTOne{}/\GPTEvolvedTTwo{}/\GPTEvolvedTThree{}): a high-quality working end-to-end exploit with full runtime proof. This is an extremely high bar: the playbook has learned to reject anything it cannot fully materialize. \(\Eglm\) matches more targets in absolute terms (\GLMEvolvedTargetMatch{} vs.\ \GPTEvolvedTargetMatch{}) by accepting a broader evidence range (\GLMEvolvedTOne{}/\GLMEvolvedTTwo{}/\GLMEvolvedTThree{}): more coverage, lower per-finding confidence.
\end{insight}

\begin{insight}{4}
The playbooks diverge in search breadth. \(\Pglmstar\) is \textbf{explicitly designed to resist stopping}: its workflow contains eight ``CRITICAL: continue auditing ALL other vulnerability classes'' directives, three mandatory re-sweep gates (DoS, misconfiguration, build pipeline), and a required anchoring-break checklist after every finding. \(\Pgptstar\) instead caps active families at 3--6 and treats audit-budget control as a first-class design goal. Finding volume reflects this: \(\Eglm\) generates \GLMEvolvedJudgedFindings{} judged findings versus \(\Egpt\)'s \GPTEvolvedJudgedFindings{}, with target-match rates of \GLMEvolvedTargetMatchRate{} and \GPTEvolvedTargetMatchRate{} respectively.
\end{insight}

\begin{designremark}{Does GLM5.1 outperform GPT5.4-xhigh?}
Not straightforwardly. The comparison is complicated by a harness confound: GPT5.4-xhigh runs under Codex and GLM5.1 under OpenCode, so model and environment cannot be fully separated. That said, \GLMMinTargetMatch{} vs.\ \GPTMinTargetMatch{} target matches on the empty-playbook baseline suggests the difference is partly intrinsic to the model. As Insights 3 and 4 show, the two optimize for different objectives: GLM5.1 sweeps broadly, accepts varied evidence tiers, and is explicitly designed to resist early stopping; GPT5.4-xhigh demands a working end-to-end exploit on every match and caps search breadth. Neither profile dominates: broad recall favors GLM5.1; high-confidence T1 evidence favors GPT5.4-xhigh. The practical choice depends on triage capacity: if reviewers have limited time and want low-noise, immediately actionable findings, GPT5.4-xhigh's all-T1 profile is preferable; if broad vulnerability coverage matters more and the team can triage mixed-confidence findings, GLM5.1 is the stronger choice. These characterizations are grounded in manual inspection of \tool traces.
\end{designremark}

\noindent\textbf{Workflow.}
The \codexgpt{} workflow is a decision system for ranking competing root causes. It frames the repository through entrypoints and trust boundaries, builds a release-surface map, records 3--6 live component families, uses repository-local security signals to keep exact primitives live, and performs sibling and boundary-fit checks before finalizing. The important action is not simply ``look at more code''; it is to explain why a central default path, named guard, or same-family sibling no longer dominates the accepted finding. The \glmopen{} workflow is a phased coverage process: reconnaissance, systematic vulnerability-class audit, PoC construction, evidence collection, and prioritization. Its strongest rules are stop-prevention rules: after SSRF, RCE, SQL injection, hardcoded credentials, or any DoS, continue scanning other classes and components.

\noindent\textbf{Knowledge.}
The knowledge layer mirrors the workflow layer. The \codexgpt{} playbook turns each vulnerability family into a small validation program with a common schema: definition, discovery focus, validation focus, false-positive traps, \acs{cvss} notes, and evidence checklist. This organization supports selective invocation: the auditor can pull the relevant guide when a component family becomes live. The \glmopen{} playbook instead accumulates a broad catalogue of long-tail failure modes, including MyBatis mapper SQL injection, PHP deserialization variants, Python class pollution, HTTP transport DoS, and storage override path traversal, among others. The design intent is broader recall across rare patterns.

The deeper distinction is not workflow versus knowledge: in both histories, the two co-evolve. What is learned is procedural security knowledge: rules for allocating audit attention, keeping competing hypotheses live, rejecting real but off-target findings, and demanding runtime evidence before finalization. RQ2 asks whether that learned procedure, developed by a stronger agent, transfers to weaker models.

\noindent\textbf{Non-monotone improvement and breadth cost.}
Playbook evolution improves aggregate performance. However, in approximately 2--7\% of cases, the empty-playbook run recovers the target vulnerability while the evolved playbook run does not. The likely mechanism is search focus: a learned procedure directs audit budget toward prioritized component families and validation gates, which can cause the agent to miss lower-complexity bugs that a less directed run would encounter opportunistically. These case-level regressions should be read alongside aggregate gains. We also note that each playbook is evolved in a single training run due to cost constraints; randomness in the evolution loop means results could plausibly differ under reruns.

\subsection{RQ2: Procedure Transfer}\label{sec:rq2}

Transfer works for both student models, and teacher-student compatibility matters. For the \qwendense{} student, the GPT teacher raises target matches from \QwenDenseMinTargetMatch{}/\EvalCaseCount{} to \QwenDenseTransferTargetMatch{}/\EvalCaseCount{} (\QwenDenseMinTargetMatchRate{} $\to$ \QwenDenseTransferTargetMatchRate{}), while the GLM teacher raises them further to \GLMDenseTransferTargetMatch{}/\EvalCaseCount{} (\GLMDenseTransferTargetMatchRate{}); qualification rate rises from \QwenDenseMinQualifiedRate{} to \QwenDenseTransferQualifiedRate{} under GPT transfer and to \GLMDenseTransferQualifiedRate{} under GLM transfer. For the \qwenmoe{} student, the GPT teacher raises target matches from \QwenMoEMinTargetMatch{} to \QwenMoETransferTargetMatch{}; the GLM teacher raises them to \GLMMoETransferTargetMatch{} (\GLMMoETransferTargetMatchRate{}), outperforming the GPT teacher by $2.4{\times}$ on this student. Transfer comparisons are paired: each condition is compared to the same student's empty-playbook baseline. Table~\ref{tab:rq2-finding-outcomes} reports the post-judge aggregate.

\begin{table}[t]
\caption{RQ2 Post-Judge High/Critical Finding Aggregate}
\label{tab:rq2-finding-outcomes}
\centering
\scriptsize
\resizebox{\columnwidth}{!}{%
\begin{tabular}{@{}lrrrrrr@{}}
\midrule
Condition & \circledtab{1} & \circledtab{2} & T1 & T2 & T3 & \circledtab{4} \\
\midrule
\(\StudentBaseline{A3B}\) & \QwenMoEMinQualifiedFindings{}/\QwenMoEMinJudgedFindings{} (\QwenMoEMinQualifiedRate{}) & \QwenMoEMinTargetMatch{}/\EvalCaseCount{} (\QwenMoEMinTargetMatchRate{}) & \QwenMoEMinTOne{} & \QwenMoEMinTTwo{} & \QwenMoEMinTThree{} & \QwenMoEMinNoMatchFindings{} \\
\(\StudentTransfer{A3B}\) & \QwenMoETransferQualifiedFindings{}/\QwenMoETransferJudgedFindings{} (\QwenMoETransferQualifiedRate{}) & \QwenMoETransferTargetMatch{}/\EvalCaseCount{} (\QwenMoETransferTargetMatchRate{}) & \QwenMoETransferTOne{} & \QwenMoETransferTTwo{} & \QwenMoETransferTThree{} & \QwenMoETransferNoMatchFindings{} \\
\(\StudentGLMTransfer{A3B}\) & \GLMMoETransferQualifiedFindings{}/\GLMMoETransferJudgedFindings{} (\GLMMoETransferQualifiedRate{}) & \GLMMoETransferTargetMatch{}/\EvalCaseCount{} (\GLMMoETransferTargetMatchRate{}) & \GLMMoETransferTOne{} & \GLMMoETransferTTwo{} & \GLMMoETransferTThree{} & \GLMMoETransferNoMatchFindings{} \\
\midrule
\(\StudentBaseline{27B}\) & \QwenDenseMinQualifiedFindings{}/\QwenDenseMinJudgedFindings{} (\QwenDenseMinQualifiedRate{}) & \QwenDenseMinTargetMatch{}/\EvalCaseCount{} (\QwenDenseMinTargetMatchRate{}) & \QwenDenseMinTOne{} & \QwenDenseMinTTwo{} & \QwenDenseMinTThree{} & \QwenDenseMinNoMatchFindings{} \\
\(\StudentTransfer{27B}\) & \QwenDenseTransferQualifiedFindings{}/\QwenDenseTransferJudgedFindings{} (\QwenDenseTransferQualifiedRate{}) & \QwenDenseTransferTargetMatch{}/\EvalCaseCount{} (\QwenDenseTransferTargetMatchRate{}) & \QwenDenseTransferTOne{} & \QwenDenseTransferTTwo{} & \QwenDenseTransferTThree{} & \QwenDenseTransferNoMatchFindings{} \\
\(\StudentGLMTransfer{27B}\) & \GLMDenseTransferQualifiedFindings{}/\GLMDenseTransferJudgedFindings{} (\GLMDenseTransferQualifiedRate{}) & \GLMDenseTransferTargetMatch{}/\EvalCaseCount{} (\GLMDenseTransferTargetMatchRate{}) & \GLMDenseTransferTOne{} & \GLMDenseTransferTTwo{} & \GLMDenseTransferTThree{} & \GLMDenseTransferNoMatchFindings{} \\
\bottomrule
\end{tabular}
}
\par\vspace{3pt}
\begin{minipage}{\columnwidth}
\scriptsize
\emph{Note:} Paired comparison: each transfer condition vs.\ the same student's empty-playbook baseline. \circledtab{1}: retained high/critical qualified findings / judged findings. \circledtab{2}: target-matched / \EvalCaseCount{} held-out cases. T1/T2/T3: end-to-end exploit, PoC activation, speculative. \circledtab{4}: retained qualified no-match findings.
\end{minipage}
\end{table}

\begin{insight}{5}
\textbf{The teacher's search strategy propagates to the student.} The GPT teacher condenses student output: both students generate fewer judged findings under GPT transfer (\qwenmoe{}: \QwenMoEMinJudgedFindings{}$\to$\QwenMoETransferJudgedFindings{}; \qwendense{}: \QwenDenseMinJudgedFindings{}$\to$\QwenDenseTransferJudgedFindings{}) while qualification rate climbs. The GLM teacher expands output for both students: \qwenmoe{} rises to \GLMMoETransferJudgedFindings{} judged findings and \qwendense{} to \GLMDenseTransferJudgedFindings{}, with larger qualification gains in both cases. Evidence tier signatures mirror the teacher: \qwendense{} under GPT transfer produces T1/T2/T3 of \QwenDenseTransferTOne{}/\QwenDenseTransferTTwo{}/\QwenDenseTransferTThree{}, matching the GPT teacher's precision style; \qwendense{} under GLM transfer produces \GLMDenseTransferTOne{}/\GLMDenseTransferTTwo{}/\GLMDenseTransferTThree{}, and \qwenmoe{} under GLM transfer produces \GLMMoETransferTOne{}/\GLMMoETransferTTwo{}/\GLMMoETransferTThree{}, mirroring the GLM teacher's mixed-tier acceptance.
\end{insight}

\begin{insight}{6}
Teacher choice matters more than teacher quality for the A3B student. The GLM teacher gives \GLMMoETransferTargetMatch{} target matches (\GLMMoETransferTargetMatchRate{}); the GPT teacher gives the same student only \QwenMoETransferTargetMatch{} (\QwenMoETransferTargetMatchRate{}), a $2.4{\times}$ gap. The exhaustive coverage strategy of \(\Pglmstar\) is a better fit for a model that benefits from being told what to check rather than what to skip. The qualification rate lift tells the same story: the GLM teacher raises \qwenmoe{} qualification by $+24.3$ pp versus the GPT teacher's $+3.4$ pp.
\end{insight}

\begin{insight}{7}
Signal quality transfers more efficiently than raw discovery. For the 27B student, the GPT teacher lifts qualification rate by $+18.7$ pp (\QwenDenseMinQualifiedRate{} $\to$ \QwenDenseTransferQualifiedRate{}), slightly exceeding the teacher's own self-evolution lift ($+16.4$ pp); the GLM teacher lifts the same student by $+43.9$ pp (\QwenDenseMinQualifiedRate{} $\to$ \GLMDenseTransferQualifiedRate{}). A student model following the right procedure can absorb the teacher's filtering discipline even when it does not replicate the teacher's full target-match profile. This suggests that explicit rules for comparison, pruning, and evidence thresholds are highly portable; the residual gap to the teacher reflects model-intrinsic capabilities (context management, reasoning depth, and tool use) that procedure cannot close.
\end{insight}

\vspace{-\smallskipamount}
\begin{designremark}{Why does the GLM5.1 playbook transfer more coverage to weaker students, while the GPT5.4-xhigh playbook transfers better precision?}
The two playbooks encode fundamentally different search policies. The GPT-evolved playbook is pruning-oriented: it asks the agent to maintain a bounded checklist of 3--6 first-class component families, to make active decisions about which families to clear or prune, and to prioritize central attack surfaces over adjacent ones. This strategy works well for a strong model that can execute those implicit judgments reliably (which families are worth keeping, when to stop, how to upgrade a finding into a qualifying report). The GLM-evolved playbook instead externalizes those judgments as explicit procedural steps: it repeats ``enumerate ALL'' and ``continue auditing'' over thirty times in total, requires per-endpoint and per-parameter audit passes, mandates anti-anchoring after any discovered bug, and specifies root-cause alignment as a required final step before submission. For a weaker student model, this explicitness matters: the agent does not need to infer the audit schedule or stopping rule on its own. The cost is lower per-finding precision and larger no-match counts, which reflects the same breadth-versus-precision tradeoff visible in the teachers themselves (cf.\ Insight~3). In short: the GPT playbook is a frontier-model policy that requires the student to fill in the gaps; the GLM playbook is an enumeration protocol that works even when the student cannot. These characterizations are grounded in manual inspection of \tool traces across both student models.
\end{designremark}

\subsection{False Positive Rate}\label{sec:fpr}

To estimate precision, we sampled \circledtab{4} judged no-match qualified findings from the ten \tool-run conditions, excluding the \(\Bcs\) product baseline, using a 90\% confidence level and 15\% margin of error. This resulted in 222 sampled findings: \(\Bgpt\) ($n{=}17$), \(\Egpt\) ($n{=}25$), \(\Bglm\) ($n{=}27$), \(\Eglm\) ($n{=}29$), \(\StudentBaseline{A3B}\) ($n{=}17$), \(\StudentTransfer{A3B}\) ($n{=}18$), \(\StudentGLMTransfer{A3B}\) ($n{=}26$), \(\StudentBaseline{27B}\) ($n{=}15$), \(\StudentTransfer{27B}\) ($n{=}21$), and \(\StudentGLMTransfer{27B}\) ($n{=}27$). Three authors participated; each sampled finding was independently cross-validated by two reviewers, with disagreements reconciled before labeling true or false positive. The false positive rates are: \(\Bgpt\), \FPRCodexDryrun{}; \(\Egpt\), \FPRCodexRun{}; \(\Bglm\), \FPRGLMDryrun{}; \(\Eglm\), \FPRGLMRun{}; \(\StudentBaseline{A3B}\), \FPRQwenAThreeBDryrun{}; \(\StudentTransfer{A3B}\), \FPRQwenAThreeBGPTRun{}; \(\StudentGLMTransfer{A3B}\), \FPRQwenAThreeBGLMRun{}; \(\StudentBaseline{27B}\), \FPRQwenTwentySevenBDryrun{}; \(\StudentTransfer{27B}\), \FPRQwenTwentySevenBGPTRun{}; and \(\StudentGLMTransfer{27B}\), \FPRQwenTwentySevenBGLMRun{}.

\textbf{The most striking result is \(\Egpt\) at \FPRCodexRun{}, down from \FPRCodexDryrun{} for \(\Bgpt\)}: with the evolved playbook in place, zero of 25 sampled findings were false positives. This suggests that the T1-only evidence standard and qualification gates encoded by the GPT-evolved playbook filter out speculative findings before they reach the reporting stage, rather than shifting them to a later review step.

\section{Discussion}\label{sec:discussion}

\tool does not claim procedure replaces model capability; it measures how much security auditing moves into an inspectable artifact. Separating procedure from model and harness matters because the two are often changed together, making it unclear why performance improves.

\noindent\textbf{What playbook transfer means.} If an evolved playbook improves the same fixed agent over an empty playbook, the result supports procedure acquisition: repeated attempts and grounded evaluation can distill audit behavior into an external artifact. If the same playbook improves weaker students, the result supports procedure transfer: at least part of the teacher's advantage can be expressed as guidance that another model can follow.

\noindent\textbf{Procedure as an alternative to fine-tuning.} Fine-tuning specializes a model by updating weights on domain-specific data, producing a checkpoint coupled to a specific architecture and training framework. Playbook evolution differs in three structural respects. \emph{Model agnosticism}: the playbook is a text artifact that loads into any compatible agent; switching base models requires no retraining. \emph{Continuous deployment}: because the playbook is versioned, any intermediate revision is immediately deployable (no training-convergence barrier), and regressions can be rolled back by reverting a commit. \emph{Training-inference cost decoupling}: evolution uses a frontier model only to generate the procedure; inference can then be served by cheaper open-source models that execute the produced playbook, separating the cost of procedure quality from the cost of deployment at scale.

\noindent\textbf{Capability transfer at low marginal cost.} The transfer results suggest a practical cost model: a frontier model runs the evolution loop once to produce a procedure; smaller open-source models then execute that procedure repeatedly at low marginal cost. A meaningful fraction of the frontier model\'s audit advantage is procedure-encodable, expressible as explicit text rules rather than tacit, weight-encoded capability. This differs from classical knowledge distillation, which requires training compute and labeled data to compress capability into smaller weights. The distilled artifact here is a text playbook: inspectable, versioned, and transferable across model generations without gradient updates. Organizations with limited inference budgets can invest in a one-time evolution campaign and amortize that cost across many open-source deployments.

\noindent\textbf{Evolution campaign cost.} The playbook evolution campaigns ran four GPT-side accounts (GPT Pro 20x) and four GLM-side accounts (GLM Coding Plan Max) concurrently for one calendar month. Accounting for mid-month token-quota resets, the total training expenditure was approximately \$200 per GPT account and \$144 per GLM account, for a combined evolution budget of roughly \$1{,}400. Three properties of this budget merit emphasis. First, the cost is a ceiling rather than a per-token bill: subscription-based access caps spending regardless of how many advisory rollouts the evolution loop completes. Second, the loop is restartable: each accepted playbook revision is a git commit, so an interrupted campaign resumes from the last accepted state rather than from scratch. Third, the training implementation is decoupled from the produced artifact: a future campaign could substitute a different harness, teacher model, or training corpus without discarding the evolved playbook.

\noindent\textbf{Per-case inference cost.} Figure~\ref{fig:token-cost-4panel} shows per-case cost distributions across conditions. Median per-case cost is approximately \resval{\$2} for \codexgpt{} (\(\Egpt\)) and \resval{\$1.6} for \glmopen{} (\(\Eglm\)); Qwen student conditions are estimated at \empirical{\$0.8} (\qwendense{}) and \empirical{\$0.5} (\qwenmoe{}) from provider pricing. Transfer adds playbook tokens without extra API calls, so per-case cost is nearly identical to the baseline. Deploying the GPT-evolved playbook on \qwendense{} (\(\StudentTransfer{27B}\)) recovers \QwenDenseTransferTargetMatchRate{} target-match rate at roughly $3{\times}$ lower cost than the teacher. Switching to Qwen inference reduces per-token cost by $2{\times}$ for \qwendense{} (\$0.45/\$2.70 vs.\ \$1.40/\$4.40 per million input/output) and $3.5{\times}$ for \qwenmoe{} (\$0.20/\$1.60), at a performance trade-off from \GLMEvolvedTargetMatchRate{} (\(\Eglm\)) to \GLMDenseTransferTargetMatchRate{} (\(\StudentGLMTransfer{27B}\)) or \GLMMoETransferTargetMatchRate{} (\(\StudentGLMTransfer{A3B}\)).

\noindent\textbf{Safety and responsible use.} \tool studies vulnerability discovery and validation, so the system can generate exploit hypotheses and reproduction artifacts. The experiments are scoped to local, source-available targets and known advisories, and the benchmark excludes settings that require live third-party systems or production infrastructure. Local reproduction constraints are important for measurement and safety: they let the evaluator demand concrete evidence without encouraging uncontrolled exploitation. The same constraints should apply when releasing artifacts, where playbooks, traces, and reproduction scripts should be reviewed to avoid publishing operational exploit details beyond what is already disclosed for known advisories.

\section{Scope and Limitations}\label{sec:limitations}

\noindent\textbf{Benchmark scope.} The benchmark is deliberately narrower than security auditing in practice. We focus on source-available repositories, high and critical advisories, \acs{cvss}-qualified reachability, and cases investigable locally, excluding vulnerability classes that require live cloud state, privileged infrastructure, or conditions not locally reproducible.

\noindent\textbf{Off-target findings.} A run is scored as a failure when it finds a valid, security-relevant vulnerability different from the designated benchmark target. Such off-target findings are counted as misses, so the benchmark can underestimate broader auditing value when the agent discovers a different vulnerability family. A concrete example: in \href{https://github.com/advisories/GHSA-g7j6-fmwx-7vp8}{GHSA-g7j6-fmwx-7vp8}, a run failed to recover the benchmark target but discovered a GraphQL asset-upload SSRF that matches the later-confirmed advisory \href{https://github.com/advisories/GHSA-3m9m-24vh-39wx}{GHSA-3m9m-24vh-39wx}, a case outside the benchmark due to the dataset cutoff date. Benchmark failure does not equal no useful finding.

\noindent\textbf{Threat models and disclosure outcomes.} Disclosure outcomes do not collapse cleanly into accepted versus incorrect. Some reports expose risk whose fix belongs to an upstream dependency rather than the audited application. Others reflect threat-model variance: an agent may identify a capability that a maintainer considers intentional for a trusted-user deployment, development tool, or administrative platform. A third category remains pending. Appendix~\ref{app:reported-vulnerabilities} therefore reports only the externally trackable subset.

\noindent\textbf{Evaluator constraints.} Any evaluator can misjudge borderline cases, especially when a finding is adjacent to the ground truth; this risk is sharper because our adjudicator includes an \ac{llm}. We reduce it by decomposing judgment into target-match and runtime-evidence axes, requiring evidence for top tiers, and separating off-target findings from target credit. These controls do not eliminate evaluator error.

\noindent\textbf{Distribution shift.} Playbook learning can overfit to the advisory distribution. A temporal split reduces direct leakage, and project overlap is measured rather than hidden, but the evolved procedure may still specialize to common ecosystems, common \acs{cwe} classes, or the reporting style of the GitHub Advisory Database. Transfer experiments partially address this risk because a teacher playbook must remain useful under different student models, but broader future work should evaluate across additional advisory sources, private audits, and prospective undisclosed vulnerability tasks.

\noindent\textbf{Single training run.} Each playbook is evolved once; cost constraints make repeated runs prohibitively expensive, so we cannot assess how much the outcome depends on randomness in the loop. We treat the reported playbooks as existence proofs that the loop can produce useful procedures, not characterizations of the typical outcome.

\noindent\textbf{Hyperparameter choices.} Hyperparameters (replay ratio, batch size, scoring weights, revision gate limits) are set by judgment rather than systematic search. Full LLM audit rollouts make grid search infeasible, and optimizing against the advisory benchmark risks overfitting its distribution. We treat systematic hyperparameter selection as future work.

\noindent\textbf{Threats to validity.} Model and execution environment are not fully disentangled: \codexgpt{} runs under Codex while \glmopen{} and both Qwen students run under OpenCode, so observed differences between these groups may reflect harness effects rather than model or playbook effects alone. Additionally, evolved playbooks are calibrated to specific model checkpoints; silent provider-side updates may alter model behavior and reduce playbook effectiveness without a version identifier change. Finally, advisory details in our benchmark may appear in the pre-training corpora of evaluated models, making it difficult to fully rule out partial memorization on specific cases. Our scoring also relies on an \ac{llm} judge, and manual validation is limited to a consistency scan of judge-qualified findings plus a sampled deep review, so borderline target-match, severity, and evidence-tier decisions may still contain errors.

\section{Related Work}\label{sec:related-work}

Our work sits at the intersection of agentic LLM systems for security auditing and the hypothesis that audit procedures can be externalized and evolved independently of the underlying model and harness. We survey five related threads: agentic security testing in specialized domains, repository-level vulnerability detection and code auditing, analysis and validation scaffolds, benchmarks, and approaches leveraging external knowledge or learned procedures.

\noindent\textbf{Agentic security testing and specialized domains.} Security is a natural target for agentic \ac{llm}s because effective analysis requires long-horizon tool interaction, source code inspection, and evidence construction. PentestGPT and web-focused multi-agent work organize \ac{llm}s for penetration testing~\cite{deng2024pentestgpt,david2025multi}; other systems target vulnerability management, Android discovery, web reconstruction, blockchain postmortems, multi-agent exploitation, protocol fuzzing, and IoT firmware~\cite{liu2024chatgptvulnmgmt,wang2025a2,allen2024webrr,wang2026txray,zhu2026teams,meng2024chatafl,song2026protocolguard,ji2026firmagent}. Complementary work finds that architecture choice meaningfully affects coverage and efficiency~\cite{david2026towards}. These systems typically bake in a domain-specific workflow; \tool instead studies whether a reusable audit procedure can be learned from validated attempts while model / harness remain fixed.

\noindent\textbf{Repository-level vulnerability detection and code auditing.} The closest direct line treats vulnerability discovery as a repository auditing problem rather than a single-function classification task. PrimeVul, ReposVul, VulEval, and JITVul construct datasets with repository-level context and agent settings~\cite{ding2025primevul,wang2024reposvul,wen2024vuleval,yildiz2025jitvul}. IRIS, LLMxCPG, RepoAudit, and ReasonVul add static analysis, graph context, memory, or multi-agent reasoning for repository-level auditing~\cite{li2025iris,lekssays2025llmxcpg,guo2025repoaudit,peng2026reasonvul}. \tool differs by treating the audit procedure as the learned object, guided by a versioned playbook rather than a fixed system design.

\noindent\textbf{LLM assisted analysis and validation.} Another line decomposes security reasoning into program analysis and validation subtasks. LLMDFA and LLMSAN use \ac{llm}s for data-flow analysis and hallucination reduction~\cite{wang2024llmdfa,wang2024llmsan}; GPTAid and Midas extend decomposition to API security rules and resource management misuse~\cite{liu2025gptaid,yang2025midas}. \tool applies the same externalization principle but targets the full audit procedure rather than a single analysis pass.

\noindent\textbf{Benchmark design and off-target findings.} Recent cybersecurity benchmarks bind success to the specific target vulnerability rather than any observable security effect. CyberGym and ExploitGym verify PoCs against pre/post-patch behavior and distinguish flag capture from intended exploit~\cite{wang2026cybergym,wang2026exploitgym}. ExploitBench and RealVuln grade progress through capability-ladder oracles or bind scanner output to ground truth~\cite{lee2026exploitbench,pellew2026realvuln}. \tool follows the same principle: the primary score counts only root-cause-equivalent matches, and off-target findings are recorded separately.

\noindent\textbf{External knowledge and learned procedure.} Other work externalizes knowledge or experience for later agent behavior. Vul-RAG retrieves historical CVE knowledge for vulnerability detection~\cite{du2024vulrag}. BugScope derives detection prompts from historical bug reports; its learned artifact is a checker-style procedure for selected anti-patterns rather than a versioned playbook tested for transfer~\cite{guo2025bugscope}. Agent learning work (ExpeL, Memento, Agent Lightning) improves agents from experience without replacing the base model~\cite{zhao2023expel,zhou2025memento,luo2025agentlightning}. \tool differs in that the learned artifact is a versioned playbook repository selected through branch tournaments and tested for cross-model transfer, where a stronger teacher's procedure is transferred to weaker student models.

\section{Conclusion}\label{sec:conclusion}

\tool keeps the model / harness fixed and evolves a versioned audit playbook. Our results show that procedure alone can lift auditing performance, and that a procedure ``distilled'' by a stronger teacher transfers to benefit weaker student models. Security auditing is not only a race toward stronger models; disciplined coverage, validation, and evidence construction are teachable procedures that can be learned from validated attempts, versioned as external artifacts, and transferred across agents. The results offer an analogue to the bitter lesson: durable gains come from learning and search, not from freezing expert knowledge.

\bibliographystyle{IEEEtran}
\bibliography{ref}

\appendices

\section{Project-Reported Vulnerabilities}
\label{app:reported-vulnerabilities}

Table~\ref{tab:reported-vulnerabilities} lists the \ConfirmedVulnerabilityCount{} confirmed vulnerabilities.

\begingroup
\setlength{\tabcolsep}{2pt}
\renewcommand{\arraystretch}{1.06}
\begin{table}[!htbp]
\centering
\caption{\ConfirmedVulnerabilityCount{} Confirmed Vulnerabilities}
\label{tab:reported-vulnerabilities}
\scriptsize
\begin{tabular}{@{}>{\centering\arraybackslash}l>{\raggedright\arraybackslash}l>{\raggedright\arraybackslash}l>{\raggedright\arraybackslash}l@{}}
\toprule
CVSS & Advisory / CVE & Status & Class \\
\midrule
8.7 & GHSA-j8vf-vqr9-\maskid{} & Fixed & DoS \\
5.4 & GHSA-jfc7-64v2-\maskid{} / CVE-2026-4\maskid{} & Published & Integrity Bypass \\
7.5 & GHSA-52v5-jr5w-\maskid{} / CVE-2026-4\maskid{} & Published & Verification Bypass \\
6.5 & GHSA-xgjw-pm74-\maskid{} / CVE-2026-4\maskid{} & Published & Auth Bypass \\
8.2 & GHSA-p2f4-r6v6-\maskid{} & Accepted & Credential Leakage \\
6.9 & GHSA-64mm-vxmg-\maskid{} & Accepted & Routing Bypass \\
5.3 & GHSA-hgw6-g7cp-\maskid{} & Accepted & DoS \\
7.0 & GHSA-c4c3-pg64-\maskid{} & Accepted & Prototype Pollution \\
7.5 & GHSA-cgwc-pv48-\maskid{} / CVE-2026-4\maskid{} & Published & DoS \\
7.5 & GHSA-m9gh-vj53-\maskid{} / CVE-2026-4\maskid{} & Published & DoS \\
7.5 & GHSA-5w7q-77mv-\maskid{} / CVE-2026-4\maskid{} & Published & DoS \\
5.0 & GHSA-p43p-whwx-\maskid{} & Fixing & DoS \\
7.5 & GHSA-38rv-x7px-\maskid{} / CVE-2026-\maskid{} & Fixing & DoS \\
7.5 & GHSA-836r-79rf-\maskid{} / CVE-2026-4\maskid{} & Published & ReDoS \\
7.5 & GHSA-2wc2-fm75-\maskid{} / CVE-2026-4\maskid{} & Published & DoS \\
7.5 & GHSA-j5g9-f88f-\maskid{} & Fixing & DoS \\
5.1 & CAN-2026-203\maskid{} & Fixed & DoS \\
6.5 & GHSA-4xfr-4p46-\maskid{} & Fixed & IDOR \\
7.1 & GHSA-8jj7-4v57-\maskid{} & Fixed & DoS \\
6.5 & GHSA-vgxm-h9gx-\maskid{} & Fixed & Auth Bypass \\
4.8 & GHSA-fwjf-m4qw-\maskid{} & Fixed & Cache Poisoning \\
5.2 & GHSA-cpgj-f7g3-\maskid{} / CVE-2026-4\maskid{} & Fixed & Sandbox Bypass \\
7.1 & Pending CVE & Fixed & SQL Injection \\
5.1 & GHSA-7gcf-g7xr-\maskid{} & Fixed & DoS \\
5.1 & GHSA-vx77-6w57-\maskid{} & Accepted & Path Traversal \\
6.8 & GHSA-vx77-6w57-\maskid{} & Accepted & DoS \\
8.7 & RUSTSEC-2026-01\maskid{} & Fixed & DoS \\
7.5 & Binance Bug Bounty (P2) & Awarded & DoS \\
\bottomrule
\end{tabular}
\end{table}
\endgroup

\section{Covered Advisory Cases by Vulnerability Type}
\label{app:coverage-table}

Table~\ref{tab:coverage-by-vuln-type} lists advisory cases covered under our system and threat models. Model columns use the experimental condition symbols from Table~\ref{tab:conditions}; the Orig. column gives the advisory CVSS 4.0 score, and covered cells show validated target tiers with rerated CVSS scores in parentheses.

\section{Playbook Pointer Histories}
\label{app:playbook-pointer-histories}

Figure~\ref{fig:playbook-pointer-histories} compares the paper-normalized pointer histories for both evolved playbooks. Each trace is rendered as a single chronological column from P01 to P40; the left panel shows \codexgpt{}, the middle panel shows \glmopen{}, and the caption summarizes the evolution on the right and below the plots.

\pagebreak

\begingroup
\setlength{\tabcolsep}{0.45pt}
\renewcommand{\arraystretch}{0.45}
\begin{table*}[p]
\centering
\caption{Advisory Cases Covered by Model--Scaffold Combinations}
\label{tab:coverage-by-vuln-type}
\fontsize{5.5}{5.55}\selectfont
\newcommand{\coverageRule}{\specialrule{0.25pt}{0.95pt}{0.95pt}}
\resizebox{\textwidth}{!}{%
\begin{tabular}{llccccccccccccc}
\toprule
Type & GHSA & Orig. & \(\Bcs\) & \(\Bgpt\) & \(\Egpt\) & \(\Bglm\) & \(\Eglm\) & \(\StudentBaseline{27B}\) & \(\StudentTransfer{27B}\) & \(\StudentGLMTransfer{27B}\) & \(\StudentBaseline{A3B}\) & \(\StudentTransfer{A3B}\) & \(\StudentGLMTransfer{A3B}\) & CWE \\
\coverageRule
\multirow{8}{*}{AuthN Bypass} & \ghsa{GHSA-6h7w-v2xr-mqvw} & 8.8 & T3(9.3) & -- & T1(9.3) & -- & -- & -- & -- & -- & -- & -- & -- & 306\\
  & \ghsa{GHSA-7g56-fwxj-cm23} & 8.1 & T3(8.8) & -- & T1(8.8) & -- & T1(9.3) & -- & -- & -- & T1(8.8) & -- & T1(9.3) & 306,434\\
  & \ghsa{GHSA-pchf-49fh-w34r} & 8.1 & T1(9.3) & -- & -- & -- & -- & -- & -- & -- & -- & -- & -- & 289\\
  & \ghsa{GHSA-m2cq-xjgm-f668} & 9.2 & -- & -- & -- & -- & -- & -- & -- & T1(8.2) & -- & -- & -- & 306\\
  & \ghsa{GHSA-r5m2-fqcf-qrf7} & 8.0 & -- & -- & -- & T3(9.3) & T1(9.3) & -- & T1(9.3) & T1(9.3) & -- & -- & -- & 1188,306\\
  & \ghsa{GHSA-qwc3-h9mg-4582} & 9.9 & -- & -- & -- & -- & T2(8.2) & -- & -- & T2(9.4) & -- & -- & -- & 306\\
  & \ghsa{GHSA-rwp9-5g7q-73q3} & 9.3 & -- & -- & -- & -- & -- & -- & -- & -- & -- & T1(9.3) & -- & 306,425\\
  & \ghsa{GHSA-xv8g-fj9h-6gmv} & 9.3 & -- & -- & -- & -- & T1(8.8) & -- & -- & -- & -- & -- & -- & 306\\
\coverageRule
\multirow{11}{*}{AuthZ/Access} & \ghsa{GHSA-228v-wc5r-j8m7} & 7.1 & T3(7.1) & -- & -- & -- & -- & -- & -- & -- & -- & -- & -- & 284,863\\
  & \ghsa{GHSA-5448-v74m-7mv7} & 8.7 & T3(8.7) & -- & -- & -- & -- & -- & -- & -- & -- & -- & -- & 915\\
  & \ghsa{GHSA-ggxw-g3cp-mgf8} & 9.3 & T3(9.3) & -- & -- & -- & -- & -- & -- & -- & -- & -- & -- & 862\\
  & \ghsa{GHSA-hmqr-wjmj-376c} & 8.6 & T3(8.6) & -- & -- & -- & -- & -- & -- & -- & -- & -- & -- & 863\\
  & \ghsa{GHSA-hwr4-mq23-wcv5} & 7.1 & T2(7.1) & -- & -- & -- & -- & -- & -- & -- & -- & T3(8.6) & -- & 1289\\
  & \ghsa{GHSA-m5wg-cjgh-223j} & 10.0 & -- & -- & -- & -- & T1(10.0) & -- & T2(9.3) & T3(9.3) & -- & -- & T1(10.0) & 284\\
  & \ghsa{GHSA-p52w-7rhw-9m67} & 7.1 & T3(7.2) & -- & -- & -- & -- & -- & -- & -- & -- & -- & -- & 862\\
  & \ghsa{GHSA-x5r2-r74c-3w28} & 8.7 & -- & -- & -- & T1(8.8) & -- & -- & -- & -- & -- & -- & -- & 285,636\\
  & \ghsa{GHSA-xgxp-f695-6vrp} & 7.1 & -- & -- & T1(7.1) & -- & -- & -- & T1(7.1) & -- & -- & -- & -- & 200,862\\
  & \ghsa{GHSA-87fh-rc96-6fr6} & 7.7 & -- & -- & -- & -- & -- & -- & -- & -- & -- & -- & T3(8.7) & 284,639\\
  & \ghsa{GHSA-8c4j-f57c-35cf} & 8.7 & -- & -- & -- & -- & -- & -- & -- & -- & -- & T3(8.7) & -- & 639,862\\
\coverageRule
\multirow{8}{*}{Cmd Injection} & \ghsa{GHSA-25fp-8w8p-mx36} & 9.4 & -- & -- & -- & T1(9.4) & T1(8.7) & -- & T1(8.7) & -- & -- & -- & -- & 78\\
  & \ghsa{GHSA-4wwf-f7w3-94f5} & 8.7 & T1(8.7) & -- & -- & -- & -- & -- & -- & -- & -- & -- & -- & 78\\
  & \ghsa{GHSA-5r63-q8hg-p8qx} & 8.1 & -- & -- & -- & T3(9.3) & -- & -- & -- & T1(9.3) & -- & -- & -- & 78,94\\
  & \ghsa{GHSA-7fv4-fmmc-86g2} & 8.7 & -- & -- & -- & -- & -- & -- & T1(8.7) & -- & -- & -- & -- & 94\\
  & \ghsa{GHSA-gjw9-34gf-rp6m} & 8.7 & -- & -- & -- & -- & -- & T1(8.7) & -- & -- & -- & -- & -- & 78\\
  & \ghsa{GHSA-gv8f-wpm2-m5wr} & 8.7 & -- & -- & T1(9.3) & T3(9.3) & T1(9.3) & -- & T1(9.3) & -- & -- & -- & T3(9.3) & 1188,287,78\\
  & \ghsa{GHSA-mwr6-3gp8-9jmj} & 9.3 & T3(8.7) & -- & -- & T1(7.1) & -- & -- & -- & -- & -- & -- & -- & 77\\
  & \ghsa{GHSA-8q4h-8crm-5cvc} & 9.8 & -- & -- & -- & -- & T1(9.2) & -- & -- & -- & -- & -- & -- & 78\\
\coverageRule
\multirow{19}{*}{Code Execution} & \ghsa{GHSA-2679-6mx9-h9xc} & 9.3 & T3(9.3) & -- & T1(9.3) & -- & -- & -- & -- & T1(9.3) & -- & -- & -- & 306\\
  & \ghsa{GHSA-5882-5rx9-xgxp} & 10.0 & -- & -- & -- & T1(10.0) & -- & -- & -- & -- & T1(9.3) & -- & -- & 94\\
  & \ghsa{GHSA-6fmw-82m7-jq6p} & 8.8 & -- & -- & -- & -- & -- & T1(7.1) & -- & -- & -- & -- & -- & 94\\
  & \ghsa{GHSA-77rh-m34w-rv36} & 9.3 & -- & -- & -- & T1(9.3) & T1(9.3) & -- & T1(9.3) & T1(9.2) & -- & -- & -- & 95\\
  & \ghsa{GHSA-8645-p2v4-73r2} & 8.7 & -- & T2(8.7) & T1(8.7) & -- & -- & -- & -- & -- & -- & -- & -- & 770\\
  & \ghsa{GHSA-9c4h-pwmf-m6fj} & 9.4 & -- & -- & -- & T3(9.2) & T3(8.4) & -- & -- & T3(7.5) & -- & -- & -- & 15,20,269,695,754,94\\
  & \ghsa{GHSA-c87c-78rc-vmv2} & 8.1 & -- & T1(9.3) & -- & -- & T2(9.2) & -- & -- & -- & -- & -- & T3(9.3) & 74\\
  & \ghsa{GHSA-g22f-v6f7-2hrh} & 8.9 & -- & -- & -- & T3(9.3) & T2(9.3) & -- & -- & -- & -- & -- & -- & 829\\
  & \ghsa{GHSA-v52c-386h-88mc} & 8.7 & -- & -- & T1(8.7) & -- & -- & -- & -- & -- & -- & -- & -- & 772\\
  & \ghsa{GHSA-wccx-j62j-r448} & 8.9 & T1(9.3) & -- & -- & -- & T1(9.3) & -- & -- & -- & -- & -- & -- & 693\\
  & \ghsa{GHSA-wxx7-mcgf-j869} & 9.4 & -- & -- & -- & -- & -- & -- & -- & -- & -- & T1(8.7) & T2(9.4) & 89,94\\
  & \ghsa{GHSA-x34r-63hx-w57f} & 9.4 & -- & -- & T1(8.6) & -- & -- & T1(8.7) & -- & -- & -- & -- & -- & 94\\
  & \ghsa{GHSA-xg9w-vg3g-6m68} & 8.7 & T1(7.1) & -- & -- & T3(7.1) & -- & -- & T1(7.1) & -- & -- & -- & -- & 22\\
  & \ghsa{GHSA-xv3r-vr59-95rg} & 9.4 & -- & -- & -- & T3(8.7) & -- & -- & -- & -- & -- & -- & -- & 22\\
  & \ghsa{GHSA-69v7-xpr6-6gjm} & 10.0 & -- & -- & -- & -- & T1(7.9) & -- & -- & -- & -- & -- & -- & 284,693\\
  & \ghsa{GHSA-prf8-cf2x-rhx7} & 9.3 & -- & -- & -- & -- & T2(9.2) & -- & -- & T3(7.5) & -- & -- & -- & 502\\
  & \ghsa{GHSA-q5qq-mvfm-j35x} & 8.9 & -- & -- & -- & -- & T1(9.1) & -- & -- & -- & -- & -- & -- & 184,502\\
  & \ghsa{GHSA-vg9h-jx4v-cwx2} & 9.3 & -- & -- & -- & -- & T1(9.3) & -- & -- & -- & -- & -- & -- & 489\\
  & \ghsa{GHSA-xw7x-h9fj-p2c7} & 9.3 & -- & -- & -- & -- & T2(9.2) & -- & -- & -- & -- & -- & -- & 502\\
\coverageRule
\multirow{5}{*}{Crypto/Signature} & \ghsa{GHSA-2w8x-224x-785m} & 7.7 & -- & -- & -- & -- & -- & -- & T2(8.7) & -- & -- & -- & -- & 325,347\\
  & \ghsa{GHSA-796p-j2gh-9m2q} & 9.3 & T3(8.7) & -- & -- & -- & -- & -- & -- & -- & -- & -- & -- & 295,347\\
  & \ghsa{GHSA-88q6-jcjg-hvmw} & 8.8 & -- & T1(8.7) & T1(9.3) & T3(9.3) & -- & -- & T1(9.3) & T1(9.3) & T1(9.3) & -- & T2(9.3) & 327\\
  & \ghsa{GHSA-8x4m-qw58-3pcx} & 9.3 & T3(8.8) & -- & -- & -- & -- & -- & -- & -- & -- & -- & -- & 288,294,345\\
  & \ghsa{GHSA-qpv2-rwc8-c993} & 9.2 & -- & T1(8.8) & T1(8.8) & -- & -- & T1(8.7) & T1(8.7) & -- & -- & -- & -- & 347\\
\coverageRule
\multirow{21}{*}{DoS} & \ghsa{GHSA-2gjw-fg97-vg3r} & 8.7 & T2(8.7) & -- & -- & -- & -- & -- & -- & -- & -- & -- & -- & 20\\
  & \ghsa{GHSA-2v35-w6hq-6mfw} & 8.7 & T1(8.7) & -- & T1(8.7) & T1(8.7) & -- & -- & T1(8.7) & -- & -- & -- & -- & 674\\
  & \ghsa{GHSA-353c-v8x9-v7c3} & 8.7 & -- & T2(8.7) & T1(8.7) & T1(8.7) & T1(8.7) & T1(8.7) & -- & T3(8.7) & -- & -- & -- & 770\\
  & \ghsa{GHSA-3ppc-4f35-3m26} & 8.7 & T1(8.7) & -- & -- & -- & -- & -- & -- & -- & -- & -- & -- & 1333\\
  & \ghsa{GHSA-5528-5vmv-3xc2} & 8.7 & -- & -- & T1(8.7) & T3(8.7) & -- & -- & T1(8.7) & -- & -- & -- & -- & 674\\
  & \ghsa{GHSA-5j86-7r7m-p8h6} & 8.8 & -- & -- & -- & -- & -- & T1(8.7) & -- & -- & -- & -- & -- & 1321\\
  & \ghsa{GHSA-5jg4-p4qw-cgfr} & 8.7 & -- & -- & T1(8.7) & -- & T1(8.7) & -- & T1(8.7) & T1(8.7) & -- & -- & -- & 674\\
  & \ghsa{GHSA-677m-j7p3-52f9} & 8.7 & T1(8.7) & -- & -- & -- & -- & -- & -- & -- & -- & -- & -- & 754\\
  & \ghsa{GHSA-6g43-577r-wf4x} & 7.1 & -- & -- & -- & T3(7.1) & -- & -- & -- & -- & -- & -- & -- & 129\\
  & \ghsa{GHSA-67pg-wm7f-q7fj} & 8.7 & -- & -- & -- & -- & -- & -- & -- & T1(8.2) & -- & -- & -- & 770\\
  & \ghsa{GHSA-95fx-jjr5-f39c} & 8.7 & T3(8.7) & -- & -- & -- & -- & -- & -- & -- & -- & -- & -- & 20,400,770\\
  & \ghsa{GHSA-fmwg-qcqh-m992} & 8.7 & T1(8.7) & -- & -- & -- & -- & -- & -- & -- & -- & -- & -- & 1333\\
  & \ghsa{GHSA-h5qv-qjv4-pc5m} & 8.7 & -- & -- & T1(8.7) & -- & T3(8.7) & -- & -- & -- & -- & -- & T1(8.7) & 400,409\\
  & \ghsa{GHSA-jg4p-7fhp-p32p} & 8.7 & -- & -- & -- & T1(8.7) & -- & -- & -- & -- & T1(8.7) & -- & -- & 1333\\
  & \ghsa{GHSA-mq3p-rrmp-79jg} & 7.1 & T2(8.7) & -- & -- & -- & -- & -- & -- & -- & -- & -- & -- & 20,400\\
  & \ghsa{GHSA-rhr9-hgcm-x289} & 8.7 & -- & -- & -- & T3(8.7) & T1(8.7) & T1(8.7) & T1(8.7) & T3(8.7) & -- & -- & -- & 404\\
  & \ghsa{GHSA-xf7r-hgr6-v32p} & 8.7 & -- & -- & -- & T1(8.7) & -- & -- & -- & -- & -- & -- & -- & 459\\
  & \ghsa{GHSA-2phg-qgmm-r638} & 7.7 & -- & -- & -- & -- & -- & -- & -- & -- & -- & -- & T2(8.7) & 409\\
  & \ghsa{GHSA-6v53-7c9g-w56r} & 8.7 & -- & -- & -- & -- & T2(8.7) & -- & -- & T2(8.2) & -- & -- & T2(8.7) & 770\\
  & \ghsa{GHSA-7g27-v5wj-jr75} & 8.7 & -- & -- & -- & -- & T2(8.7) & -- & -- & -- & -- & -- & -- & 476,478\\
  & \ghsa{GHSA-7h2j-956f-4vf2} & 8.7 & -- & -- & -- & -- & T1(8.7) & -- & -- & -- & -- & -- & -- & 1333\\
\coverageRule
\multirow{2}{*}{Info Leak} & \ghsa{GHSA-gh4x-f7cq-wwx6} & 8.7 & -- & -- & T1(8.7) & -- & T1(8.7) & -- & -- & T1(8.7) & -- & -- & T1(8.7) & 200\\
  & \ghsa{GHSA-wvxv-4j8q-4wjq} & 8.7 & -- & -- & -- & -- & -- & -- & -- & T1(8.7) & -- & -- & -- & 200\\
\coverageRule
\multirow{10}{*}{Injection} & \ghsa{GHSA-2453-mppf-46cj} & 8.7 & -- & -- & -- & -- & -- & T3(7.1) & -- & -- & -- & -- & -- & 89\\
  & \ghsa{GHSA-2r2p-4cgf-hv7h} & 8.6 & T3(7.1) & -- & -- & T1(8.6) & T1(8.6) & -- & -- & -- & -- & -- & -- & 1188,306,352,942\\
  & \ghsa{GHSA-g7j6-fmwx-7vp8} & 8.7 & -- & T1(7.1) & -- & -- & -- & -- & -- & T3(7.1) & -- & -- & -- & 89\\
  & \ghsa{GHSA-hm9j-cgmm-2w36} & 8.8 & -- & -- & -- & T3(8.8) & -- & -- & -- & -- & -- & -- & -- & 89\\
  & \ghsa{GHSA-j759-j44w-7fr8} & 8.7 & -- & -- & -- & T1(8.7) & -- & -- & -- & -- & -- & -- & -- & 91\\
  & \ghsa{GHSA-q6g3-fv43-m2w6} & 8.7 & T3(7.1) & -- & -- & T3(8.6) & -- & -- & -- & -- & -- & -- & -- & 89\\
  & \ghsa{GHSA-rg7c-g689-fr3x} & 9.3 & -- & -- & T1(9.3) & -- & T3(9.3) & -- & -- & -- & -- & -- & -- & 306\\
  & \ghsa{GHSA-x6wf-f3px-wcqx} & 8.7 & -- & -- & T1(8.7) & T1(8.7) & -- & -- & -- & -- & -- & -- & -- & 91\\
  & \ghsa{GHSA-p864-fqgv-92q4} & 8.7 & -- & -- & -- & -- & -- & -- & -- & -- & -- & T3(8.6) & T3(7.1) & 89\\
  & \ghsa{GHSA-qp2j-v5jg-hg68} & 7.1 & -- & -- & -- & -- & T2(7.1) & -- & -- & -- & -- & -- & -- & 89\\
\coverageRule
Memory Safety & \ghsa{GHSA-4w32-2493-32g7} & 8.7 & T1(8.7) & -- & T1(8.7) & T3(8.7) & T1(8.7) & -- & -- & T1(8.7) & -- & -- & T1(8.7) & 190\\
\coverageRule
\multirow{5}{*}{Object Pollution} & \ghsa{GHSA-hf2r-9gf9-rwch} & 9.4 & -- & -- & -- & T1(7.0) & -- & -- & -- & -- & -- & -- & -- & 1321\\
  & \ghsa{GHSA-rf6f-7fwh-wjgh} & 8.9 & -- & -- & -- & T1(8.8) & -- & -- & -- & T1(8.2) & -- & -- & -- & 1321\\
  & \ghsa{GHSA-wf6x-7x77-mvgw} & 8.7 & -- & -- & -- & -- & -- & -- & -- & T3(8.2) & -- & -- & -- & 1321\\
  & \ghsa{GHSA-wfq2-52f7-7qvj} & 8.9 & -- & -- & -- & T1(9.3) & T1(9.3) & -- & -- & -- & -- & -- & -- & 184,502\\
  & \ghsa{GHSA-m272-9rp6-32mc} & 9.3 & -- & -- & -- & -- & T1(8.8) & -- & -- & -- & -- & -- & -- & 1321\\
\coverageRule
\multirow{21}{*}{Path/File} & \ghsa{GHSA-2657-3c98-63jq} & 7.7 & T1(8.8) & -- & -- & -- & T1(8.8) & -- & T1(8.7) & T1(8.7) & -- & -- & -- & 22\\
  & \ghsa{GHSA-5h6h-7rc9-3824} & 8.7 & -- & -- & T1(8.7) & -- & -- & -- & -- & -- & -- & -- & -- & 22\\
  & \ghsa{GHSA-6v48-fcq6-ff23} & 7.1 & -- & -- & -- & T3(8.7) & -- & -- & -- & -- & -- & -- & -- & 22\\
  & \ghsa{GHSA-9ppj-qmqm-q256} & 8.2 & T3(8.2) & -- & -- & -- & -- & -- & -- & -- & -- & -- & -- & 22\\
  & \ghsa{GHSA-f8cm-6447-x5h2} & 9.2 & -- & -- & -- & T3(7.7) & T1(9.3) & -- & -- & -- & -- & -- & -- & 22,35,73\\
  & \ghsa{GHSA-hjh7-r5w8-5872} & 7.1 & T3(7.1) & -- & -- & -- & -- & -- & -- & -- & -- & -- & -- & 22\\
  & \ghsa{GHSA-hqjg-pww4-pcgq} & 8.7 & -- & -- & -- & -- & -- & -- & -- & T1(7.1) & -- & -- & -- & 22\\
  & \ghsa{GHSA-p3h2-2j4p-p83g} & 7.2 & T1(7.1) & -- & -- & T1(7.2) & T1(7.2) & -- & -- & -- & -- & -- & -- & 22\\
  & \ghsa{GHSA-v2xr-wvrv-p969} & 7.7 & T1(8.7) & -- & -- & T1(8.7) & T1(8.7) & -- & T1(8.7) & -- & -- & T1(8.7) & -- & 22,770,918\\
  & \ghsa{GHSA-v92g-xgxw-vvmm} & 7.7 & -- & -- & T1(7.0) & -- & -- & -- & -- & -- & -- & -- & -- & 22\\
  & \ghsa{GHSA-vmwq-8g8c-jm79} & 8.7 & T2(8.8) & -- & -- & -- & -- & -- & -- & -- & -- & -- & -- & 22\\
  & \ghsa{GHSA-vv7q-7jx5-f767} & 10.0 & -- & -- & T1(7.1) & -- & T1(7.9) & -- & -- & -- & -- & -- & -- & 918\\
  & \ghsa{GHSA-w789-49fc-v8hr} & 8.7 & T1(8.7) & -- & -- & -- & -- & -- & -- & -- & -- & -- & T1(8.7) & 20,918\\
  & \ghsa{GHSA-wmfp-5q7x-987x} & 8.7 & -- & -- & -- & T1(8.7) & T1(8.7) & -- & -- & -- & -- & -- & -- & 22\\
  & \ghsa{GHSA-5458-7hh9-v7p4} & 8.7 & -- & -- & -- & -- & T1(8.7) & -- & -- & -- & -- & -- & -- & 22,23\\
  & \ghsa{GHSA-8x9r-hvwg-c55h} & 8.7 & -- & -- & -- & -- & T1(7.1) & -- & -- & -- & -- & -- & -- & 22\\
  & \ghsa{GHSA-94c7-g2fj-7682} & 8.3 & -- & -- & -- & -- & -- & -- & -- & -- & -- & T3(8.6) & -- & 22\\
  & \ghsa{GHSA-h7cj-j2vv-qw8r} & 8.7 & -- & -- & -- & -- & -- & -- & -- & -- & -- & -- & T3(8.7) & 22\\
  & \ghsa{GHSA-m6w7-qv66-g3mf} & 8.6 & -- & -- & -- & -- & -- & -- & -- & -- & -- & -- & T1(8.6) & 22,59\\
  & \ghsa{GHSA-mw96-cpmx-2vgc} & 8.8 & -- & -- & -- & -- & T1(8.7) & -- & -- & -- & -- & -- & -- & 22\\
  & \ghsa{GHSA-pq29-69jg-9mxc} & 8.8 & -- & -- & -- & -- & -- & -- & -- & -- & -- & -- & T3(8.6) & 22\\
\coverageRule
\multirow{2}{*}{XSS} & \ghsa{GHSA-qpq4-pw7f-pp8w} & 8.5 & T3(8.5) & -- & -- & -- & -- & -- & -- & -- & -- & -- & -- & 79\\
  & \ghsa{GHSA-w8x4-x68c-m6fc} & 8.7 & -- & -- & T1(8.7) & T3(8.7) & T1(8.7) & -- & T1(8.7) & T1(8.7) & -- & -- & -- & 79\\
\coverageRule
\multirow{3}{*}{Other} & \ghsa{GHSA-4hjq-9h5c-252j} & 7.7 & -- & -- & -- & -- & -- & T1(8.7) & -- & -- & -- & -- & -- & 476\\
  & \ghsa{GHSA-9fjp-q3c4-6w3j} & 8.7 & -- & -- & -- & -- & -- & -- & T1(8.7) & -- & -- & -- & -- & 674\\
  & \ghsa{GHSA-jpcj-7wfg-mqxv} & 8.7 & -- & -- & -- & T3(10.0) & -- & -- & -- & -- & -- & -- & -- & 20\\
\coverageRule
\textbf{Total} & \textbf{116} & -- & \textbf{34} & \textbf{6} & \textbf{23} & \textbf{36} & \textbf{42} & \textbf{9} & \textbf{19} & \textbf{24} & \textbf{4} & \textbf{7} & \textbf{17} & --\\
\bottomrule
\end{tabular}
}
\end{table*}
\endgroup

\clearpage

\begin{figure*}[!t]
\centering
\vspace*{-0.55in}
\refstepcounter{figure}\label{fig:playbook-pointer-histories}
\begin{minipage}[t]{0.32\textwidth}
\vspace{0pt}
\centering
{\scriptsize\bfseries \codexgpt{} playbook}\par\vspace{2pt}
\resizebox{\linewidth}{!}{\input{figures/gpt_playbook_pointer_normalized_tikz.tex}}
\end{minipage}\hfill
\begin{minipage}[t]{0.32\textwidth}
\vspace{0pt}
\centering
{\scriptsize\bfseries \glmopen{} playbook}\par\vspace{2pt}
\resizebox{\linewidth}{!}{\input{figures/glm_playbook_pointer_normalized_tikz.tex}}
\end{minipage}\hfill
\begin{minipage}[t]{0.32\textwidth}
\vspace{0pt}
\raggedright
\footnotesize

\textbf{\glmopen{} evolution.}
P02--P05 bootstrap workflow/knowledge-base candidates, broaden vulnerability classes, and prioritize core components. P06--P13 add subsystem enumeration plus SQL/YAML/header DoS, native overflow, crypto/BOLA/deserialize, builder traversal, GraphQL/XML/CLI tracing, parameter precision, and ReDoS/XPath matching. P14--P20 add internal-function audits, no-vulnerability safeguards, blind-spot coverage, endpoint sweeps, RCE/CORS/SSRF checks, binary overflow, ORM SQL injection, auth-flow checks, and per-endpoint authorization. P21--P30 add PoC path validation, polyglot/eval coverage, denylist and LLM-parameter SSRF checks, middleware/path/race checks, CRUD, access-control matrices, weak crypto, DB permission bypasses, DoS gates, PR:L anchoring fixes, and MyBatis/native-code/RCE coverage. P31--P40 tighten warning/wiki checks, REST anchoring, Ruby/Python/IPv6 patterns, root-cause matching, dependency/upload/storage heuristics, CWE/DoS distinctions, attack-surface anchoring, and transport-DoS separation. 

\medskip

\textbf{\codexgpt{} evolution.}
P02--P05 turn a README-only seed into a structured workflow with CVSS gates, root-cause hypotheses, DoS evidence, XXE/data-exposure checks, and drift control. P06--P13 sharpen auth disclosure, hidden security primitives, output-injection versus DoS separation, token lifecycle checks, vulnerable-path tracing, crypto misuse, and metadata drift. P14--P20 distinguish eval/state bugs, shipped parser/render roots, state and secret boundaries, sibling root causes, guard bypasses, wrapper sinks, and breadth-first ranking. P21--P30 add file/output/SQL sinks, parser trust boundaries, real-path evidence, validator/state bypasses, prototype/evaluator state, upload/write sinks, default entry/config checks, workflow-versus-SSRF separation, meta-object/plugin/body-DoS checks, and read/auth/SQL/DoS triage. P31--P40 converge on default-boundary clearing, security-signal-guided families, launcher/SQL validation paths, runtime surfaces, artifact lifecycle, named-guard fit, sibling roots, exception/DoS/leak separation, framework router/path/prototype roots, and final root-cause comparison.
\end{minipage}
\vspace{3pt}

\begin{minipage}{0.965\textwidth}
\footnotesize
\textbf{Figure~\thefigure. Compact comparison of normalized playbook pointer histories.}
Blue nodes are selected incumbents; orange nodes are retained alternatives; purple outlines and arrows mark candidates promoted to the next incumbent. Horizontal fan-out shows alternatives evaluated under the same incumbent, not candidate-to-candidate ancestry.

\end{minipage}
\end{figure*}
\clearpage

\input{case_study/rusqlite_walkthrough.tex}

\end{document}

%% file: figures/gpt_playbook_pointer_normalized_tikz.tex
% Compact single-column rendering of the normalized GPT playbook pointer table.
\begin{tikzpicture}[
  x=1cm,y=1cm,
  pointerlabel/.style={font=\fontsize{6.4}{7}\selectfont\bfseries, anchor=east, text=black!75},
  bestnode/.style={rounded corners=1.2pt, fill=legendBlue!90, text=white, font=\fontsize{4.9}{5.2}\selectfont\ttfamily, minimum width=0.82cm, minimum height=0.22cm, inner sep=0.8pt},
  candnode/.style={rounded corners=1.2pt, fill=legendOrange!95, text=white, font=\fontsize{4.9}{5.2}\selectfont\ttfamily, minimum width=0.82cm, minimum height=0.22cm, inner sep=0.8pt},
  promotednode/.style={candnode, draw=legendPurple!95, line width=0.55pt},
  branch/.style={-{Latex[length=1.0mm]}, line width=0.22pt, draw=black!55, opacity=0.72},
  promote/.style={-{Latex[length=1.1mm]}, line width=0.28pt, draw=legendPurple!60, opacity=0.45},
]
\draw[black!7] (0.00,-0.680) -- (8.45,-0.680);
\node[pointerlabel] at (0.62,-0.450) {P01};
\node[bestnode] (GPTb01) at (1.55,-0.450) {cf450ad};
\draw[black!7] (0.00,-1.240) -- (8.45,-1.240);
\node[pointerlabel] at (0.62,-1.010) {P02};
\node[bestnode] (GPTb02) at (1.55,-1.010) {cf450ad};
\node[promotednode] (GPTc02_1) at (2.75,-1.010) {92bfe76};
\draw[branch] (GPTb02.east) -- (GPTc02_1.west);
\draw[black!7] (0.00,-1.800) -- (8.45,-1.800);
\node[pointerlabel] at (0.62,-1.570) {P03};
\node[bestnode] (GPTb03) at (1.55,-1.570) {92bfe76};
\node[candnode] (GPTc03_1) at (2.75,-1.570) {684eb3a};
\node[promotednode] (GPTc03_2) at (8.05,-1.570) {791aa27};
\draw[branch] (GPTb03.east) .. controls +(0.45,0.162) and +(-0.45,0.162) .. (GPTc03_1.west);
\draw[branch] (GPTb03.east) .. controls +(0.45,0.174) and +(-0.45,0.174) .. (GPTc03_2.west);
\draw[black!7] (0.00,-2.360) -- (8.45,-2.360);
\node[pointerlabel] at (0.62,-2.130) {P04};
\node[bestnode] (GPTb04) at (1.55,-2.130) {791aa27};
\node[candnode] (GPTc04_1) at (2.75,-2.130) {183ad84};
\node[candnode] (GPTc04_2) at (4.08,-2.130) {458634e};
\node[candnode] (GPTc04_3) at (5.40,-2.130) {68366f0};
\node[candnode] (GPTc04_4) at (6.73,-2.130) {9e7eb25};
\node[promotednode] (GPTc04_5) at (8.05,-2.130) {0babf84};
\draw[branch] (GPTb04.east) .. controls +(0.45,0.162) and +(-0.45,0.162) .. (GPTc04_1.west);
\draw[branch] (GPTb04.east) .. controls +(0.45,0.174) and +(-0.45,0.174) .. (GPTc04_2.west);
\draw[branch] (GPTb04.east) .. controls +(0.45,0.186) and +(-0.45,0.186) .. (GPTc04_3.west);
\draw[branch] (GPTb04.east) .. controls +(0.45,0.198) and +(-0.45,0.198) .. (GPTc04_4.west);
\draw[branch] (GPTb04.east) .. controls +(0.45,0.210) and +(-0.45,0.210) .. (GPTc04_5.west);
\draw[black!7] (0.00,-2.920) -- (8.45,-2.920);
\node[pointerlabel] at (0.62,-2.690) {P05};
\node[bestnode] (GPTb05) at (1.55,-2.690) {0babf84};
\node[promotednode] (GPTc05_1) at (2.75,-2.690) {bac5a9c};
\draw[branch] (GPTb05.east) -- (GPTc05_1.west);
\draw[black!7] (0.00,-3.480) -- (8.45,-3.480);
\node[pointerlabel] at (0.62,-3.250) {P06};
\node[bestnode] (GPTb06) at (1.55,-3.250) {bac5a9c};
\node[candnode] (GPTc06_1) at (2.75,-3.250) {8f2f80e};
\node[promotednode] (GPTc06_2) at (8.05,-3.250) {293eeda};
\draw[branch] (GPTb06.east) .. controls +(0.45,0.162) and +(-0.45,0.162) .. (GPTc06_1.west);
\draw[branch] (GPTb06.east) .. controls +(0.45,0.174) and +(-0.45,0.174) .. (GPTc06_2.west);
\draw[black!7] (0.00,-4.040) -- (8.45,-4.040);
\node[pointerlabel] at (0.62,-3.810) {P07};
\node[bestnode] (GPTb07) at (1.55,-3.810) {293eeda};
\node[candnode] (GPTc07_1) at (2.75,-3.810) {c4c959e};
\node[promotednode] (GPTc07_2) at (8.05,-3.810) {b4d8871};
\draw[branch] (GPTb07.east) .. controls +(0.45,0.162) and +(-0.45,0.162) .. (GPTc07_1.west);
\draw[branch] (GPTb07.east) .. controls +(0.45,0.174) and +(-0.45,0.174) .. (GPTc07_2.west);
\draw[black!7] (0.00,-4.600) -- (8.45,-4.600);
\node[pointerlabel] at (0.62,-4.370) {P08};
\node[bestnode] (GPTb08) at (1.55,-4.370) {b4d8871};
\node[candnode] (GPTc08_1) at (2.75,-4.370) {4fd004e};
\node[candnode] (GPTc08_2) at (5.40,-4.370) {1d0c4f4};
\node[promotednode] (GPTc08_3) at (8.05,-4.370) {a8652e0};
\draw[branch] (GPTb08.east) .. controls +(0.45,0.162) and +(-0.45,0.162) .. (GPTc08_1.west);
\draw[branch] (GPTb08.east) .. controls +(0.45,0.174) and +(-0.45,0.174) .. (GPTc08_2.west);
\draw[branch] (GPTb08.east) .. controls +(0.45,0.186) and +(-0.45,0.186) .. (GPTc08_3.west);
\draw[black!7] (0.00,-5.160) -- (8.45,-5.160);
\node[pointerlabel] at (0.62,-4.930) {P09};
\node[bestnode] (GPTb09) at (1.55,-4.930) {a8652e0};
\node[candnode] (GPTc09_1) at (2.75,-4.930) {61000f1};
\node[promotednode] (GPTc09_2) at (8.05,-4.930) {3d045c8};
\draw[branch] (GPTb09.east) .. controls +(0.45,0.162) and +(-0.45,0.162) .. (GPTc09_1.west);
\draw[branch] (GPTb09.east) .. controls +(0.45,0.174) and +(-0.45,0.174) .. (GPTc09_2.west);
\draw[black!7] (0.00,-5.720) -- (8.45,-5.720);
\node[pointerlabel] at (0.62,-5.490) {P10};
\node[bestnode] (GPTb10) at (1.55,-5.490) {3d045c8};
\node[candnode] (GPTc10_1) at (2.75,-5.490) {5d0e2a9};
\node[promotednode] (GPTc10_2) at (8.05,-5.490) {9ee9092};
\draw[branch] (GPTb10.east) .. controls +(0.45,0.162) and +(-0.45,0.162) .. (GPTc10_1.west);
\draw[branch] (GPTb10.east) .. controls +(0.45,0.174) and +(-0.45,0.174) .. (GPTc10_2.west);
\draw[black!7] (0.00,-6.280) -- (8.45,-6.280);
\node[pointerlabel] at (0.62,-6.050) {P11};
\node[bestnode] (GPTb11) at (1.55,-6.050) {9ee9092};
\node[candnode] (GPTc11_1) at (2.75,-6.050) {b3561e5};
\node[candnode] (GPTc11_2) at (5.40,-6.050) {3300f23};
\node[promotednode] (GPTc11_3) at (8.05,-6.050) {decfef8};
\draw[branch] (GPTb11.east) .. controls +(0.45,0.162) and +(-0.45,0.162) .. (GPTc11_1.west);
\draw[branch] (GPTb11.east) .. controls +(0.45,0.174) and +(-0.45,0.174) .. (GPTc11_2.west);
\draw[branch] (GPTb11.east) .. controls +(0.45,0.186) and +(-0.45,0.186) .. (GPTc11_3.west);
\draw[black!7] (0.00,-6.840) -- (8.45,-6.840);
\node[pointerlabel] at (0.62,-6.610) {P12};
\node[bestnode] (GPTb12) at (1.55,-6.610) {decfef8};
\node[candnode] (GPTc12_1) at (2.75,-6.610) {e321133};
\node[candnode] (GPTc12_2) at (5.40,-6.610) {ec86458};
\node[promotednode] (GPTc12_3) at (8.05,-6.610) {7a0833d};
\draw[branch] (GPTb12.east) .. controls +(0.45,0.162) and +(-0.45,0.162) .. (GPTc12_1.west);
\draw[branch] (GPTb12.east) .. controls +(0.45,0.174) and +(-0.45,0.174) .. (GPTc12_2.west);
\draw[branch] (GPTb12.east) .. controls +(0.45,0.186) and +(-0.45,0.186) .. (GPTc12_3.west);
\draw[black!7] (0.00,-7.400) -- (8.45,-7.400);
\node[pointerlabel] at (0.62,-7.170) {P13};
\node[bestnode] (GPTb13) at (1.55,-7.170) {7a0833d};
\node[candnode] (GPTc13_1) at (2.75,-7.170) {b08b004};
\node[candnode] (GPTc13_2) at (5.40,-7.170) {f9daed9};
\node[promotednode] (GPTc13_3) at (8.05,-7.170) {b7b0b57};
\draw[branch] (GPTb13.east) .. controls +(0.45,0.162) and +(-0.45,0.162) .. (GPTc13_1.west);
\draw[branch] (GPTb13.east) .. controls +(0.45,0.174) and +(-0.45,0.174) .. (GPTc13_2.west);
\draw[branch] (GPTb13.east) .. controls +(0.45,0.186) and +(-0.45,0.186) .. (GPTc13_3.west);
\draw[black!7] (0.00,-7.960) -- (8.45,-7.960);
\node[pointerlabel] at (0.62,-7.730) {P14};
\node[bestnode] (GPTb14) at (1.55,-7.730) {b7b0b57};
\node[candnode] (GPTc14_1) at (2.75,-7.730) {0aa0565};
\node[promotednode] (GPTc14_2) at (8.05,-7.730) {793d13b};
\draw[branch] (GPTb14.east) .. controls +(0.45,0.162) and +(-0.45,0.162) .. (GPTc14_1.west);
\draw[branch] (GPTb14.east) .. controls +(0.45,0.174) and +(-0.45,0.174) .. (GPTc14_2.west);
\draw[black!7] (0.00,-8.520) -- (8.45,-8.520);
\node[pointerlabel] at (0.62,-8.290) {P15};
\node[bestnode] (GPTb15) at (1.55,-8.290) {793d13b};
\node[candnode] (GPTc15_1) at (2.75,-8.290) {9bf76d7};
\node[candnode] (GPTc15_2) at (5.40,-8.290) {05bd393};
\node[promotednode] (GPTc15_3) at (8.05,-8.290) {84c05b5};
\draw[branch] (GPTb15.east) .. controls +(0.45,0.162) and +(-0.45,0.162) .. (GPTc15_1.west);
\draw[branch] (GPTb15.east) .. controls +(0.45,0.174) and +(-0.45,0.174) .. (GPTc15_2.west);
\draw[branch] (GPTb15.east) .. controls +(0.45,0.186) and +(-0.45,0.186) .. (GPTc15_3.west);
\draw[black!7] (0.00,-9.080) -- (8.45,-9.080);
\node[pointerlabel] at (0.62,-8.850) {P16};
\node[bestnode] (GPTb16) at (1.55,-8.850) {84c05b5};
\node[candnode] (GPTc16_1) at (2.75,-8.850) {8173250};
\node[promotednode] (GPTc16_2) at (8.05,-8.850) {0554346};
\draw[branch] (GPTb16.east) .. controls +(0.45,0.162) and +(-0.45,0.162) .. (GPTc16_1.west);
\draw[branch] (GPTb16.east) .. controls +(0.45,0.174) and +(-0.45,0.174) .. (GPTc16_2.west);
\draw[black!7] (0.00,-9.640) -- (8.45,-9.640);
\node[pointerlabel] at (0.62,-9.410) {P17};
\node[bestnode] (GPTb17) at (1.55,-9.410) {0554346};
\node[candnode] (GPTc17_1) at (2.75,-9.410) {c9dc425};
\node[promotednode] (GPTc17_2) at (8.05,-9.410) {fcb294b};
\draw[branch] (GPTb17.east) .. controls +(0.45,0.162) and +(-0.45,0.162) .. (GPTc17_1.west);
\draw[branch] (GPTb17.east) .. controls +(0.45,0.174) and +(-0.45,0.174) .. (GPTc17_2.west);
\draw[black!7] (0.00,-10.200) -- (8.45,-10.200);
\node[pointerlabel] at (0.62,-9.970) {P18};
\node[bestnode] (GPTb18) at (1.55,-9.970) {fcb294b};
\node[promotednode] (GPTc18_1) at (2.75,-9.970) {c659e51};
\draw[branch] (GPTb18.east) -- (GPTc18_1.west);
\draw[black!7] (0.00,-10.760) -- (8.45,-10.760);
\node[pointerlabel] at (0.62,-10.530) {P19};
\node[bestnode] (GPTb19) at (1.55,-10.530) {c659e51};
\node[promotednode] (GPTc19_1) at (2.75,-10.530) {6387772};
\draw[branch] (GPTb19.east) -- (GPTc19_1.west);
\draw[black!7] (0.00,-11.320) -- (8.45,-11.320);
\node[pointerlabel] at (0.62,-11.090) {P20};
\node[bestnode] (GPTb20) at (1.55,-11.090) {6387772};
\node[candnode] (GPTc20_1) at (2.75,-11.090) {6c6c5df};
\node[promotednode] (GPTc20_2) at (8.05,-11.090) {e435eff};
\draw[branch] (GPTb20.east) .. controls +(0.45,0.162) and +(-0.45,0.162) .. (GPTc20_1.west);
\draw[branch] (GPTb20.east) .. controls +(0.45,0.174) and +(-0.45,0.174) .. (GPTc20_2.west);
\draw[black!7] (0.00,-11.880) -- (8.45,-11.880);
\node[pointerlabel] at (0.62,-11.650) {P21};
\node[bestnode] (GPTb21) at (1.55,-11.650) {e435eff};
\node[candnode] (GPTc21_1) at (2.75,-11.650) {889275c};
\node[candnode] (GPTc21_2) at (4.52,-11.650) {550960f};
\node[candnode] (GPTc21_3) at (6.28,-11.650) {278160d};
\node[promotednode] (GPTc21_4) at (8.05,-11.650) {46af80c};
\draw[branch] (GPTb21.east) .. controls +(0.45,0.162) and +(-0.45,0.162) .. (GPTc21_1.west);
\draw[branch] (GPTb21.east) .. controls +(0.45,0.174) and +(-0.45,0.174) .. (GPTc21_2.west);
\draw[branch] (GPTb21.east) .. controls +(0.45,0.186) and +(-0.45,0.186) .. (GPTc21_3.west);
\draw[branch] (GPTb21.east) .. controls +(0.45,0.198) and +(-0.45,0.198) .. (GPTc21_4.west);
\draw[black!7] (0.00,-12.440) -- (8.45,-12.440);
\node[pointerlabel] at (0.62,-12.210) {P22};
\node[bestnode] (GPTb22) at (1.55,-12.210) {46af80c};
\node[candnode] (GPTc22_1) at (2.75,-12.210) {0aacfab};
\node[promotednode] (GPTc22_2) at (8.05,-12.210) {0519be3};
\draw[branch] (GPTb22.east) .. controls +(0.45,0.162) and +(-0.45,0.162) .. (GPTc22_1.west);
\draw[branch] (GPTb22.east) .. controls +(0.45,0.174) and +(-0.45,0.174) .. (GPTc22_2.west);
\draw[black!7] (0.00,-13.000) -- (8.45,-13.000);
\node[pointerlabel] at (0.62,-12.770) {P23};
\node[bestnode] (GPTb23) at (1.55,-12.770) {0519be3};
\node[candnode] (GPTc23_1) at (2.75,-12.770) {83d2573};
\node[promotednode] (GPTc23_2) at (8.05,-12.770) {a9c2875};
\draw[branch] (GPTb23.east) .. controls +(0.45,0.162) and +(-0.45,0.162) .. (GPTc23_1.west);
\draw[branch] (GPTb23.east) .. controls +(0.45,0.174) and +(-0.45,0.174) .. (GPTc23_2.west);
\draw[black!7] (0.00,-13.560) -- (8.45,-13.560);
\node[pointerlabel] at (0.62,-13.330) {P24};
\node[bestnode] (GPTb24) at (1.55,-13.330) {a9c2875};
\node[candnode] (GPTc24_1) at (2.75,-13.330) {8cb2be7};
\node[promotednode] (GPTc24_2) at (8.05,-13.330) {e12c613};
\draw[branch] (GPTb24.east) .. controls +(0.45,0.162) and +(-0.45,0.162) .. (GPTc24_1.west);
\draw[branch] (GPTb24.east) .. controls +(0.45,0.174) and +(-0.45,0.174) .. (GPTc24_2.west);
\draw[black!7] (0.00,-14.120) -- (8.45,-14.120);
\node[pointerlabel] at (0.62,-13.890) {P25};
\node[bestnode] (GPTb25) at (1.55,-13.890) {e12c613};
\node[promotednode] (GPTc25_1) at (2.75,-13.890) {58f858e};
\draw[branch] (GPTb25.east) -- (GPTc25_1.west);
\draw[black!7] (0.00,-14.680) -- (8.45,-14.680);
\node[pointerlabel] at (0.62,-14.450) {P26};
\node[bestnode] (GPTb26) at (1.55,-14.450) {58f858e};
\node[candnode] (GPTc26_1) at (2.75,-14.450) {ff2b30b};
\node[candnode] (GPTc26_2) at (5.40,-14.450) {2af5854};
\node[promotednode] (GPTc26_3) at (8.05,-14.450) {5f7db5f};
\draw[branch] (GPTb26.east) .. controls +(0.45,0.162) and +(-0.45,0.162) .. (GPTc26_1.west);
\draw[branch] (GPTb26.east) .. controls +(0.45,0.174) and +(-0.45,0.174) .. (GPTc26_2.west);
\draw[branch] (GPTb26.east) .. controls +(0.45,0.186) and +(-0.45,0.186) .. (GPTc26_3.west);
\draw[black!7] (0.00,-15.240) -- (8.45,-15.240);
\node[pointerlabel] at (0.62,-15.010) {P27};
\node[bestnode] (GPTb27) at (1.55,-15.010) {5f7db5f};
\node[candnode] (GPTc27_1) at (2.75,-15.010) {65cfbfc};
\node[promotednode] (GPTc27_2) at (8.05,-15.010) {6a170da};
\draw[branch] (GPTb27.east) .. controls +(0.45,0.162) and +(-0.45,0.162) .. (GPTc27_1.west);
\draw[branch] (GPTb27.east) .. controls +(0.45,0.174) and +(-0.45,0.174) .. (GPTc27_2.west);
\draw[black!7] (0.00,-15.800) -- (8.45,-15.800);
\node[pointerlabel] at (0.62,-15.570) {P28};
\node[bestnode] (GPTb28) at (1.55,-15.570) {6a170da};
\node[promotednode] (GPTc28_1) at (2.75,-15.570) {f13a9fc};
\draw[branch] (GPTb28.east) -- (GPTc28_1.west);
\draw[black!7] (0.00,-16.360) -- (8.45,-16.360);
\node[pointerlabel] at (0.62,-16.130) {P29};
\node[bestnode] (GPTb29) at (1.55,-16.130) {f13a9fc};
\node[candnode] (GPTc29_1) at (2.75,-16.130) {fc6a0fa};
\node[promotednode] (GPTc29_2) at (8.05,-16.130) {df30b0f};
\draw[branch] (GPTb29.east) .. controls +(0.45,0.162) and +(-0.45,0.162) .. (GPTc29_1.west);
\draw[branch] (GPTb29.east) .. controls +(0.45,0.174) and +(-0.45,0.174) .. (GPTc29_2.west);
\draw[black!7] (0.00,-16.920) -- (8.45,-16.920);
\node[pointerlabel] at (0.62,-16.690) {P30};
\node[bestnode] (GPTb30) at (1.55,-16.690) {df30b0f};
\node[promotednode] (GPTc30_1) at (2.75,-16.690) {dc300f0};
\draw[branch] (GPTb30.east) -- (GPTc30_1.west);
\draw[black!7] (0.00,-17.480) -- (8.45,-17.480);
\node[pointerlabel] at (0.62,-17.250) {P31};
\node[bestnode] (GPTb31) at (1.55,-17.250) {dc300f0};
\node[candnode] (GPTc31_1) at (2.75,-17.250) {7184dce};
\node[promotednode] (GPTc31_2) at (8.05,-17.250) {a3d306a};
\draw[branch] (GPTb31.east) .. controls +(0.45,0.162) and +(-0.45,0.162) .. (GPTc31_1.west);
\draw[branch] (GPTb31.east) .. controls +(0.45,0.174) and +(-0.45,0.174) .. (GPTc31_2.west);
\draw[black!7] (0.00,-18.040) -- (8.45,-18.040);
\node[pointerlabel] at (0.62,-17.810) {P32};
\node[bestnode] (GPTb32) at (1.55,-17.810) {a3d306a};
\node[candnode] (GPTc32_1) at (2.75,-17.810) {5f36f8f};
\node[candnode] (GPTc32_2) at (5.40,-17.810) {4d97c59};
\node[promotednode] (GPTc32_3) at (8.05,-17.810) {e48fec5};
\draw[branch] (GPTb32.east) .. controls +(0.45,0.162) and +(-0.45,0.162) .. (GPTc32_1.west);
\draw[branch] (GPTb32.east) .. controls +(0.45,0.174) and +(-0.45,0.174) .. (GPTc32_2.west);
\draw[branch] (GPTb32.east) .. controls +(0.45,0.186) and +(-0.45,0.186) .. (GPTc32_3.west);
\draw[black!7] (0.00,-18.600) -- (8.45,-18.600);
\node[pointerlabel] at (0.62,-18.370) {P33};
\node[bestnode] (GPTb33) at (1.55,-18.370) {e48fec5};
\node[candnode] (GPTc33_1) at (2.75,-18.370) {e041af6};
\node[promotednode] (GPTc33_2) at (8.05,-18.370) {44e032e};
\draw[branch] (GPTb33.east) .. controls +(0.45,0.162) and +(-0.45,0.162) .. (GPTc33_1.west);
\draw[branch] (GPTb33.east) .. controls +(0.45,0.174) and +(-0.45,0.174) .. (GPTc33_2.west);
\draw[black!7] (0.00,-19.160) -- (8.45,-19.160);
\node[pointerlabel] at (0.62,-18.930) {P34};
\node[bestnode] (GPTb34) at (1.55,-18.930) {44e032e};
\node[candnode] (GPTc34_1) at (2.75,-18.930) {5a852c5};
\node[candnode] (GPTc34_2) at (5.40,-18.930) {e871562};
\node[promotednode] (GPTc34_3) at (8.05,-18.930) {296c3ef};
\draw[branch] (GPTb34.east) .. controls +(0.45,0.162) and +(-0.45,0.162) .. (GPTc34_1.west);
\draw[branch] (GPTb34.east) .. controls +(0.45,0.174) and +(-0.45,0.174) .. (GPTc34_2.west);
\draw[branch] (GPTb34.east) .. controls +(0.45,0.186) and +(-0.45,0.186) .. (GPTc34_3.west);
\draw[black!7] (0.00,-19.720) -- (8.45,-19.720);
\node[pointerlabel] at (0.62,-19.490) {P35};
\node[bestnode] (GPTb35) at (1.55,-19.490) {296c3ef};
\node[candnode] (GPTc35_1) at (2.75,-19.490) {51b9eb4};
\node[candnode] (GPTc35_2) at (5.40,-19.490) {9ce5603};
\node[promotednode] (GPTc35_3) at (8.05,-19.490) {4d4f04f};
\draw[branch] (GPTb35.east) .. controls +(0.45,0.162) and +(-0.45,0.162) .. (GPTc35_1.west);
\draw[branch] (GPTb35.east) .. controls +(0.45,0.174) and +(-0.45,0.174) .. (GPTc35_2.west);
\draw[branch] (GPTb35.east) .. controls +(0.45,0.186) and +(-0.45,0.186) .. (GPTc35_3.west);
\draw[black!7] (0.00,-20.280) -- (8.45,-20.280);
\node[pointerlabel] at (0.62,-20.050) {P36};
\node[bestnode] (GPTb36) at (1.55,-20.050) {4d4f04f};
\node[candnode] (GPTc36_1) at (2.75,-20.050) {db3d19b};
\node[promotednode] (GPTc36_2) at (8.05,-20.050) {a34e6e9};
\draw[branch] (GPTb36.east) .. controls +(0.45,0.162) and +(-0.45,0.162) .. (GPTc36_1.west);
\draw[branch] (GPTb36.east) .. controls +(0.45,0.174) and +(-0.45,0.174) .. (GPTc36_2.west);
\draw[black!7] (0.00,-20.840) -- (8.45,-20.840);
\node[pointerlabel] at (0.62,-20.610) {P37};
\node[bestnode] (GPTb37) at (1.55,-20.610) {a34e6e9};
\node[promotednode] (GPTc37_1) at (2.75,-20.610) {ff90d2a};
\draw[branch] (GPTb37.east) -- (GPTc37_1.west);
\draw[black!7] (0.00,-21.400) -- (8.45,-21.400);
\node[pointerlabel] at (0.62,-21.170) {P38};
\node[bestnode] (GPTb38) at (1.55,-21.170) {ff90d2a};
\node[candnode] (GPTc38_1) at (2.75,-21.170) {c9f5922};
\node[promotednode] (GPTc38_2) at (8.05,-21.170) {3ea2143};
\draw[branch] (GPTb38.east) .. controls +(0.45,0.162) and +(-0.45,0.162) .. (GPTc38_1.west);
\draw[branch] (GPTb38.east) .. controls +(0.45,0.174) and +(-0.45,0.174) .. (GPTc38_2.west);
\draw[black!7] (0.00,-21.960) -- (8.45,-21.960);
\node[pointerlabel] at (0.62,-21.730) {P39};
\node[bestnode] (GPTb39) at (1.55,-21.730) {3ea2143};
\node[promotednode] (GPTc39_1) at (2.75,-21.730) {fb7a727};
\draw[branch] (GPTb39.east) -- (GPTc39_1.west);
\draw[black!7] (0.00,-22.520) -- (8.45,-22.520);
\node[pointerlabel] at (0.62,-22.290) {P40};
\node[bestnode] (GPTb40) at (1.55,-22.290) {fb7a727};
\node[candnode] (GPTc40_1) at (2.75,-22.290) {494bb04};
\draw[branch] (GPTb40.east) -- (GPTc40_1.west);
\draw[promote] (GPTc02_1.south) .. controls +(0,-0.22) and +(0,0.22) .. (GPTb03.north);
\draw[promote] (GPTc03_2.south) .. controls +(0,-0.22) and +(0,0.22) .. (GPTb04.north);
\draw[promote] (GPTc04_5.south) .. controls +(0,-0.22) and +(0,0.22) .. (GPTb05.north);
\draw[promote] (GPTc05_1.south) .. controls +(0,-0.22) and +(0,0.22) .. (GPTb06.north);
\draw[promote] (GPTc06_2.south) .. controls +(0,-0.22) and +(0,0.22) .. (GPTb07.north);
\draw[promote] (GPTc07_2.south) .. controls +(0,-0.22) and +(0,0.22) .. (GPTb08.north);
\draw[promote] (GPTc08_3.south) .. controls +(0,-0.22) and +(0,0.22) .. (GPTb09.north);
\draw[promote] (GPTc09_2.south) .. controls +(0,-0.22) and +(0,0.22) .. (GPTb10.north);
\draw[promote] (GPTc10_2.south) .. controls +(0,-0.22) and +(0,0.22) .. (GPTb11.north);
\draw[promote] (GPTc11_3.south) .. controls +(0,-0.22) and +(0,0.22) .. (GPTb12.north);
\draw[promote] (GPTc12_3.south) .. controls +(0,-0.22) and +(0,0.22) .. (GPTb13.north);
\draw[promote] (GPTc13_3.south) .. controls +(0,-0.22) and +(0,0.22) .. (GPTb14.north);
\draw[promote] (GPTc14_2.south) .. controls +(0,-0.22) and +(0,0.22) .. (GPTb15.north);
\draw[promote] (GPTc15_3.south) .. controls +(0,-0.22) and +(0,0.22) .. (GPTb16.north);
\draw[promote] (GPTc16_2.south) .. controls +(0,-0.22) and +(0,0.22) .. (GPTb17.north);
\draw[promote] (GPTc17_2.south) .. controls +(0,-0.22) and +(0,0.22) .. (GPTb18.north);
\draw[promote] (GPTc18_1.south) .. controls +(0,-0.22) and +(0,0.22) .. (GPTb19.north);
\draw[promote] (GPTc19_1.south) .. controls +(0,-0.22) and +(0,0.22) .. (GPTb20.north);
\draw[promote] (GPTc20_2.south) .. controls +(0,-0.22) and +(0,0.22) .. (GPTb21.north);
\draw[promote] (GPTc21_4.south) .. controls +(0,-0.22) and +(0,0.22) .. (GPTb22.north);
\draw[promote] (GPTc22_2.south) .. controls +(0,-0.22) and +(0,0.22) .. (GPTb23.north);
\draw[promote] (GPTc23_2.south) .. controls +(0,-0.22) and +(0,0.22) .. (GPTb24.north);
\draw[promote] (GPTc24_2.south) .. controls +(0,-0.22) and +(0,0.22) .. (GPTb25.north);
\draw[promote] (GPTc25_1.south) .. controls +(0,-0.22) and +(0,0.22) .. (GPTb26.north);
\draw[promote] (GPTc26_3.south) .. controls +(0,-0.22) and +(0,0.22) .. (GPTb27.north);
\draw[promote] (GPTc27_2.south) .. controls +(0,-0.22) and +(0,0.22) .. (GPTb28.north);
\draw[promote] (GPTc28_1.south) .. controls +(0,-0.22) and +(0,0.22) .. (GPTb29.north);
\draw[promote] (GPTc29_2.south) .. controls +(0,-0.22) and +(0,0.22) .. (GPTb30.north);
\draw[promote] (GPTc30_1.south) .. controls +(0,-0.22) and +(0,0.22) .. (GPTb31.north);
\draw[promote] (GPTc31_2.south) .. controls +(0,-0.22) and +(0,0.22) .. (GPTb32.north);
\draw[promote] (GPTc32_3.south) .. controls +(0,-0.22) and +(0,0.22) .. (GPTb33.north);
\draw[promote] (GPTc33_2.south) .. controls +(0,-0.22) and +(0,0.22) .. (GPTb34.north);
\draw[promote] (GPTc34_3.south) .. controls +(0,-0.22) and +(0,0.22) .. (GPTb35.north);
\draw[promote] (GPTc35_3.south) .. controls +(0,-0.22) and +(0,0.22) .. (GPTb36.north);
\draw[promote] (GPTc36_2.south) .. controls +(0,-0.22) and +(0,0.22) .. (GPTb37.north);
\draw[promote] (GPTc37_1.south) .. controls +(0,-0.22) and +(0,0.22) .. (GPTb38.north);
\draw[promote] (GPTc38_2.south) .. controls +(0,-0.22) and +(0,0.22) .. (GPTb39.north);
\draw[promote] (GPTc39_1.south) .. controls +(0,-0.22) and +(0,0.22) .. (GPTb40.north);
\end{tikzpicture}

%% file: figures/glm_playbook_pointer_normalized_tikz.tex
% Compact single-column rendering of the normalized GLM playbook pointer table.
\begin{tikzpicture}[
  x=1cm,y=1cm,
  pointerlabel/.style={font=\fontsize{6.4}{7}\selectfont\bfseries, anchor=east, text=black!75},
  bestnode/.style={rounded corners=1.2pt, fill=legendBlue!90, text=white, font=\fontsize{4.9}{5.2}\selectfont\ttfamily, minimum width=0.82cm, minimum height=0.22cm, inner sep=0.8pt},
  candnode/.style={rounded corners=1.2pt, fill=legendOrange!95, text=white, font=\fontsize{4.9}{5.2}\selectfont\ttfamily, minimum width=0.82cm, minimum height=0.22cm, inner sep=0.8pt},
  promotednode/.style={candnode, draw=legendPurple!95, line width=0.55pt},
  branch/.style={-{Latex[length=1.0mm]}, line width=0.22pt, draw=black!55, opacity=0.72},
  promote/.style={-{Latex[length=1.1mm]}, line width=0.28pt, draw=legendPurple!60, opacity=0.45},
]
\draw[black!7] (0.00,-0.680) -- (8.45,-0.680);
\node[pointerlabel] at (0.62,-0.450) {P01};
\node[bestnode] (GLMb01) at (1.55,-0.450) {a324657};
\draw[black!7] (0.00,-1.240) -- (8.45,-1.240);
\node[pointerlabel] at (0.62,-1.010) {P02};
\node[bestnode] (GLMb02) at (1.55,-1.010) {a324657};
\node[candnode] (GLMc02_1) at (2.75,-1.010) {dd378c4};
\node[candnode] (GLMc02_2) at (5.40,-1.010) {126b8fb};
\node[promotednode] (GLMc02_3) at (8.05,-1.010) {fba6b99};
\draw[branch] (GLMb02.east) .. controls +(0.45,0.162) and +(-0.45,0.162) .. (GLMc02_1.west);
\draw[branch] (GLMb02.east) .. controls +(0.45,0.174) and +(-0.45,0.174) .. (GLMc02_2.west);
\draw[branch] (GLMb02.east) .. controls +(0.45,0.186) and +(-0.45,0.186) .. (GLMc02_3.west);
\draw[black!7] (0.00,-1.800) -- (8.45,-1.800);
\node[pointerlabel] at (0.62,-1.570) {P03};
\node[bestnode] (GLMb03) at (1.55,-1.570) {fba6b99};
\node[promotednode] (GLMc03_1) at (2.75,-1.570) {074acfb};
\draw[branch] (GLMb03.east) -- (GLMc03_1.west);
\draw[black!7] (0.00,-2.360) -- (8.45,-2.360);
\node[pointerlabel] at (0.62,-2.130) {P04};
\node[bestnode] (GLMb04) at (1.55,-2.130) {074acfb};
\node[candnode] (GLMc04_1) at (2.75,-2.130) {ba8c970};
\node[candnode] (GLMc04_2) at (4.52,-2.130) {65ca0c0};
\node[candnode] (GLMc04_3) at (6.28,-2.130) {f8b75a2};
\node[promotednode] (GLMc04_4) at (8.05,-2.130) {43ff77f};
\draw[branch] (GLMb04.east) .. controls +(0.45,0.162) and +(-0.45,0.162) .. (GLMc04_1.west);
\draw[branch] (GLMb04.east) .. controls +(0.45,0.174) and +(-0.45,0.174) .. (GLMc04_2.west);
\draw[branch] (GLMb04.east) .. controls +(0.45,0.186) and +(-0.45,0.186) .. (GLMc04_3.west);
\draw[branch] (GLMb04.east) .. controls +(0.45,0.198) and +(-0.45,0.198) .. (GLMc04_4.west);
\draw[black!7] (0.00,-2.920) -- (8.45,-2.920);
\node[pointerlabel] at (0.62,-2.690) {P05};
\node[bestnode] (GLMb05) at (1.55,-2.690) {43ff77f};
\node[candnode] (GLMc05_1) at (2.75,-2.690) {5a609d0};
\node[candnode] (GLMc05_2) at (4.08,-2.690) {51e2cbe};
\node[candnode] (GLMc05_3) at (5.40,-2.690) {571c990};
\node[candnode] (GLMc05_4) at (6.73,-2.690) {66ef62b};
\node[promotednode] (GLMc05_5) at (8.05,-2.690) {91321dc};
\draw[branch] (GLMb05.east) .. controls +(0.45,0.162) and +(-0.45,0.162) .. (GLMc05_1.west);
\draw[branch] (GLMb05.east) .. controls +(0.45,0.174) and +(-0.45,0.174) .. (GLMc05_2.west);
\draw[branch] (GLMb05.east) .. controls +(0.45,0.186) and +(-0.45,0.186) .. (GLMc05_3.west);
\draw[branch] (GLMb05.east) .. controls +(0.45,0.198) and +(-0.45,0.198) .. (GLMc05_4.west);
\draw[branch] (GLMb05.east) .. controls +(0.45,0.210) and +(-0.45,0.210) .. (GLMc05_5.west);
\draw[black!7] (0.00,-3.480) -- (8.45,-3.480);
\node[pointerlabel] at (0.62,-3.250) {P06};
\node[bestnode] (GLMb06) at (1.55,-3.250) {91321dc};
\node[promotednode] (GLMc06_1) at (2.75,-3.250) {3fe2d81};
\draw[branch] (GLMb06.east) -- (GLMc06_1.west);
\draw[black!7] (0.00,-4.040) -- (8.45,-4.040);
\node[pointerlabel] at (0.62,-3.810) {P07};
\node[bestnode] (GLMb07) at (1.55,-3.810) {3fe2d81};
\node[candnode] (GLMc07_1) at (2.75,-3.810) {acb2626};
\node[candnode] (GLMc07_2) at (4.52,-3.810) {6a5641f};
\node[candnode] (GLMc07_3) at (6.28,-3.810) {b4576f6};
\node[promotednode] (GLMc07_4) at (8.05,-3.810) {910f58c};
\draw[branch] (GLMb07.east) .. controls +(0.45,0.162) and +(-0.45,0.162) .. (GLMc07_1.west);
\draw[branch] (GLMb07.east) .. controls +(0.45,0.174) and +(-0.45,0.174) .. (GLMc07_2.west);
\draw[branch] (GLMb07.east) .. controls +(0.45,0.186) and +(-0.45,0.186) .. (GLMc07_3.west);
\draw[branch] (GLMb07.east) .. controls +(0.45,0.198) and +(-0.45,0.198) .. (GLMc07_4.west);
\draw[black!7] (0.00,-4.600) -- (8.45,-4.600);
\node[pointerlabel] at (0.62,-4.370) {P08};
\node[bestnode] (GLMb08) at (1.55,-4.370) {910f58c};
\node[promotednode] (GLMc08_1) at (2.75,-4.370) {18ed55d};
\draw[branch] (GLMb08.east) -- (GLMc08_1.west);
\draw[black!7] (0.00,-5.160) -- (8.45,-5.160);
\node[pointerlabel] at (0.62,-4.930) {P09};
\node[bestnode] (GLMb09) at (1.55,-4.930) {18ed55d};
\node[candnode] (GLMc09_1) at (2.75,-4.930) {64da137};
\node[candnode] (GLMc09_2) at (5.40,-4.930) {7bb6807};
\node[promotednode] (GLMc09_3) at (8.05,-4.930) {e74a5e4};
\draw[branch] (GLMb09.east) .. controls +(0.45,0.162) and +(-0.45,0.162) .. (GLMc09_1.west);
\draw[branch] (GLMb09.east) .. controls +(0.45,0.174) and +(-0.45,0.174) .. (GLMc09_2.west);
\draw[branch] (GLMb09.east) .. controls +(0.45,0.186) and +(-0.45,0.186) .. (GLMc09_3.west);
\draw[black!7] (0.00,-5.720) -- (8.45,-5.720);
\node[pointerlabel] at (0.62,-5.490) {P10};
\node[bestnode] (GLMb10) at (1.55,-5.490) {e74a5e4};
\node[promotednode] (GLMc10_1) at (2.75,-5.490) {e6c0108};
\draw[branch] (GLMb10.east) -- (GLMc10_1.west);
\draw[black!7] (0.00,-6.280) -- (8.45,-6.280);
\node[pointerlabel] at (0.62,-6.050) {P11};
\node[bestnode] (GLMb11) at (1.55,-6.050) {e6c0108};
\node[candnode] (GLMc11_1) at (2.75,-6.050) {70b1184};
\node[promotednode] (GLMc11_2) at (8.05,-6.050) {61b7e04};
\draw[branch] (GLMb11.east) .. controls +(0.45,0.162) and +(-0.45,0.162) .. (GLMc11_1.west);
\draw[branch] (GLMb11.east) .. controls +(0.45,0.174) and +(-0.45,0.174) .. (GLMc11_2.west);
\draw[black!7] (0.00,-6.840) -- (8.45,-6.840);
\node[pointerlabel] at (0.62,-6.610) {P12};
\node[bestnode] (GLMb12) at (1.55,-6.610) {61b7e04};
\node[promotednode] (GLMc12_1) at (2.75,-6.610) {6f38653};
\draw[branch] (GLMb12.east) -- (GLMc12_1.west);
\draw[black!7] (0.00,-7.400) -- (8.45,-7.400);
\node[pointerlabel] at (0.62,-7.170) {P13};
\node[bestnode] (GLMb13) at (1.55,-7.170) {6f38653};
\node[promotednode] (GLMc13_1) at (2.75,-7.170) {3738759};
\draw[branch] (GLMb13.east) -- (GLMc13_1.west);
\draw[black!7] (0.00,-7.960) -- (8.45,-7.960);
\node[pointerlabel] at (0.62,-7.730) {P14};
\node[bestnode] (GLMb14) at (1.55,-7.730) {3738759};
\node[promotednode] (GLMc14_1) at (2.75,-7.730) {d38ea7c};
\draw[branch] (GLMb14.east) -- (GLMc14_1.west);
\draw[black!7] (0.00,-8.520) -- (8.45,-8.520);
\node[pointerlabel] at (0.62,-8.290) {P15};
\node[bestnode] (GLMb15) at (1.55,-8.290) {d38ea7c};
\node[promotednode] (GLMc15_1) at (2.75,-8.290) {38fe223};
\draw[branch] (GLMb15.east) -- (GLMc15_1.west);
\draw[black!7] (0.00,-9.080) -- (8.45,-9.080);
\node[pointerlabel] at (0.62,-8.850) {P16};
\node[bestnode] (GLMb16) at (1.55,-8.850) {38fe223};
\node[promotednode] (GLMc16_1) at (2.75,-8.850) {93a963d};
\draw[branch] (GLMb16.east) -- (GLMc16_1.west);
\draw[black!7] (0.00,-9.640) -- (8.45,-9.640);
\node[pointerlabel] at (0.62,-9.410) {P17};
\node[bestnode] (GLMb17) at (1.55,-9.410) {93a963d};
\node[promotednode] (GLMc17_1) at (2.75,-9.410) {59b54ca};
\draw[branch] (GLMb17.east) -- (GLMc17_1.west);
\draw[black!7] (0.00,-10.200) -- (8.45,-10.200);
\node[pointerlabel] at (0.62,-9.970) {P18};
\node[bestnode] (GLMb18) at (1.55,-9.970) {59b54ca};
\node[candnode] (GLMc18_1) at (2.75,-9.970) {c3685ed};
\node[promotednode] (GLMc18_2) at (8.05,-9.970) {ac41539};
\draw[branch] (GLMb18.east) .. controls +(0.45,0.162) and +(-0.45,0.162) .. (GLMc18_1.west);
\draw[branch] (GLMb18.east) .. controls +(0.45,0.174) and +(-0.45,0.174) .. (GLMc18_2.west);
\draw[black!7] (0.00,-10.760) -- (8.45,-10.760);
\node[pointerlabel] at (0.62,-10.530) {P19};
\node[bestnode] (GLMb19) at (1.55,-10.530) {ac41539};
\node[promotednode] (GLMc19_1) at (2.75,-10.530) {2be3a55};
\draw[branch] (GLMb19.east) -- (GLMc19_1.west);
\draw[black!7] (0.00,-11.320) -- (8.45,-11.320);
\node[pointerlabel] at (0.62,-11.090) {P20};
\node[bestnode] (GLMb20) at (1.55,-11.090) {2be3a55};
\node[promotednode] (GLMc20_1) at (2.75,-11.090) {634fed2};
\draw[branch] (GLMb20.east) -- (GLMc20_1.west);
\draw[black!7] (0.00,-11.880) -- (8.45,-11.880);
\node[pointerlabel] at (0.62,-11.650) {P21};
\node[bestnode] (GLMb21) at (1.55,-11.650) {634fed2};
\node[promotednode] (GLMc21_1) at (2.75,-11.650) {29f289c};
\draw[branch] (GLMb21.east) -- (GLMc21_1.west);
\draw[black!7] (0.00,-12.440) -- (8.45,-12.440);
\node[pointerlabel] at (0.62,-12.210) {P22};
\node[bestnode] (GLMb22) at (1.55,-12.210) {29f289c};
\node[candnode] (GLMc22_1) at (2.75,-12.210) {27ad2cc};
\node[candnode] (GLMc22_2) at (5.40,-12.210) {1500d8e};
\node[promotednode] (GLMc22_3) at (8.05,-12.210) {cad7659};
\draw[branch] (GLMb22.east) .. controls +(0.45,0.162) and +(-0.45,0.162) .. (GLMc22_1.west);
\draw[branch] (GLMb22.east) .. controls +(0.45,0.174) and +(-0.45,0.174) .. (GLMc22_2.west);
\draw[branch] (GLMb22.east) .. controls +(0.45,0.186) and +(-0.45,0.186) .. (GLMc22_3.west);
\draw[black!7] (0.00,-13.000) -- (8.45,-13.000);
\node[pointerlabel] at (0.62,-12.770) {P23};
\node[bestnode] (GLMb23) at (1.55,-12.770) {cad7659};
\node[promotednode] (GLMc23_1) at (2.75,-12.770) {9809e14};
\draw[branch] (GLMb23.east) -- (GLMc23_1.west);
\draw[black!7] (0.00,-13.560) -- (8.45,-13.560);
\node[pointerlabel] at (0.62,-13.330) {P24};
\node[bestnode] (GLMb24) at (1.55,-13.330) {9809e14};
\node[promotednode] (GLMc24_1) at (2.75,-13.330) {ccd87ed};
\draw[branch] (GLMb24.east) -- (GLMc24_1.west);
\draw[black!7] (0.00,-14.120) -- (8.45,-14.120);
\node[pointerlabel] at (0.62,-13.890) {P25};
\node[bestnode] (GLMb25) at (1.55,-13.890) {ccd87ed};
\node[promotednode] (GLMc25_1) at (2.75,-13.890) {6d735b9};
\draw[branch] (GLMb25.east) -- (GLMc25_1.west);
\draw[black!7] (0.00,-14.680) -- (8.45,-14.680);
\node[pointerlabel] at (0.62,-14.450) {P26};
\node[bestnode] (GLMb26) at (1.55,-14.450) {6d735b9};
\node[candnode] (GLMc26_1) at (2.75,-14.450) {1d6ac18};
\node[candnode] (GLMc26_2) at (5.40,-14.450) {e7ae710};
\node[promotednode] (GLMc26_3) at (8.05,-14.450) {5218d06};
\draw[branch] (GLMb26.east) .. controls +(0.45,0.162) and +(-0.45,0.162) .. (GLMc26_1.west);
\draw[branch] (GLMb26.east) .. controls +(0.45,0.174) and +(-0.45,0.174) .. (GLMc26_2.west);
\draw[branch] (GLMb26.east) .. controls +(0.45,0.186) and +(-0.45,0.186) .. (GLMc26_3.west);
\draw[black!7] (0.00,-15.240) -- (8.45,-15.240);
\node[pointerlabel] at (0.62,-15.010) {P27};
\node[bestnode] (GLMb27) at (1.55,-15.010) {5218d06};
\node[candnode] (GLMc27_1) at (2.75,-15.010) {8c96723};
\node[promotednode] (GLMc27_2) at (8.05,-15.010) {1d25368};
\draw[branch] (GLMb27.east) .. controls +(0.45,0.162) and +(-0.45,0.162) .. (GLMc27_1.west);
\draw[branch] (GLMb27.east) .. controls +(0.45,0.174) and +(-0.45,0.174) .. (GLMc27_2.west);
\draw[black!7] (0.00,-15.800) -- (8.45,-15.800);
\node[pointerlabel] at (0.62,-15.570) {P28};
\node[bestnode] (GLMb28) at (1.55,-15.570) {1d25368};
\node[promotednode] (GLMc28_1) at (2.75,-15.570) {254d03b};
\draw[branch] (GLMb28.east) -- (GLMc28_1.west);
\draw[black!7] (0.00,-16.360) -- (8.45,-16.360);
\node[pointerlabel] at (0.62,-16.130) {P29};
\node[bestnode] (GLMb29) at (1.55,-16.130) {254d03b};
\node[candnode] (GLMc29_1) at (2.75,-16.130) {3c113cb};
\node[candnode] (GLMc29_2) at (3.63,-16.130) {6919dd8};
\node[candnode] (GLMc29_3) at (4.52,-16.130) {1527c97};
\node[candnode] (GLMc29_4) at (5.40,-16.130) {360f619};
\node[candnode] (GLMc29_5) at (6.28,-16.130) {f30a5ce};
\node[candnode] (GLMc29_6) at (7.17,-16.130) {bc69e17};
\node[promotednode] (GLMc29_7) at (8.05,-16.130) {f6c13d3};
\draw[branch] (GLMb29.east) .. controls +(0.45,0.162) and +(-0.45,0.162) .. (GLMc29_1.west);
\draw[branch] (GLMb29.east) .. controls +(0.45,0.174) and +(-0.45,0.174) .. (GLMc29_2.west);
\draw[branch] (GLMb29.east) .. controls +(0.45,0.186) and +(-0.45,0.186) .. (GLMc29_3.west);
\draw[branch] (GLMb29.east) .. controls +(0.45,0.198) and +(-0.45,0.198) .. (GLMc29_4.west);
\draw[branch] (GLMb29.east) .. controls +(0.45,0.210) and +(-0.45,0.210) .. (GLMc29_5.west);
\draw[branch] (GLMb29.east) .. controls +(0.45,0.222) and +(-0.45,0.222) .. (GLMc29_6.west);
\draw[branch] (GLMb29.east) .. controls +(0.45,0.222) and +(-0.45,0.222) .. (GLMc29_7.west);
\draw[black!7] (0.00,-16.920) -- (8.45,-16.920);
\node[pointerlabel] at (0.62,-16.690) {P30};
\node[bestnode] (GLMb30) at (1.55,-16.690) {f6c13d3};
\node[candnode] (GLMc30_1) at (2.75,-16.690) {8d13533};
\node[candnode] (GLMc30_2) at (4.07,-16.690) {5cb81a2};
\node[candnode] (GLMc30_3) at (5.40,-16.690) {60b2e3c};
\node[candnode] (GLMc30_4) at (6.73,-16.690) {e36b47b};
\node[promotednode] (GLMc30_5) at (8.05,-16.690) {3fe7179};
\draw[branch] (GLMb30.east) .. controls +(0.45,0.162) and +(-0.45,0.162) .. (GLMc30_1.west);
\draw[branch] (GLMb30.east) .. controls +(0.45,0.174) and +(-0.45,0.174) .. (GLMc30_2.west);
\draw[branch] (GLMb30.east) .. controls +(0.45,0.186) and +(-0.45,0.186) .. (GLMc30_3.west);
\draw[branch] (GLMb30.east) .. controls +(0.45,0.198) and +(-0.45,0.198) .. (GLMc30_4.west);
\draw[branch] (GLMb30.east) .. controls +(0.45,0.210) and +(-0.45,0.210) .. (GLMc30_5.west);
\draw[black!7] (0.00,-17.480) -- (8.45,-17.480);
\node[pointerlabel] at (0.62,-17.250) {P31};
\node[bestnode] (GLMb31) at (1.55,-17.250) {3fe7179};
\node[promotednode] (GLMc31_1) at (2.75,-17.250) {19d251e};
\draw[branch] (GLMb31.east) -- (GLMc31_1.west);
\draw[black!7] (0.00,-18.040) -- (8.45,-18.040);
\node[pointerlabel] at (0.62,-17.810) {P32};
\node[bestnode] (GLMb32) at (1.55,-17.810) {19d251e};
\node[promotednode] (GLMc32_1) at (2.75,-17.810) {78d7115};
\draw[branch] (GLMb32.east) -- (GLMc32_1.west);
\draw[black!7] (0.00,-18.600) -- (8.45,-18.600);
\node[pointerlabel] at (0.62,-18.370) {P33};
\node[bestnode] (GLMb33) at (1.55,-18.370) {78d7115};
\node[promotednode] (GLMc33_1) at (2.75,-18.370) {fce08c6};
\draw[branch] (GLMb33.east) -- (GLMc33_1.west);
\draw[black!7] (0.00,-19.160) -- (8.45,-19.160);
\node[pointerlabel] at (0.62,-18.930) {P34};
\node[bestnode] (GLMb34) at (1.55,-18.930) {fce08c6};
\node[candnode] (GLMc34_1) at (2.75,-18.930) {e5abc23};
\node[candnode] (GLMc34_2) at (4.52,-18.930) {9195526};
\node[candnode] (GLMc34_3) at (6.28,-18.930) {7d45ce6};
\node[promotednode] (GLMc34_4) at (8.05,-18.930) {877a333};
\draw[branch] (GLMb34.east) .. controls +(0.45,0.162) and +(-0.45,0.162) .. (GLMc34_1.west);
\draw[branch] (GLMb34.east) .. controls +(0.45,0.174) and +(-0.45,0.174) .. (GLMc34_2.west);
\draw[branch] (GLMb34.east) .. controls +(0.45,0.186) and +(-0.45,0.186) .. (GLMc34_3.west);
\draw[branch] (GLMb34.east) .. controls +(0.45,0.198) and +(-0.45,0.198) .. (GLMc34_4.west);
\draw[black!7] (0.00,-19.720) -- (8.45,-19.720);
\node[pointerlabel] at (0.62,-19.490) {P35};
\node[bestnode] (GLMb35) at (1.55,-19.490) {877a333};
\node[candnode] (GLMc35_1) at (2.75,-19.490) {e11a52d};
\node[promotednode] (GLMc35_2) at (8.05,-19.490) {44bd13b};
\draw[branch] (GLMb35.east) .. controls +(0.45,0.162) and +(-0.45,0.162) .. (GLMc35_1.west);
\draw[branch] (GLMb35.east) .. controls +(0.45,0.174) and +(-0.45,0.174) .. (GLMc35_2.west);
\draw[black!7] (0.00,-20.280) -- (8.45,-20.280);
\node[pointerlabel] at (0.62,-20.050) {P36};
\node[bestnode] (GLMb36) at (1.55,-20.050) {44bd13b};
\node[candnode] (GLMc36_1) at (2.75,-20.050) {72ec2a1};
\node[promotednode] (GLMc36_2) at (8.05,-20.050) {84ecce3};
\draw[branch] (GLMb36.east) .. controls +(0.45,0.162) and +(-0.45,0.162) .. (GLMc36_1.west);
\draw[branch] (GLMb36.east) .. controls +(0.45,0.174) and +(-0.45,0.174) .. (GLMc36_2.west);
\draw[black!7] (0.00,-20.840) -- (8.45,-20.840);
\node[pointerlabel] at (0.62,-20.610) {P37};
\node[bestnode] (GLMb37) at (1.55,-20.610) {84ecce3};
\node[candnode] (GLMc37_1) at (2.75,-20.610) {9acf29a};
\node[candnode] (GLMc37_2) at (4.52,-20.610) {a5cdbad};
\node[candnode] (GLMc37_3) at (6.28,-20.610) {7134256};
\node[promotednode] (GLMc37_4) at (8.05,-20.610) {e2ffe8b};
\draw[branch] (GLMb37.east) .. controls +(0.45,0.162) and +(-0.45,0.162) .. (GLMc37_1.west);
\draw[branch] (GLMb37.east) .. controls +(0.45,0.174) and +(-0.45,0.174) .. (GLMc37_2.west);
\draw[branch] (GLMb37.east) .. controls +(0.45,0.186) and +(-0.45,0.186) .. (GLMc37_3.west);
\draw[branch] (GLMb37.east) .. controls +(0.45,0.198) and +(-0.45,0.198) .. (GLMc37_4.west);
\draw[black!7] (0.00,-21.400) -- (8.45,-21.400);
\node[pointerlabel] at (0.62,-21.170) {P38};
\node[bestnode] (GLMb38) at (1.55,-21.170) {e2ffe8b};
\node[candnode] (GLMc38_1) at (2.75,-21.170) {4c352ae};
\node[candnode] (GLMc38_2) at (5.40,-21.170) {ad83fed};
\node[promotednode] (GLMc38_3) at (8.05,-21.170) {7b711f0};
\draw[branch] (GLMb38.east) .. controls +(0.45,0.162) and +(-0.45,0.162) .. (GLMc38_1.west);
\draw[branch] (GLMb38.east) .. controls +(0.45,0.174) and +(-0.45,0.174) .. (GLMc38_2.west);
\draw[branch] (GLMb38.east) .. controls +(0.45,0.186) and +(-0.45,0.186) .. (GLMc38_3.west);
\draw[black!7] (0.00,-21.960) -- (8.45,-21.960);
\node[pointerlabel] at (0.62,-21.730) {P39};
\node[bestnode] (GLMb39) at (1.55,-21.730) {7b711f0};
\node[candnode] (GLMc39_1) at (2.75,-21.730) {bed3118};
\node[candnode] (GLMc39_2) at (5.40,-21.730) {934951f};
\node[promotednode] (GLMc39_3) at (8.05,-21.730) {e33522e};
\draw[branch] (GLMb39.east) .. controls +(0.45,0.162) and +(-0.45,0.162) .. (GLMc39_1.west);
\draw[branch] (GLMb39.east) .. controls +(0.45,0.174) and +(-0.45,0.174) .. (GLMc39_2.west);
\draw[branch] (GLMb39.east) .. controls +(0.45,0.186) and +(-0.45,0.186) .. (GLMc39_3.west);
\draw[black!7] (0.00,-22.520) -- (8.45,-22.520);
\node[pointerlabel] at (0.62,-22.290) {P40};
\node[bestnode] (GLMb40) at (1.55,-22.290) {e33522e};
\node[candnode] (GLMc40_1) at (2.75,-22.290) {2ba7552};
\draw[branch] (GLMb40.east) -- (GLMc40_1.west);
\draw[promote] (GLMc02_3.south) .. controls +(0,-0.22) and +(0,0.22) .. (GLMb03.north);
\draw[promote] (GLMc03_1.south) .. controls +(0,-0.22) and +(0,0.22) .. (GLMb04.north);
\draw[promote] (GLMc04_4.south) .. controls +(0,-0.22) and +(0,0.22) .. (GLMb05.north);
\draw[promote] (GLMc05_5.south) .. controls +(0,-0.22) and +(0,0.22) .. (GLMb06.north);
\draw[promote] (GLMc06_1.south) .. controls +(0,-0.22) and +(0,0.22) .. (GLMb07.north);
\draw[promote] (GLMc07_4.south) .. controls +(0,-0.22) and +(0,0.22) .. (GLMb08.north);
\draw[promote] (GLMc08_1.south) .. controls +(0,-0.22) and +(0,0.22) .. (GLMb09.north);
\draw[promote] (GLMc09_3.south) .. controls +(0,-0.22) and +(0,0.22) .. (GLMb10.north);
\draw[promote] (GLMc10_1.south) .. controls +(0,-0.22) and +(0,0.22) .. (GLMb11.north);
\draw[promote] (GLMc11_2.south) .. controls +(0,-0.22) and +(0,0.22) .. (GLMb12.north);
\draw[promote] (GLMc12_1.south) .. controls +(0,-0.22) and +(0,0.22) .. (GLMb13.north);
\draw[promote] (GLMc13_1.south) .. controls +(0,-0.22) and +(0,0.22) .. (GLMb14.north);
\draw[promote] (GLMc14_1.south) .. controls +(0,-0.22) and +(0,0.22) .. (GLMb15.north);
\draw[promote] (GLMc15_1.south) .. controls +(0,-0.22) and +(0,0.22) .. (GLMb16.north);
\draw[promote] (GLMc16_1.south) .. controls +(0,-0.22) and +(0,0.22) .. (GLMb17.north);
\draw[promote] (GLMc17_1.south) .. controls +(0,-0.22) and +(0,0.22) .. (GLMb18.north);
\draw[promote] (GLMc18_2.south) .. controls +(0,-0.22) and +(0,0.22) .. (GLMb19.north);
\draw[promote] (GLMc19_1.south) .. controls +(0,-0.22) and +(0,0.22) .. (GLMb20.north);
\draw[promote] (GLMc20_1.south) .. controls +(0,-0.22) and +(0,0.22) .. (GLMb21.north);
\draw[promote] (GLMc21_1.south) .. controls +(0,-0.22) and +(0,0.22) .. (GLMb22.north);
\draw[promote] (GLMc22_3.south) .. controls +(0,-0.22) and +(0,0.22) .. (GLMb23.north);
\draw[promote] (GLMc23_1.south) .. controls +(0,-0.22) and +(0,0.22) .. (GLMb24.north);
\draw[promote] (GLMc24_1.south) .. controls +(0,-0.22) and +(0,0.22) .. (GLMb25.north);
\draw[promote] (GLMc25_1.south) .. controls +(0,-0.22) and +(0,0.22) .. (GLMb26.north);
\draw[promote] (GLMc26_3.south) .. controls +(0,-0.22) and +(0,0.22) .. (GLMb27.north);
\draw[promote] (GLMc27_2.south) .. controls +(0,-0.22) and +(0,0.22) .. (GLMb28.north);
\draw[promote] (GLMc28_1.south) .. controls +(0,-0.22) and +(0,0.22) .. (GLMb29.north);
\draw[promote] (GLMc29_7.south) .. controls +(0,-0.22) and +(0,0.22) .. (GLMb30.north);
\draw[promote] (GLMc30_5.south) .. controls +(0,-0.22) and +(0,0.22) .. (GLMb31.north);
\draw[promote] (GLMc31_1.south) .. controls +(0,-0.22) and +(0,0.22) .. (GLMb32.north);
\draw[promote] (GLMc32_1.south) .. controls +(0,-0.22) and +(0,0.22) .. (GLMb33.north);
\draw[promote] (GLMc33_1.south) .. controls +(0,-0.22) and +(0,0.22) .. (GLMb34.north);
\draw[promote] (GLMc34_4.south) .. controls +(0,-0.22) and +(0,0.22) .. (GLMb35.north);
\draw[promote] (GLMc35_2.south) .. controls +(0,-0.22) and +(0,0.22) .. (GLMb36.north);
\draw[promote] (GLMc36_2.south) .. controls +(0,-0.22) and +(0,0.22) .. (GLMb37.north);
\draw[promote] (GLMc37_4.south) .. controls +(0,-0.22) and +(0,0.22) .. (GLMb38.north);
\draw[promote] (GLMc38_3.south) .. controls +(0,-0.22) and +(0,0.22) .. (GLMb39.north);
\draw[promote] (GLMc39_3.south) .. controls +(0,-0.22) and +(0,0.22) .. (GLMb40.north);
\end{tikzpicture}

%% file: case_study/rusqlite_walkthrough.tex
\newcommand{\walkbox}[1]{%
    \par
\noindent%
    \begingroup%
    \setlength{\fboxsep}{2pt}%
    \setlength{\fboxrule}{0.8pt}%
    \begin{tikzpicture}
        \node[
            draw=black,
            rounded corners=2pt,
            line width=\fboxrule,
            inner sep=\fboxsep,
            text width=\dimexpr\linewidth-2\fboxsep-2\fboxrule\relax
        ] {%
            \raggedright\fontsize{8}{8.8}\selectfont
            \setlength{\parindent}{0pt}%
            \setlength{\parskip}{0pt}%
            \lineskip=0pt\lineskiplimit=0pt%
            \noindent\ignorespaces#1\unskip%
        };
    \end{tikzpicture}%
    \endgroup%
    \par%
}

\newcommand{\walkentry}[3]{%
    \par\medskip
    \noindent\textbf{#1.}\enspace #2\par\vspace{2pt}%
    \walkbox{#3}%
}
\newcommand{\turnplaybook}{\textbf{\textcolor{legendBlue}{Playbook read.}}}
\newcommand{\turnreasoning}{\textbf{\textcolor{legendOrange}{Reasoning.}}}
\newcommand{\turnoperation}{\textbf{\textcolor{legendPurple}{Operation.}}}
\newcommand{\turnsignal}{\textbf{\textcolor{legendTeal}{Signal.}}}

\section{Case Walkthrough: \texttt{rusqlite}}
\label{app:rusqlite-walkthrough}

\noindent\texttt{rusqlite} is a Rust SQLite binding library with public transaction and savepoint APIs. \tool identified a previously unknown SQL injection in the savepoint-name path: caller-controlled identifier text is interpolated verbatim into \texttt{SAVEPOINT}, \texttt{RELEASE}, and \texttt{ROLLBACK TO} commands executed via \texttt{execute\_\allowbreak batch}, with no quoting, escaping, or allowlisting. \textbf{The flaw dates to May~2016 (ten years ago)}, when commit \texttt{703cf22b5} (\emph{``Separate Savepoint out from Transaction''}) first introduced the savepoint-name parameter---baked in from the API's inception without SQL guards and undetected for nearly a decade until fixed after responsible disclosure. Critically, Rust's type system offered no protection: \texttt{T: Into<String>} is safe at the language boundary but not at the SQL-semantics level. The five-turn run used 52 tool calls and 3.2M tokens (95\% cached), and produced a passing \acs{poc}. We highlight three aspects where, in our view, playbook guidance changes what the agent does---Turn~3's source-to-sink trace reflects methodology any skilled analyst would apply regardless of procedure; the three below do not:
\begin{itemize}
\setlength{\topsep}{2pt}\setlength{\itemsep}{1pt}\setlength{\parsep}{0pt}
\item \textbf{Structured scoping.} Turns~1--2 show the playbook directing the agent to build a bounded release-surface map before any deep validation, steering it away from isolated sinks toward repository-owned construction boundaries.
\item \textbf{Sibling clearing.} Turn~4 illustrates checklist discipline: adjacent families (PRAGMA builders, \texttt{Name} helpers, raw SQL APIs) must each reach a terminal state before the savepoint path is accepted, preventing symptom drift.
\item \textbf{Executable, pinned \acs{poc}.} Turn~5 shows the playbook mandating a replayable proof with a verifier signal---the same ground truth used to score all benchmark results.
\end{itemize}

\par\medskip
\noindent\textbf{Turn 1: Release-Surface Scoping.}\enspace The playbook directs \tool to analyze \texttt{rusqlite} from the perspective of its callers, not its internals: attacker influence enters not through implementation details but through the safe, typed APIs that downstream applications invoke with lower-trust values. Reading the manifest, README, and exported modules, \tool anchors the run to five families on a live checklist---savepoints, PRAGMA builders, raw SQL APIs, \texttt{Name} helpers, and optional extension/vtab/blob paths---establishing a bounded surface map before deep validations.

\par\medskip
\noindent\textbf{Turn 2: SQL Triage.}\enspace \tool applies the SQL-injection guide and draws a critical distinction: caller-controlled SQL APIs (intended for arbitrary queries) are not a bug class, but repository-owned wrappers that silently translate non-SQL input---names, identifiers, fragments---into executable SQL text are. PRAGMA builders pass this test; they validate keywords and escape identifiers before execution. Savepoint-name handling does not: the library accepts any string and renders it directly as SQL syntax, with no comparable guard. Raw SQL and PRAGMA stay open as comparison families.

\walkentry{Turn 3: Savepoint Source-to-Sink Trace}{\tool confirms that the savepoint-name parameter travels from the public API to the SQL executor.}{%
\turnplaybook\ \emph{``attacker-controlled input source''}; \emph{``query-construction path''}; \emph{``exact unsafe SQL boundary''}; \emph{``prove parameterization is absent, partial, or bypassed''}.\par
\turnreasoning\ I follow the savepoint-name parameter because \texttt{T: Into<String>} is type-safe at the Rust boundary---the type system accepts any string as a savepoint name---but not SQL-safe: once that string is rendered as SQL syntax, it can carry arbitrary statements that SQLite will execute.\par
\turnoperation\ \tool traces \texttt{Connection::\allowbreak savepoint\_\allowbreak with\_\allowbreak name} and \texttt{Transaction::\allowbreak savepoint\_\allowbreak with\_\allowbreak name} into \texttt{Savepoint::\allowbreak with\_\allowbreak name\_}, where \texttt{SAVEPOINT \{name\}} reaches \texttt{execute\_\allowbreak batch}; the same name is reused verbatim in \texttt{RELEASE} and \texttt{ROLLBACK TO}, yielding three unsanitized SQL entry points from one parameter.\par
\turnsignal\ One public wrapper passes attacker-controlled text into three unguarded SQL commands.
}

\walkentry{Turn 4: Sibling Clearing and Root-Cause Fit}{\tool does not accept the savepoint finding until every family on the checklist reaches a terminal state with repository-grounded justification, ruling out the possibility that a stronger or more central boundary was missed.}{%
\turnplaybook\ \emph{``Every family in the checklist must end in a terminal state''}; \emph{``Do not start writing a final report or reproduction bundle until the checklist names the accepted component family, at least one sibling primitive in that family that you directly cleared, and any higher-signal family you deliberately pruned with repository-grounded justification''}.\par
\turnreasoning\ I need the savepoint path to remain the strongest repository-owned boundary after comparing PRAGMA---the most plausible competitor, since it also converts non-SQL identifiers into SQL---along with \texttt{Name} helpers, raw SQL APIs, and optional feature paths.\par
\turnoperation\ \tool clears PRAGMA through its own keyword allowlist and identifier/literal escaping; clears \texttt{Name} helpers because they pass C strings to SQLite rather than constructing SQL text; prunes raw SQL as intended caller capability; and prunes optional feature surfaces as lower-centrality or explicitly unsafe.\par
\turnsignal\ Savepoints remain the only unprotected boundary after all sibling families are cleared or pruned.
}

\walkentry{Turn 5: PoC Validation and Artifact Finalization}{\tool materializes the accepted path as a pinned, replayable proof. The payload exploits a property of the savepoint lifecycle: because the name appears in both \texttt{SAVEPOINT} and \texttt{RELEASE}, a single injected \texttt{INSERT} executes twice.}{%
\turnplaybook\ \emph{``Do not start writing a reproduction bundle until you know the exact pinned revision, dependency installation path, and proof signal''}; \emph{``keep the reproduction as narrow as the proof''}; \emph{``rerun it end to end from the produced \texttt{run.sh}''}; \emph{``concrete security impact tied to the reachable query''}.\par
\turnreasoning\ I use an in-process SQLite harness because the decisive proof is a database mutation caused by the savepoint name alone, not by surrounding application behavior.\par
\turnoperation\ \tool invokes \texttt{savepoint\_\allowbreak with\_\allowbreak name} with payload \texttt{sp1;\allowbreak\ INSERT INTO audit\allowbreak(message)\allowbreak\ VALUES\allowbreak('owned');\allowbreak\ --}, packages the harness under \texttt{sql\_\allowbreak injection\_\allowbreak savepoint\_\allowbreak name/\allowbreak reproduction/}, and reruns \texttt{./\allowbreak reproduction/\allowbreak run.sh} against the pinned vulnerable checkout.\par
\turnsignal\ The verifier prints \texttt{count=2} and \texttt{messages=\allowbreak owned,\allowbreak owned}; both reports pass.
}

%% file: main.bbl
\begin{thebibliography}{10}
\providecommand{\url}[1]{#1}
\csname url@samestyle\endcsname
\providecommand{\newblock}{\relax}
\providecommand{\bibinfo}[2]{#2}
\providecommand{\BIBentrySTDinterwordspacing}{\spaceskip=0pt\relax}
\providecommand{\BIBentryALTinterwordstretchfactor}{4}
\providecommand{\BIBentryALTinterwordspacing}{\spaceskip=\fontdimen2\font plus
\BIBentryALTinterwordstretchfactor\fontdimen3\font minus \fontdimen4\font\relax}
\providecommand{\BIBforeignlanguage}[2]{{%
\expandafter\ifx\csname l@#1\endcsname\relax
\typeout{** WARNING: IEEEtran.bst: No hyphenation pattern has been}%
\typeout{** loaded for the language `#1'. Using the pattern for}%
\typeout{** the default language instead.}%
\else
\language=\csname l@#1\endcsname
\fi
#2}}
\providecommand{\BIBdecl}{\relax}
\BIBdecl

\bibitem{zhang2025cybench}
\BIBentryALTinterwordspacing
A.~K. Zhang, N.~Perry, R.~Dulepet, J.~Ji, C.~Menders, J.~Lin, E.~Jones, G.~Hussein, S.~Liu, D.~Jasper, P.~Peetathawatchai, A.~Glenn, V.~Sivashankar, D.~Zamoshchin, L.~Glikbarg, D.~Askaryar, H.~Yang, A.~Zhang, R.~Alluri, N.~Tran, R.~Sangpisit, K.~Oseleononmen, D.~Boneh, D.~Ho, and P.~Liang, ``Cybench: A framework for evaluating cybersecurity capabilities and risks of language models,'' in \emph{The Thirteenth International Conference on Learning Representations (ICLR 2025)}, 2025. [Online]. Available: \url{https://proceedings.iclr.cc/paper_files/paper/2025/hash/3e9412a9c1d93810ef3ef7825115016b-Abstract-Conference.html}
\BIBentrySTDinterwordspacing

\bibitem{wang2026cybergym}
\BIBentryALTinterwordspacing
Z.~Wang, T.~Shi, J.~He, M.~Cai, J.~Zhang, and D.~Song, ``Cybergym: Evaluating {AI} agents' real-world cybersecurity capabilities at scale,'' in \emph{The Fourteenth International Conference on Learning Representations (ICLR 2026)}, 2026. [Online]. Available: \url{https://openreview.net/forum?id=2YvbLQEdYt}
\BIBentrySTDinterwordspacing

\bibitem{wang2026exploitgym}
\BIBentryALTinterwordspacing
Z.~Wang, N.~Schiller, H.~Li, S.~S. Narayana, M.~Nasr, N.~Carlini, X.~Qi, E.~Wallace, E.~Bursztein, L.~Invernizzi, K.~Thomas, Y.~Shoshitaishvili, W.~Guo, J.~He, T.~Holz, and D.~Song, ``Exploitgym: Can {AI} agents turn security vulnerabilities into real attacks?'' 2026. [Online]. Available: \url{https://arxiv.org/abs/2605.11086}
\BIBentrySTDinterwordspacing

\bibitem{deng2024pentestgpt}
\BIBentryALTinterwordspacing
G.~Deng, Y.~Liu, V.~Mayoral{-}Vilches, P.~Liu, Y.~Li, Y.~Xu, T.~Zhang, Y.~Liu, M.~Pinzger, and S.~Rass, ``{PentestGPT}: Evaluating and harnessing large language models for automated penetration testing,'' in \emph{33rd USENIX Security Symposium (USENIX Security 24)}.\hskip 1em plus 0.5em minus 0.4em\relax USENIX Association, 2024, pp. 847--864. [Online]. Available: \url{https://www.usenix.org/conference/usenixsecurity24/presentation/deng}
\BIBentrySTDinterwordspacing

\bibitem{wang2025a2}
\BIBentryALTinterwordspacing
Z.~Wang and L.~Zhou, ``Agentic discovery and validation of android app vulnerabilities,'' 2025. [Online]. Available: \url{https://arxiv.org/abs/2508.21579}
\BIBentrySTDinterwordspacing

\bibitem{allen2024webrr}
\BIBentryALTinterwordspacing
J.~Allen, Z.~Yang, F.~Xiao, M.~Landen, R.~Perdisci, and W.~Lee, ``{WEBRR}: A forensic system for replaying and investigating {Web-Based} attacks in the modern web,'' in \emph{33rd USENIX Security Symposium (USENIX Security 24)}.\hskip 1em plus 0.5em minus 0.4em\relax USENIX Association, 2024, pp. 1669--1686. [Online]. Available: \url{https://www.usenix.org/conference/usenixsecurity24/presentation/allen}
\BIBentrySTDinterwordspacing

\bibitem{song2026protocolguard}
\BIBentryALTinterwordspacing
X.~Song, L.~Pei, J.~Wu, Y.~Zeng, G.~He, C.~Zuo, X.~Liu, Q.~Zhao, and S.~Guo, ``Protocolguard: Detecting protocol non-compliance bugs via {LLM-guided} static analysis and dynamic verification,'' in \emph{Proceedings of the 33rd Annual Network and Distributed System Security Symposium (NDSS 2026)}, 2026. [Online]. Available: \url{https://www.ndss-symposium.org/ndss-paper/protocolguard-detecting-protocol-non-compliance-bugs-via-llm-guided-static-analysis-and-dynamic-verification/}
\BIBentrySTDinterwordspacing

\bibitem{ji2026firmagent}
\BIBentryALTinterwordspacing
J.~Ji, C.~Zhang, S.~Gan, L.~Jian, H.~Liu, T.~Liu, L.~Zheng, and Z.~Jia, ``Firmagent: Leveraging fuzzing to assist {LLM} agents with {IoT} firmware vulnerability discovery,'' in \emph{Proceedings of the 33rd Annual Network and Distributed System Security Symposium (NDSS 2026)}, 2026. [Online]. Available: \url{https://www.ndss-symposium.org/wp-content/uploads/2026-s1943-paper.pdf}
\BIBentrySTDinterwordspacing

\bibitem{ding2025primevul}
\BIBentryALTinterwordspacing
Y.~Ding, Y.~Fu, O.~Ibrahim, C.~Sitawarin, X.~Chen, B.~Alomair, D.~Wagner, B.~Ray, and Y.~Chen, ``Vulnerability detection with code language models: How far are we?'' in \emph{2025 IEEE/ACM 47th International Conference on Software Engineering (ICSE)}, 2025, pp. 1729--1741. [Online]. Available: \url{https://arxiv.org/abs/2403.18624}
\BIBentrySTDinterwordspacing

\bibitem{wang2024reposvul}
\BIBentryALTinterwordspacing
X.~Wang, R.~Hu, C.~Gao, X.-C. Wen, Y.~Chen, and Q.~Liao, ``{ReposVul}: A repository-level high-quality vulnerability dataset,'' in \emph{Proceedings of the 2024 IEEE/ACM 46th International Conference on Software Engineering: Companion Proceedings}, 2024, pp. 472--483. [Online]. Available: \url{https://conf.researchr.org/details/icse-2024/icse-2024-industry-challenge-track/2/ReposVul-A-Repository-Level-High-Quality-Vulnerability-Dataset}
\BIBentrySTDinterwordspacing

\bibitem{guo2025repoaudit}
\BIBentryALTinterwordspacing
J.~Guo, C.~Wang, X.~Xu, Z.~Su, and X.~Zhang, ``{RepoAudit}: An autonomous {LLM-Agent} for repository-level code auditing,'' in \emph{Proceedings of the 42nd International Conference on Machine Learning}, ser. Proceedings of Machine Learning Research, vol. 267.\hskip 1em plus 0.5em minus 0.4em\relax PMLR, 2025, pp. 21\,083--21\,100. [Online]. Available: \url{https://proceedings.mlr.press/v267/guo25n.html}
\BIBentrySTDinterwordspacing

\bibitem{peng2026reasonvul}
\BIBentryALTinterwordspacing
X.~Peng, B.~Lin, J.~Wang, X.~Li, J.~Ma, J.~Yu, X.~Mao, and S.~Wang, ``Three heads are better than one: A multi-perspective reasoning framework for enhanced vulnerability detection,'' in \emph{Proceedings of the ACM on Software Engineering, FSE}, 2026, accepted paper; final proceedings entry to be verified. [Online]. Available: \url{https://conf.researchr.org/details/fse-2026/fse-2026-research-papers/175/Three-Heads-Are-Better-Than-One-A-Multi-Perspective-Reasoning-Framework-for-Enhanced}
\BIBentrySTDinterwordspacing

\bibitem{anthropic2026glasswing}
{Anthropic}, ``Project glasswing: Securing critical software for the {AI} era,'' \url{https://www.anthropic.com/glasswing}, 2026, accessed: 2026-06-02.

\bibitem{openai2026daybreak}
{OpenAI}, ``Daybreak: {OpenAI} for cybersecurity,'' \url{https://openai.com/daybreak}, 2026, accessed: 2026-06-02.

\bibitem{openai2026codexsecurity}
------, ``{Codex Security},'' \url{https://developers.openai.com/codex/security}, 2026, accessed: 2026-05-30.

\bibitem{sutton2019bitterlesson}
R.~S. Sutton, ``The bitter lesson,'' \url{http://www.incompleteideas.net/IncIdeas/BitterLesson.html}, 2019, accessed: 2026-05-28.

\bibitem{zai2026glm51}
{Z.AI}, ``{GLM5.1},'' \url{https://docs.z.ai/guides/llm/glm-5.1}, 2026, accessed: 2026-05-29.

\bibitem{opencode2026}
{Anomaly}, ``{OpenCode}: The open source {AI} coding agent,'' \url{https://opencode.ai/}, 2026, accessed: 2026-05-29.

\bibitem{qwen2026qwen36a3b}
{Qwen Team}, ``{Qwen3.6-35B-A3B},'' \url{https://qwen.ai/blog?id=qwen3.6-35b-a3b}, 2026, accessed: 2026-06-02.

\bibitem{qwen2026qwen3627b}
------, ``{Qwen3.6-27B},'' \url{https://huggingface.co/Qwen/Qwen3.6-27B}, 2026, accessed: 2026-06-02.

\bibitem{firstCVSSv40}
{Forum of Incident Response and Security Teams}, ``Common vulnerability scoring system version 4.0: Specification document,'' \url{https://www.first.org/cvss/v4.0/specification-document}, 2023, accessed: 2026-05-28.

\bibitem{mitreCWE}
{MITRE}, ``Common weakness enumeration,'' \url{https://cwe.mitre.org/}, 2026, accessed: 2026-05-30.

\bibitem{githubAdvisoryDatabase}
{GitHub}, ``Github advisory database,'' \url{https://github.com/advisories}, 2026, accessed: 2026-05-28.

\bibitem{mccloskey1989catastrophic}
M.~McCloskey and N.~J. Cohen, ``Catastrophic interference in connectionist networks: The sequential learning problem,'' in \emph{Psychology of Learning and Motivation}.\hskip 1em plus 0.5em minus 0.4em\relax Academic Press, 1989, vol.~24, pp. 109--165.

\bibitem{parisi2019continual}
G.~I. Parisi, R.~Kemker, J.~L. Part, C.~Kanan, and S.~Wermter, ``Continual lifelong learning with neural networks: A review,'' \emph{Neural Networks}, vol. 113, pp. 54--71, 2019.

\bibitem{pan2010transfer}
S.~J. Pan and Q.~Yang, ``A survey on transfer learning,'' \emph{IEEE Transactions on Knowledge and Data Engineering}, vol.~22, no.~10, pp. 1345--1359, 2010.

\bibitem{david2025multi}
I.~David and A.~Gervais, ``Multi-agent penetration testing ai for the web,'' \emph{arXiv preprint arXiv:2508.20816}, 2025.

\bibitem{liu2024chatgptvulnmgmt}
\BIBentryALTinterwordspacing
P.~Liu, J.~Liu, L.~Fu, K.~Lu, Y.~Xia, X.~Zhang, W.~Chen, H.~Weng, S.~Ji, and W.~Wang, ``Exploring {ChatGPT}'s capabilities on vulnerability management,'' in \emph{33rd USENIX Security Symposium (USENIX Security 24)}.\hskip 1em plus 0.5em minus 0.4em\relax USENIX Association, 2024, pp. 811--828. [Online]. Available: \url{https://www.usenix.org/conference/usenixsecurity24/presentation/liu-peiyu}
\BIBentrySTDinterwordspacing

\bibitem{wang2026txray}
\BIBentryALTinterwordspacing
Z.~Wang, J.~Yu, K.~Qin, D.~Song, A.~Gervais, and L.~Zhou, ``Txray: Agentic postmortem of live blockchain attacks,'' 2026. [Online]. Available: \url{https://arxiv.org/abs/2602.01317}
\BIBentrySTDinterwordspacing

\bibitem{zhu2026teams}
\BIBentryALTinterwordspacing
Y.~Zhu, A.~Kellermann, A.~Gupta, P.~Li, R.~Fang, R.~Bindu, and D.~Kang, ``Teams of {LLM} agents can exploit zero-day vulnerabilities,'' in \emph{Proceedings of the 19th Conference of the European Chapter of the Association for Computational Linguistics (Volume 1: Long Papers)}.\hskip 1em plus 0.5em minus 0.4em\relax Association for Computational Linguistics, 2026, pp. 23--35. [Online]. Available: \url{https://aclanthology.org/2026.eacl-long.2/}
\BIBentrySTDinterwordspacing

\bibitem{meng2024chatafl}
\BIBentryALTinterwordspacing
R.~Meng, M.~Mirchev, M.~B{\"o}hme, and A.~Roychoudhury, ``Large language model guided protocol fuzzing,'' in \emph{Proceedings of the 31st Annual Network and Distributed System Security Symposium (NDSS 2024)}, 2024. [Online]. Available: \url{https://www.ndss-symposium.org/ndss-paper/large-language-model-guided-protocol-fuzzing/}
\BIBentrySTDinterwordspacing

\bibitem{david2026towards}
I.~David and A.~Gervais, ``Towards optimal agentic architectures for offensive security tasks,'' \emph{arXiv preprint arXiv:2604.18718}, 2026.

\bibitem{wen2024vuleval}
\BIBentryALTinterwordspacing
X.-C. Wen, X.~Wang, Y.~Chen, R.~Hu, D.~Lo, and C.~Gao, ``{VulEval}: Towards repository-level evaluation of software vulnerability detection,'' 2024. [Online]. Available: \url{https://arxiv.org/abs/2404.15596}
\BIBentrySTDinterwordspacing

\bibitem{yildiz2025jitvul}
\BIBentryALTinterwordspacing
A.~Yildiz, S.~G. Teo, Y.~Lou, Y.~Feng, C.~Wang, and D.~M. Divakaran, ``Benchmarking {LLM}s and {LLM-Based} agents in practical vulnerability detection for code repositories,'' in \emph{Proceedings of the 63rd Annual Meeting of the Association for Computational Linguistics (Volume 1: Long Papers)}.\hskip 1em plus 0.5em minus 0.4em\relax Association for Computational Linguistics, 2025, pp. 30\,848--30\,865. [Online]. Available: \url{https://aclanthology.org/2025.acl-long.1490/}
\BIBentrySTDinterwordspacing

\bibitem{li2025iris}
\BIBentryALTinterwordspacing
Z.~Li, S.~Dutta, and M.~Naik, ``{IRIS}: {LLM-Assisted} static analysis for detecting security vulnerabilities,'' in \emph{The Thirteenth International Conference on Learning Representations (ICLR 2025)}, 2025. [Online]. Available: \url{https://openreview.net/forum?id=9LdJDU7E91}
\BIBentrySTDinterwordspacing

\bibitem{lekssays2025llmxcpg}
\BIBentryALTinterwordspacing
A.~Lekssays, H.~Mouhcine, K.~Tran, T.~Yu, and I.~Khalil, ``{LLMxCPG}: {Context-Aware} vulnerability detection through code property {Graph-Guided} large language models,'' in \emph{34th USENIX Security Symposium (USENIX Security 25)}.\hskip 1em plus 0.5em minus 0.4em\relax USENIX Association, 2025, pp. 489--507. [Online]. Available: \url{https://www.usenix.org/conference/usenixsecurity25/presentation/lekssays}
\BIBentrySTDinterwordspacing

\bibitem{wang2024llmdfa}
\BIBentryALTinterwordspacing
C.~Wang, W.~Zhang, Z.~Su, X.~Xu, X.~Xie, and X.~Zhang, ``{LLMDFA}: Analyzing dataflow in code with large language models,'' in \emph{Advances in Neural Information Processing Systems 37 (NeurIPS 2024)}, 2024. [Online]. Available: \url{https://proceedings.neurips.cc/paper_files/paper/2024/file/ed9dcde1eb9c597f68c1d375bbecf3fc-Paper-Conference.pdf}
\BIBentrySTDinterwordspacing

\bibitem{wang2024llmsan}
\BIBentryALTinterwordspacing
C.~Wang, W.~Zhang, Z.~Su, X.~Xu, and X.~Zhang, ``Sanitizing large language models in bug detection with data-flow,'' in \emph{Findings of the Association for Computational Linguistics: EMNLP 2024}.\hskip 1em plus 0.5em minus 0.4em\relax Association for Computational Linguistics, 2024, pp. 3790--3805. [Online]. Available: \url{https://aclanthology.org/2024.findings-emnlp.217/}
\BIBentrySTDinterwordspacing

\bibitem{liu2025gptaid}
\BIBentryALTinterwordspacing
J.~Liu, Y.~Yang, K.~Chen, and M.~Lin, ``Generating {API} parameter security rules with {LLM} for {API} misuse detection,'' in \emph{Proceedings of the 32nd Annual Network and Distributed System Security Symposium (NDSS 2025)}, 2025. [Online]. Available: \url{https://www.ndss-symposium.org/ndss-paper/generating-api-parameter-security-rules-with-llm-for-api-misuse-detection/}
\BIBentrySTDinterwordspacing

\bibitem{yang2025midas}
\BIBentryALTinterwordspacing
Y.~Yang, J.~Liu, K.~Chen, and M.~Lin, ``The midas touch: Triggering the capability of {LLMs} for {RM-API} misuse detection,'' in \emph{Proceedings of the 32nd Annual Network and Distributed System Security Symposium (NDSS 2025)}, 2025. [Online]. Available: \url{https://dev.ndss-symposium.org/ndss-paper/the-midas-touch-triggering-the-capability-of-llms-for-rm-api-misuse-detection/}
\BIBentrySTDinterwordspacing

\bibitem{lee2026exploitbench}
\BIBentryALTinterwordspacing
S.~Lee and D.~Brumley, ``Exploitbench: A capability ladder benchmark for {LLM} cybersecurity agents,'' 2026. [Online]. Available: \url{https://arxiv.org/abs/2605.14153}
\BIBentrySTDinterwordspacing

\bibitem{pellew2026realvuln}
\BIBentryALTinterwordspacing
J.~Pellew and F.~Raza, ``Realvuln: Benchmarking rule-based, general-purpose {LLM}, and security-specialized scanners on real-world code,'' 2026. [Online]. Available: \url{https://arxiv.org/abs/2604.13764}
\BIBentrySTDinterwordspacing

\bibitem{du2024vulrag}
\BIBentryALTinterwordspacing
X.~Du, G.~Zheng, K.~Wang, J.~Feng, W.~Deng, M.~Liu, B.~Chen, X.~Peng, T.~Ma, and Y.~Lou, ``{Vul-RAG}: Enhancing {LLM-based} vulnerability detection via knowledge-level {RAG},'' 2024. [Online]. Available: \url{https://arxiv.org/abs/2406.11147}
\BIBentrySTDinterwordspacing

\bibitem{guo2025bugscope}
\BIBentryALTinterwordspacing
J.~Guo, C.~Wang, D.~Deluca, J.~Liu, Z.~Zhang, and X.~Zhang, ``{BugScope}: Learn to find bugs like human,'' 2025. [Online]. Available: \url{https://arxiv.org/abs/2507.15671}
\BIBentrySTDinterwordspacing

\bibitem{zhao2023expel}
\BIBentryALTinterwordspacing
A.~Zhao, D.~Huang, Q.~Xu, M.~Lin, Y.-J. Liu, and G.~Huang, ``Expel: Llm agents are experiential learners,'' 2023. [Online]. Available: \url{https://arxiv.org/abs/2308.10144}
\BIBentrySTDinterwordspacing

\bibitem{zhou2025memento}
\BIBentryALTinterwordspacing
H.~Zhou, Y.~Chen, S.~Guo, X.~Yan, K.~H. Lee, Z.~Wang, K.~Y. Lee, G.~Zhang, K.~Shao, L.~Yang, and J.~Wang, ``Memento: Fine-tuning llm agents without fine-tuning llms,'' 2025. [Online]. Available: \url{https://arxiv.org/abs/2508.16153}
\BIBentrySTDinterwordspacing

\bibitem{luo2025agentlightning}
\BIBentryALTinterwordspacing
X.~Luo, Y.~Zhang, Z.~He, Z.~Wang, S.~Zhao, D.~Li, L.~K. Qiu, and Y.~Yang, ``Agent lightning: Train any ai agents with reinforcement learning,'' 2025. [Online]. Available: \url{https://arxiv.org/abs/2508.03680}
\BIBentrySTDinterwordspacing

\end{thebibliography}
